\documentclass{jpp}
\usepackage{lmodern}  
\usepackage[colorlinks=true,linkcolor=blue,citecolor=blue]{hyperref}
\usepackage{graphicx}
\graphicspath{{figs/}}
\usepackage[utf8]{inputenc}
\usepackage[T1]{fontenc}
\usepackage{amsmath}
\usepackage{dsfont}
\usepackage[section]{placeins}
\usepackage{multicol}
\usepackage{cleveref}
\usepackage{upgreek}
\usepackage{caption}
\usepackage{tikz}
\usepackage[normalem]{ulem}
\usetikzlibrary{positioning, decorations.text, decorations.pathmorphing, arrows.meta, shapes.multipart}
\crefformat{section}{\S#2#1#3} 
\crefformat{subsection}{\S#2#1#3}
\crefformat{subsubsection}{\S#2#1#3}
\crefformat{figure}{figure~#2#1#3}
\crefformat{equation}{(#2#1#3)}

\let\originalleft\left
\let\originalright\right
\renewcommand{\left}{\mathopen{}\mathclose\bgroup\originalleft}
\renewcommand{\right}{\aftergroup\egroup\originalright}

\newcommand{\vect}[1]{{\boldsymbol{#1}}}
\newcommand{\uvect}[1]{\hat{\vect{#1}}}

\newcommand{\order}[1]{\mathcal{O}\left(#1\right)}
\newcommand{\orderinline}[1]{\mathcal{O}(#1)}

\newcommand{\re}[1]{{\text{Re}{\left(#1\right)}}}
\newcommand{\reinline}[1]{{\text{Re}{(#1)}}}
\newcommand{\renobra}[1]{{\text{Re}{#1}}}
\newcommand{\im}[1]{{\text{Im}{\left(#1\right)}}}
\newcommand{\iminline}[1]{{\text{Im}{(#1)}}}

\newcommand{\lambdadisp}{{-i \omega_\vk}}
\newcommand{\omk}{\omega_\vk}
\newcommand{\omkh}{\hat{\omega}_\vk}
\newcommand{\omqh}{\hat{\omega}_\vq}
\newcommand{\gammakh}{\hat{\gamma}_\vk}
\newcommand{\kparh}{\hat{k}_\parallel}

\newcommand{\ak}{\alpha_\vk}
\newcommand{\bk}{\beta_\vk}

\newcommand{\modeleqnsnonnorm}{\cref{phiEq}--\cref{uEq}}
\newcommand{\modeleqns}{\cref{curvy_phi}--\cref{curvy_u}}

\newcommand{\modeleqnsslabavg}{\cref{curvy_slabavg_phi}--\cref{curvy_slabavg_u}}
\newcommand{\modeleqnsslabpert}{\cref{curvy_slabpert_phi}--\cref{curvy_slabpert_u}}
\newcommand{\modeleqnsslabpertsimple}{\cref{curvy_slabpert_phi_lowestorder}--\cref{curvy_slabpert_u_lowestorder}}
\newcommand{\modeleqnsslabavgsimple}{\cref{curvy_slabavg_phi_lowestorder}--\cref{curvy_slabavg_psi_lowestorder}}

\newcommand{\vti}{v_{\text{th}i}}
\newcommand{\rhoivti}{\rho_i \vti}

\newcommand{\ve}{\vect{V_\text{\textit{E}}}}
\newcommand{\exb}{\(\vect{E}\times\vect{B}\)}
\newcommand{\avgR}[1]{\left\langle #1 \right\rangle_\vect{R}}
\newcommand{\avgr}[1]{\left\langle #1 \right\rangle_\vect{r}}
\newcommand{\avgDeltat}[1]{\langle #1 \rangle_{\Delta t}}
\newcommand{\pbra}[2]{\left\{ #1, #2 \right\}}

\newcommand{\vpar}{{v_\parallel}}

\newcommand{\kperprhoisq}{k_\perp^2\rho_i^2}
\newcommand{\kperprhoisqq}{k_\perp^4\rho_i^4}
\newcommand{\rhoidelperpsq} {\rho_i^2\nabla_\perp^2}
\newcommand{\gp}{\mathcal{P}}
\newcommand{\vh}{\hat{v}}
\newcommand{\vparh}{\hat{v}_\parallel}
\newcommand{\vperph}{\hat{v}_\perp}

\newcommand{\pt}{\partial_t}
\newcommand{\py}{\partial_y}
\newcommand{\px}{\partial_x}
\newcommand{\pz}{\partial_\parallel}
\newcommand{\partd}[2]{\frac{\partial #1}{\partial #2}}

\newcommand{\del}{\vect{\nabla}_\perp}
\newcommand{\delsq}{\nabla_\perp^2}

\newcommand{\vk}{\vect{k}}

\newcommand{\kpar}{k_\parallel}
\newcommand{\kperp}{k_\perp}
\newcommand{\ksq}{{k_\perp^2}}
\newcommand{\kperpmaxflr}{k_{\perp, \text{max,FLR}}}
\newcommand{\kperpmaxcol}{k_{\perp, \text{max}, \chi}}
\newcommand{\kperpcol}{k_{\chi}}

\newcommand{\vq}{\vect{q}}
\newcommand{\vqperp}{\vect{q}_\perp}
\newcommand{\vqh}{\uvect{q}}
\newcommand{\qpar}{q_\parallel}
\newcommand{\qperp}{q_\perp}

\newcommand{\intr}{{\int d^3 \vect{r} \ }}

\newcommand{\intv}{{\int d^3 \vect{v} \ }}

\newcommand{\dr}{d^3\vect{r}}
\newcommand{\dv}{d^3\vect{v}}
\newcommand{\intdvparh}{\int_{-\infty}^{+\infty}d\vparh \ }
\newcommand{\intdvperph}{\int_0^{+\infty}d(\vperph^2)}

\newcommand{\phinorm}{\varphi}
\newcommand{\phinonnorm}{\phi}
\newcommand{\pr}{p}
\newcommand{\vt}{\kappa_T}
\newcommand{\vtvect}{\vect{\kappa}_T}
\newcommand{\vnvect}{\vect{\kappa}_n}
\newcommand{\vuvect}{\vect{\kappa}_u}
\newcommand{\deltaT}{T}
\newcommand{\deltaVpar}{u_\parallel}
\newcommand{\upar}{{u}}

\newcommand{\vtcrittd}{\vt^{c, \text{2D}}}

\newcommand{\vteff}{\vt^\text{eff}}

\newcommand{\zf}[1]{\overline{#1}}
\newcommand{\dw}[1]{#1'}

\newcommand{\achi}{a}
\newcommand{\bchi}{b}
\newcommand{\cchi}{s}

\newcommand{\phinormslab}{\hat{\varphi}}

\newcommand{\uparslab}{\hat{u}}
\newcommand{\phinormslabpert}{\slabpert{\varphi}}
\newcommand{\deltaTslabpert}{\slabpert{\deltaT}}
\newcommand{\prslabpert}{\slabpert{\pr}}
\newcommand{\uparslabpert}{\slabpert{\upar}}

\newcommand{\slabavg}[1]{\left\langle #1 \right\rangle_\parallel}

\newcommand{\slabpert}[1]{\widetilde{#1}}
\newcommand{\Lpar}{L_\parallel}
\newcommand{\Lparcrit}{L_\parallel^\text{c}}

\newcommand{\Qpert}{\slabpert{\vect{Q}}}
\newcommand{\Qpertslabavg}{\langle\Qpert\rangle_\parallel}
\newcommand{\Pipert}{\slabpert{\Uppi}}
\newcommand{\Pipertslabavg}{\langle\Pipert\rangle_\parallel}

\newcommand{\sgn}[1]{\text{sgn}\left(#1\right)}
\newcommand{\sgninline}[1]{\text{sgn}(#1)}

\shorttitle{Dimits transition in 3D}
\shortauthor{P. G. Ivanov, A. A. Schekochihin \& W. Dorland}

\title{Dimits transition in three-dimensional ion-temperature-gradient turbulence}

\author{Plamen~G.~Ivanov\aff{1,2,3}
  \corresp{\email{plamen.ivanov@physics.ox.ac.uk}},
  A.~A.~Schekochihin\aff{1,4}, and
  W.~Dorland\aff{1,5}}

\affiliation{\aff{1}Rudolf Peierls Centre for Theoretical Physics, University of Oxford, Oxford OX1 3PU, UK
\aff{2}St John's College, Oxford OX1 3JP, UK
\aff{3}EURATOM/UKAEA Fusion Association, Culham Science Centre, Abingdon, OX14 3DB, UK
\aff{4}Merton College, Oxford OX1 4JD, UK
\aff{5}Department of Physics, University of Maryland, College Park, Maryland 20740, USA}

\begin{document}

\maketitle

\begin{abstract}
We extend our previous work on the 2D Dimits transition in ion-scale turbulence \citep{ivanov2020} to include variations along the magnetic field. We consider a three-field fluid model for the perturbations of electrostatic potential, ion temperature, and ion parallel flow in a constant-magnetic-curvature geometry without magnetic shear. It is derived in the cold-ion, long-wavelength asymptotic limit of the gyrokinetic theory. Just as in the 2D model, a low-transport (Dimits) regime exists and is found to be dominated by a quasi-static staircase-like arrangement of strong zonal flows and zonal temperature. This zonal staircase is formed and maintained by a negative turbulent viscosity for the zonal flows. Unlike the 2D model, the 3D one does not suffer from an unphysical blow up beyond the Dimits threshold where the staircase becomes nonlinearly unstable. Instead, a well-defined finite-amplitude saturated state is established. This qualitative difference between 2D and 3D is due to the appearance of small-scale `parasitic' modes that exist only if we allow perturbations to vary along the magnetic field lines. These modes extract energy from the large-scale perturbations and provide an effective enhancement of large-scale thermal diffusion, thus aiding the energy transfer from large injection scales to small dissipative ones. We show that in our model, the parasitic modes always favour a zonal-flow-dominated state. In fact, a Dimits state with a zonal staircase is achieved regardless of the strength of the linear drive provided the system is sufficiently extended along the magnetic field and sufficient parallel resolution is provided. 
\end{abstract}

\section{Introduction}
\label{intro}

In our previous work \citep{ivanov2020}, we discussed the two-dimensional dynamics of ion-scale turbulence driven by the ion-temperature-gradient (ITG) instability in the plane perpendicular to the magnetic field. We identified the fundamental mechanism of the Dimits transition that demarcates saturation dominated by strong coherent zonal flows (ZFs) --- the `Dimits state' --- and the strongly turbulent regime where no coherent ZFs exist. The turbulent momentum flux of turbulence sheared by ZFs --- viz., whether the zonal `turbulent viscosity' was positive or negative --- was found to be the key to the demise of the Dimits state.

However, those findings were based on a simplified model (to which we shall here refer as the `2D model'), obtained as an asymptotic, highly collisional limit of ion gyrokinetics (GK), with the additional assumption that the dynamics were two-dimensional. This assumption cannot be justified asymptotically. In fact, GK studies of tokamak turbulence have revealed that parallel dynamics are linked to turbulence in the perpendicular plane via the `critical balance' between the nonlinear mixing time and the parallel propagation time \citep{barnes2011}. 

In this paper, we carry our work over to a more general model that is a true asymptotic limit of the GK equations by relaxing the two-dimensionality assumption to determine whether the three-dimensional Dimits transition is governed by the same mechanism as the two-dimensional one. In the highly collisional limit discussed in \citet{ivanov2020}, we obtain virtually the same equations for the perturbations of ion temperature and electric potential, with the addition of parallel dynamics and of a new equation for the perturbed parallel ion flow. These three equations (to which we refer as the `3D model') describe both of the classic ITG instabilities: one mediated by compression along the magnetic field, which we shall call the slab-ITG (sITG) instability \citep{rudakov61, coppi67, cowley91}, the other by magnetic curvature, which we shall call the curvature-driven ITG (cITG) instability \citep{pogutse68, guzdar83}. Note that we shall consider only the case of zero magnetic shear.

Our numerical results indicate that the Dimits-regime dynamics of the 3D model are essentially the same as those of the 2D model. Namely, we find that the Dimits regime is dominated by a quasi-static staircase-like arrangement of strong ZFs that rip and suppress turbulence. This zonal staircase, reminiscent of the so-called \exb{} staircase seen in global GK simulations \citep{pradalier2010, difpradalier2017, villard2013, rath2016}, slowly decays due to collisional viscosity. This viscous decay results in recurrent turbulent bursts that are triggered by localised travelling structures emerging from the ZF maxima, where they are created by a local (`tertiary') instability of the ZF profile. The turbulence that develops during a burst is sheared by the ZFs. Locally, the shear breaks the fundamental parity symmetry of GK turbulence \citep{parra2011, mfox_sym}. This gives rise to a radial flux of poloidal momentum whose sign is controlled by the sign of the zonal shear. This momentum flux consists of two parts --- the usual Reynolds stress of the \exb{} flow, which is known to generate strong ZFs \citep{diamond2005}, and a diamagnetic contribution, which is found to oppose the Reynolds stress. The distinguishing feature of the Dimits regime is that the Reynolds stress overcomes the diamagnetic one. The zonal staircase is stable to turbulent bursts because ZF-sheared turbulence provides an effective negative viscosity for the ZFs. All of these effects are found to be qualitatively identical between the 2D and 3D models.

The Dimits transition to higher turbulent transport occurs when the diamagnetic stress overcomes the Reynolds one, so the effective turbulent viscosity flips its sign and the coherent ZFs that support the Dimits state become nonlinearly unstable. The 2D model fails to reach finite-amplitude saturation in this state; instead, box-sized exponentially growing streamers emerge \citep{ivanov2020}. While such a blow up has not been observed in prior gyrokinetic studies of turbulence in a \(Z\)-pinch \citep{ricci2006, kobayashi2012}, it is not entirely unexpected in a 2D fluid system. The 3D fluid system does not suffer from such an unphysical blow up. Instead, a finite-amplitude saturated state without strong ZFs is established. This qualitative difference between the 3D and 2D models is due to the appearance of small-scale sITG modes, which exist only in the 3D model and are primarily driven by the temperature perturbations associated with the large-scale 2D perturbations (rather than by the equilibrium temperature gradient). These `parasitic' modes extract energy from those large-scale perturbations and transfer it to smaller perpendicular scales where it is dissipated, thus enabling the system to achieve saturation at finite amplitudes. The idea of such parasitic modes is hardly original \citep[see, e.g., ][]{drake88, cowley91, rath92}. We back their existence both by analytical arguments and by numerical results (\cref{sect_slabsec}) and show that their influence on the large-scale perturbations is to provide an effective enhancement to thermal diffusion (\cref{sect_largescale_resp}). 

The rest of the paper is organised as follows. In \cref{sect_3dmodel}, we discuss the 3D extension of the 2D model of \citet{ivanov2020}. Detailed derivations can be found in \cref{appendix_derivation}. \cref{sect_linear}~deals with the linear instability of the 3D model. Then, in \cref{sect_nl}, we describe the nonlinear saturated state: \cref{sect_dimits} is devoted to the 3D Dimits regime, \cref{sect_slabsec} to the small-scale sITG instability and to its role in both the Dimits and the strongly turbulent state. We summarise and discuss our results in \cref{sect_discussion}.

\section{Collisional, cold-ion \(Z\)-pinch in three dimensions}
\label{sect_3dmodel}

\subsection{Model equations}

\begin{figure}
	\centering
	\includegraphics[scale=0.35]{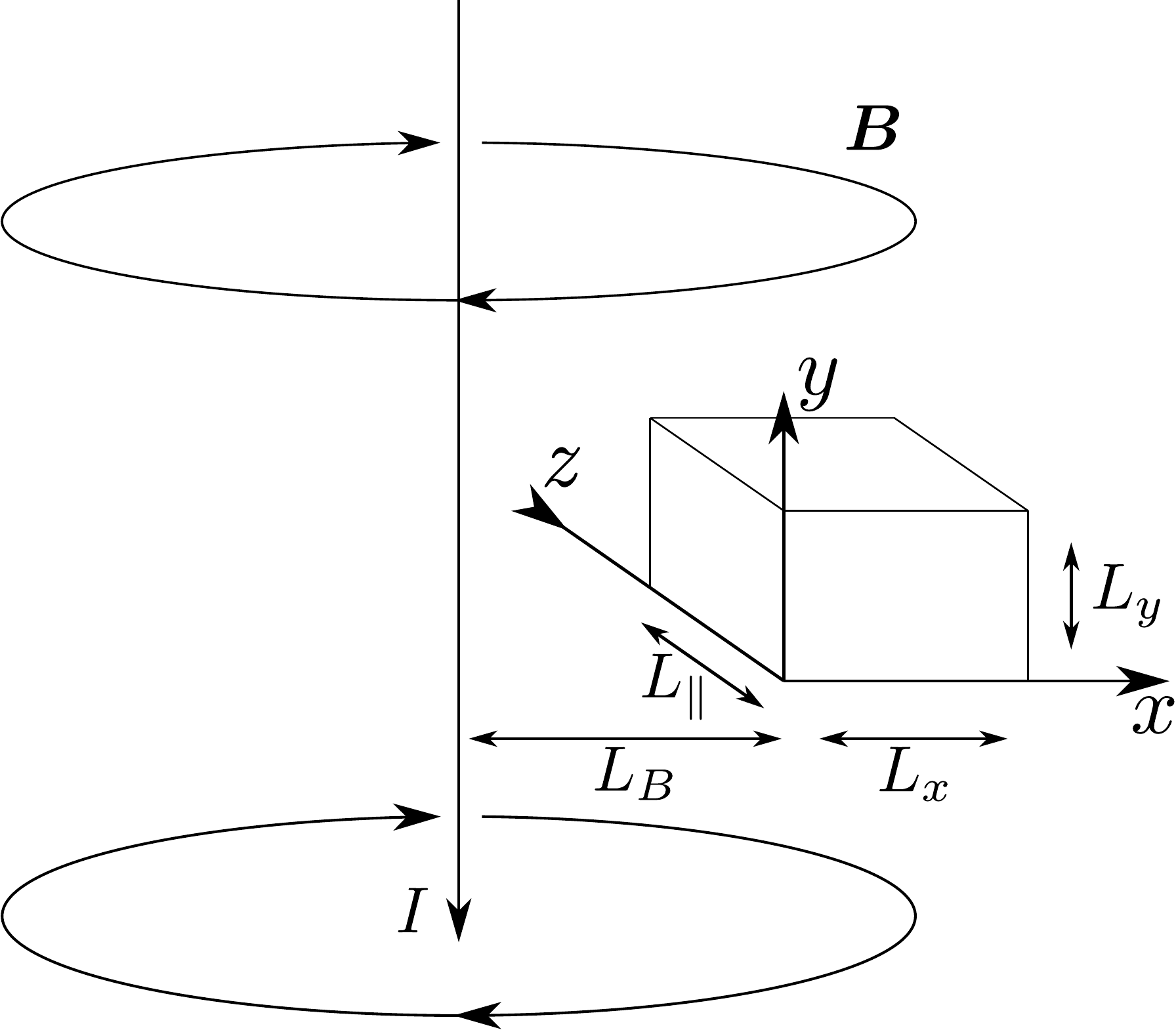}
	\caption{Illustration of the 3D \(Z\)-pinch magnetic geometry.}
	\label{fig_zpinch}
\end{figure}

The 3D model can be derived by following appendix A of \citet{ivanov2020}, with the addition of the 3D terms worked out in \cref{appendix_derivation} of the present paper. We consider a cold-ion plasma in \(Z\)-pinch magnetic geometry (shown in \cref{fig_zpinch}) with magnetic scale length \(L_B \equiv -\px \ln B \), where the magnetic field points in the \(z\) direction, \(\vect{B} = B \uvect{z}\), and \(x\) and \(y\) are the radial and poloidal coordinates, respectively. Here \(z\) is the coordinate around the current line of the \(Z\)-pinch (\(L_B\) times the azimuthal angle). The ITG scale length is defined as \(L_T \equiv - \px \ln T_i\), where \(T_i\) is the equilibrium ion temperature. We also assume a large-aspect-ratio system, viz., \(L_B \gg L_T\).\footnotemark\footnotetext{Otherwise we run into issues with the ordering of the magnetic drift in the cold-ion limit: see \mbox{equation~(A 27)} of \citet{ivanov2020}. }

The perturbed electron density \(\delta n_e\) is assumed to obey a modified adiabatic response \citep{dorland93,hammett93}
\begin{equation}
\label{eq_eresponse}
\frac{\delta n_e}{n_e} = \frac{e (\phinonnorm - \zf{\phinonnorm})}{T_e},
\end{equation}
where \(n_e\) is the equilibrium electron density, \(\phinonnorm\) is the electric potential, \(T_e\) is the electron temperature, and
\begin{equation}
\label{eq_zonal_def}
\zf{\phinonnorm}(x) \equiv \frac{1}{L_yL_z} \int dy dz \ \phinonnorm(x, y, z)
\end{equation}
is the zonal (flux-surface) spatial average of the perturbed electric potential \(\phinonnorm\). We refer to zonally averaged fields as `zonal fields'. We also define the nonzonal field \mbox{\(\dw{\phinonnorm} \equiv \phinonnorm - \zf{\phinonnorm}\)}. Even though, strictly speaking, there are no well-defined flux surfaces in a \(Z\)-pinch geometry, our aim is to model a tokamak-like system, thus our definition of a flux-surface average \cref{eq_zonal_def} is an average over both \(y\) and \(z\). This can be rationalised by the presence either of asymptotically small, but nonzero, magnetic shear \citep{ivanov2020}, or of asymptotically small irrational rotational transform. Note that neither of these is present in the final form of our equations.  

We take the density, temperature, and parallel-velocity moments of the electrostatic ion gyrokinetic equation and adopt the high-collisionality, cold-ion, long-wavelength, large-aspect-ratio ordering
\begin{equation}
\label{eq_ordering}
\frac{\pt}{\nu_i} \sim \tau \sim \kperprhoisq \sim \frac{L_T}{L_B} \ll 1, \quad \phinorm \sim \deltaT,
\end{equation}
where \(\phinorm \equiv Ze\phinonnorm / T_i \) is the normalised electric potential, \(Ze\) is the ion charge, \(\deltaT = \delta T / T_i\) is the normalised ion-temperature perturbation, \(\tau = T_i / ZT_e\) is the temperature ratio, \(\rho_i \equiv \vti / \Omega_i\) is the ion gyroradius given in terms of the ion thermal speed \(\vti \equiv \sqrt{2T_i/m_i}\) and the ion gyrofrequency \(\Omega_i \equiv ZeB/m_ic\), \(m_i\) is the ion mass, and \(\nu_i\) is the ion-ion collision frequency \citep[for an exact definition of \(\nu_i\), see appendix~A.1 of ][]{ivanov2020}. The resulting equations are 
\begin{align}
\label{phiEq}
& \frac{\partial}{\partial t} \left( \tau \phinorm' - \frac{1}{2} \rhoidelperpsq \phinorm\right) +\frac{\partial \deltaVpar}{\partial z} -\frac{\rhoivti}{L_B} \frac{\partial}{\partial y} \left( \phinorm + \deltaT \right) + \frac{\rhoivti}{2L_T} \frac{\partial}{\partial y} \left( \frac{1}{2} \rhoidelperpsq \phinorm \right)
\\& \quad + \frac{1}{2} \rhoivti \bigg( \pbra{\phinorm}{\tau\phinorm' - \frac{1}{2} \rhoidelperpsq \phinorm} + \frac{1}{2}\rho_i^2 \boldsymbol{\nabla_\perp} \bcdot \pbra{\boldsymbol{\nabla_\perp}\phinorm}{\deltaT} \bigg) \notag
\\& \quad = - \frac{1}{2} \chi \rho_i^2 \nabla_\perp^4 (\achi\phinorm - \bchi\deltaT), \notag
\label{psiEq}
\\ &\frac{\partial\deltaT}{\partial t} + \frac{5}{2} \frac{\partial \deltaVpar}{\partial z}  + \frac{\rhoivti}{2L_T} \frac{\partial\phinorm}{\partial y}  + \frac{1}{2} \rhoivti \pbra{\phinorm}{\deltaT} = \chi \nabla_\perp^2 \deltaT,
\\ &\frac{\partial \deltaVpar}{\partial t} + \frac{\vti^2}{2}\frac{\partial(\phinorm + \deltaT)}{\partial z} + \frac{1}{2} \rhoivti \pbra{\phinorm}{\deltaVpar} = \cchi \chi \nabla_\perp^2 \deltaVpar,
\label{uEq}
\end{align}
where the Poisson bracket is defined by
\begin{equation}
\label{eq_pbra_def}
\pbra{f}{g} = \uvect{b} \bcdot \left( \del f \times \del g \right) = \frac{\partial f}{\partial x} \frac{\partial g}{\partial y} - \frac{\partial f}{\partial y} \frac{\partial g}{\partial x}
\end{equation}
and \(\del \equiv \px \uvect{x} + \py \uvect{y}\) denotes the gradient operator in the perpendicular plane. The values of the thermal diffusivity \(\chi\) and the numerical constants \(\achi = 9/40\), \(\bchi = 67/160\), \(\cchi = 9/10\) are determined by the collisional operator, for which we have used the linearised Landau collision integral. We have omitted the magnetic-drift terms in \eqref{psiEq} and \cref{uEq} because those are an order \(L_T / L_B \sim \orderinline{\kperprhoisq} \ll 1\) smaller than the rest of the terms in their respective equations. The derivations of \cref{phiEq} and \cref{psiEq} can be found in \citet{ivanov2020}; the equation \cref{uEq} for the evolution of the parallel flow velocity is derived in \cref{appendix_derivation}. Note that we are yet to order the (inverse) parallel scale \(\kpar \sim \partial_z\) and flow velocity \(\deltaVpar\), so we have kept parallel streaming in all three equations.

Let us discuss briefly the physics of the `new' (compared to the 2D model) terms in \modeleqnsnonnorm{}. The terms \(\propto \partial_z \deltaVpar\) in \cref{phiEq} and \cref{psiEq} describe the compressions and rarefactions due to the parallel ion flow. Equation \cref{uEq} has a straightforward interpretation --- the parallel flow is driven by the parallel gradient of the pressure \(\pr = \phinorm + \deltaT\), advected by the \exb{} flow \(\ve = c\uvect{b}\times\del\phinonnorm / B\), and damped by the collisional viscosity \(\cchi\chi\). 

We would like to find an ordering for \(\kpar\) and \(\deltaVpar\) that allows for both sITG and cITG. The former depends on the presence of the parallel-streaming terms in \cref{phiEq} and \cref{uEq}. Thus, we require
\begin{equation}
	\label{eq_kpar_ordering_eq}
	\omega \tau \phinorm \sim \kpar \deltaVpar, \quad \omega \deltaVpar \sim \vti^2 \kpar \phinorm \implies \omega^2 \tau \sim \vti^2 \kpar^2,
\end{equation}
where \(\omega \sim \pt\) is the inverse time scale. We want to retain the curvature-driven instability, so we order \(\omega \sim \rho_s \Omega_i / L_B\), where \(\Omega_i\) is the ion gyrofrequency. Then \cref{eq_kpar_ordering_eq} implies 
\begin{equation}
	\kpar \sim L_B^{-1}, \quad \deltaVpar \sim \tau c_s \phinorm,
\end{equation}
where \(\rho_s \equiv \rho_i / \sqrt{2\tau}\) is the sound radius and \(c_s \equiv \rho_s \Omega_i\) is the sound speed.

We now introduce the following normalisations (consistent with those that we used for our 2D model):
\begin{equation}
\label{eq_normalisations}
\begin{gathered}
\hat{t} \equiv \frac{2\rho_s\Omega_i}{L_B}t, \qquad
\hat{x} \equiv \frac{x}{\rho_s}, \qquad \hat{y} \equiv \frac{y}{\rho_s}, \qquad \hat{z} \equiv \frac{2z}{L_B} \\
\hat{\phinorm} \equiv \frac{\tau L_B\phinorm}{2\rho_s}  = \frac{\tau L_B}{2\rho_s} \frac{Ze\phinonnorm}{T_i}, \qquad 
\hat{\deltaT} \equiv \frac{\tau L_B\deltaT}{2\rho_s}  = \frac{\tau L_B}{2\rho_s} \frac{\delta T}{T_i}, \qquad
\hat{u} \equiv \frac{\deltaVpar}{\rho_s \Omega_i \tau} \\
\vt \equiv \frac{\tau L_B}{2 L_T}, \qquad
\hat{\chi} \equiv \frac{L_B}{2\rho_s} \frac{\chi}{\Omega_i \rho_s^2}.
\end{gathered}
\end{equation}
All hatted quantities are ordered as \(\orderinline{1}\). Dropping hats, we obtain from \modeleqnsnonnorm{} the following equations in normalised units:

\begin{gather}
\partial_t \left( \dw{\phinorm} - \delsq \phinorm\right) + \pz \upar - \partial_y \left( \phinorm + \deltaT \right) + \vt \partial_y  \delsq \phinorm \nonumber \\ +\pbra{\phinorm}{\phinorm' - \delsq \phinorm}  
+ \del \bcdot \pbra{\del \phinorm}{\deltaT} 
=-\chi \nabla_\perp^4 (\achi\phinorm - \bchi\deltaT), \label{curvy_phi} \\
\label{curvy_psi}
\partial_t\deltaT +\vt\partial_y \phinorm +\pbra{\phinorm}{\deltaT} = \chi \delsq \deltaT, \\
\label{curvy_u}
 \pt \upar + \pz (\phinorm + \deltaT) + \pbra{\phinorm}{\upar} = \cchi \chi \delsq \upar,
\end{gather}
where \cref{curvy_psi} has lost its parallel-streaming term because it is \(\orderinline{\tau}\) smaller than the other terms, and we use \(\pz \equiv \partial_z\). These equations have two independent parameters: the normalised equilibrium temperature gradient, \(\vt\), and the normalised collisionality, \(\chi\). There are three other parameters --- \(L_x\), \(L_y\), and \(\Lpar\) that are the domain sizes in \(x\), \(y\) (in units of \(\rho_s\)), and \(z\) (in units of \(L_B/2\)), respectively. We have already seen that the physics of the 2D model is independent of \(L_x\) and \(L_y\) \citep{ivanov2020}, and that will be true for the 3D model as well, so the interesting one is \(\Lpar\). As we shall later see, the saturated state is independent of \(\Lpar\) if \(\Lpar\) is large enough, but if it is not, it will play a nontrivial role. Even though the \(Z\)-pinch geometry imposes a natural \(\Lpar\), viz., \(\Lpar = 4\pi\) (dimensionally this is \(2\pi L_B\)), we will not limit ourselves to that. By considering \(\Lpar\) as an independent parameter, we are able to model a shearless flux tube with constant magnetic drifts, periodic boundary conditions, and connection length \(\Lpar\). Varying \(\Lpar\) in our model is akin to varying the connection length \(2\pi qR\) in toroidal geometry, where \(q\) is the safety factor and \(R\) is the major radius. 

\subsection{Conservation laws}
\label{sect_cons}

The 2D cold-ion \(Z\)-pinch system has three nonlinear invariants \citep{ivanov2020}. One is the gyrokinetic free energy, while the other two result from the so-called `general 2D invariants' of GK \citep{schekochihingk2009}. The conservation law of free energy for the 3D equations \modeleqns{} is equivalent (modulo the integration domain) to that of the 2D equations. It reads
\begin{equation}
	\label{eq_free_e_cons}
	L_xL_y\Lpar\pt W \equiv \pt \int \dr \ \frac{1}{2} \deltaT^2 = -\vt \int \dr \ \deltaT \py \phinorm - \chi \int \dr \ \left( \del \deltaT \right)^2.
\end{equation}
The first term on the right-hand side of \cref{eq_free_e_cons} is proportional to the nondimensionalised radial heat flux
\begin{equation}
	\label{eq_heatflux_def}
	Q = -\frac{1}{L_xL_yL_z}\int \dr \ \deltaT \py \phinorm,
\end{equation}
whereas the second one is the collisional thermalisation. 

Surprisingly, upgrading from 2D to 3D does not eliminate both of the other two 2D invariants. One of them survives, and the following conservation law holds even in 3D:
\begin{align}
	\label{eq_sec_cons}
	L_xL_y\Lpar\pt I &\equiv \pt \int \dr \ \left[ \frac{1}{2} \left(\dw{\phinorm} + \dw{\deltaT}\right)^2 + \frac{1}{2} \zf{\deltaT}^2 + \frac{1}{2} \left(\del \deltaT + \del \phinorm\right)^2 + \frac{1}{2} u^2 \right] \nonumber \\ &= -\vt \int \dr \ \deltaT \py \phinorm  -\chi \int \dr \ \bigg[ \left(\del \dw{\phinorm}\right) \bcdot \left(\del \deltaT\right) + \left( \del \deltaT \right)^2 \nonumber \\& \quad +\achi \left( \delsq \phinorm \right)^2 +(\achi + 1 - \bchi) \left( \delsq \phinorm \right)\left( \delsq \deltaT \right) + (1-\bchi)\left( \delsq \deltaT \right)^2 + \cchi\left(\del \upar\right)^2 \bigg].
\end{align}
As expected, one recovers a corresponding 2D conservation law by setting \(u = 0\) and excluding \(z\) from the integration \citep[see \S2.7 of][]{ivanov2020}. 

Later, it will prove useful to discuss the spectra of \(W\) and \(I\). For this, we write \mbox{\(W = \sum_\vk W_\vk\)} and \mbox{\(I = \sum_\vk I_\vk\)}, where we have defined
\begin{align}
	\label{eq_Wk_def}
	&W_\vk \equiv \frac{1}{2} |\deltaT_\vk|^2, \\
	\label{eq_Ik_def}
	&I_\vk \equiv \frac{1}{2} \left( |\dw{\phinorm}_\vk + \dw{\deltaT}_\vk|^2 + |\zf{\deltaT}_{k_x}|^2 + \kperp^2 |\phinorm_\vk + \deltaT_\vk|^2 + |\upar_\vk|^2 \right).
\end{align} 
Here the \(\vk\) subscript denotes Fourier components, defined for any field \(\phinorm(\vect{r})\) as
\begin{equation}
	\phinorm(\vect{r}) = \sum_\vk \phinorm_\vk e^{i\vk\bcdot\vect{r}},
\end{equation}
and \cref{eq_Wk_def} and \cref{eq_Ik_def} follow from Parseval's theorem.

\section{Linear ITG instabilities}
\label{sect_linear}

\begin{figure}
	\centering
	\includegraphics[scale=0.26]{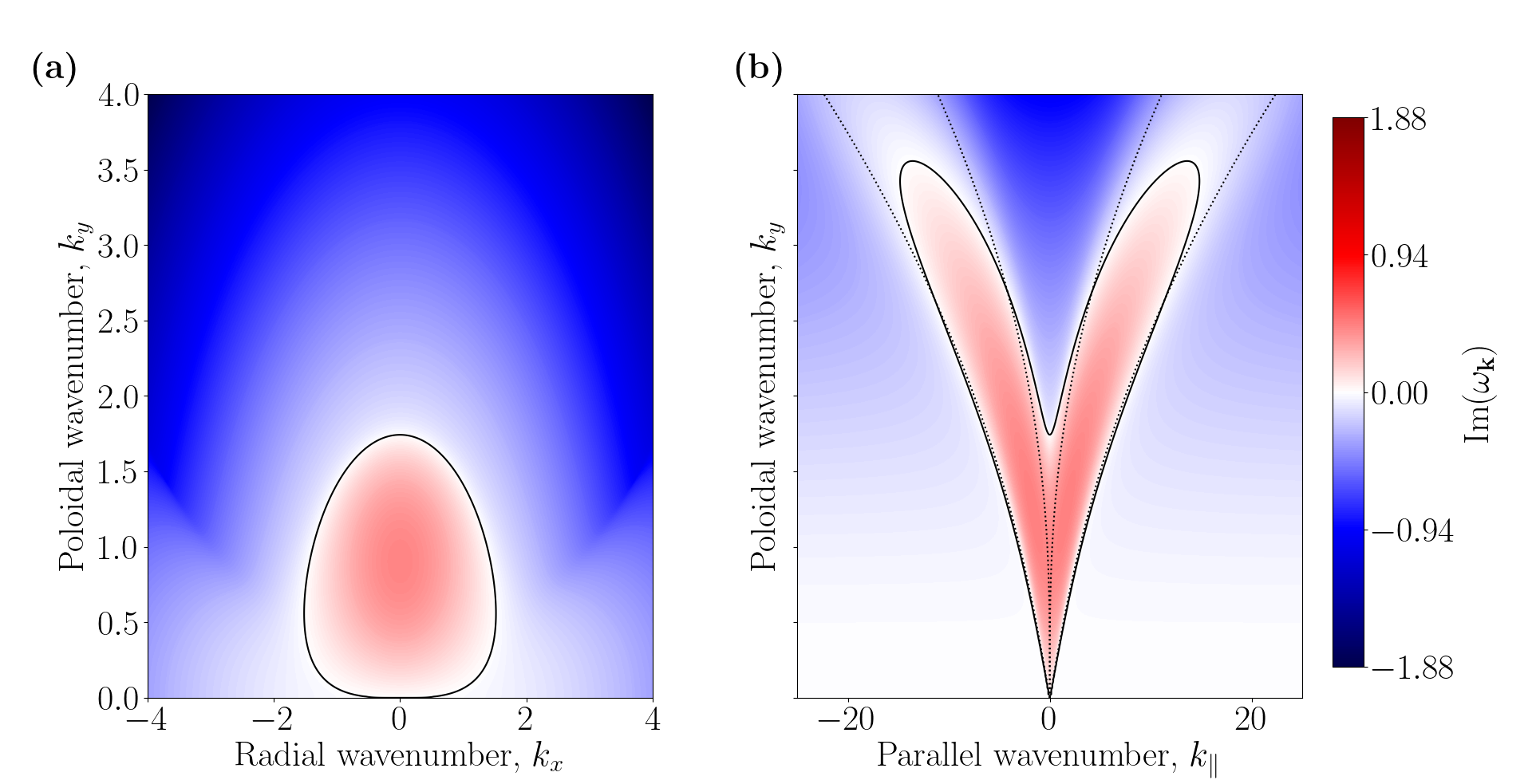}
	\caption{ A visualisation of the linear growth rate, \(\im{\omk}\), given by \cref{eq_disp}, for \(\vt = 1\) and \(\chi = 0.1\). \textbf{(a)} The linear growth rate in the \(\kpar = 0\) plane. This is the 2D cITG instability that we dealt with in \citet{ivanov2020}. \textbf{(b)} The linear growth rate in the \(k_x = 0\) plane (where it is largest). The solid black lines denote the marginal modes with \(\im{\omk} = 0\). The dotted lines outline the region of unstable collisionless (\(\chi = 0\)), pure-slab (\(L_B^{-1} = 0\)) modes, given by~\cref{eq_slab_instab_marginalmodes}. }
	\label{fig_linear3d}
\end{figure}

Equations \modeleqns{} support two distinct types of linear instability, viz., cITG and sITG. The former was studied by \citet{ivanov2020} and describes the linearly unstable 2D modes. In order to investigate the stability of the 3D modes, we drop the nonlinear terms in \modeleqns{} and look for Fourier modes \(\phinorm, \deltaT, \upar \propto \exp\left( -i\omega_\vk t + i \vect{k}\bcdot\vect{r} \right)\), where \(\re{\omega_\vk}\), \(\im{\omega_\vk}\), and \(\vk = (k_x, k_y, \kpar)\) are the real frequency, growth rate, and wavenumber of the mode, respectively. The dispersion relation can be written as
\begin{equation}
	\label{eq_disp}
	(\lambdadisp + \cchi \ksq) D_{2D} + \frac{\kpar^2}{1 + \ksq} \left( \lambdadisp + \chi \ksq - i\vt k_y\right) = 0,
\end{equation} 
where the 2D dispersion relation is given by
\begin{gather}
\label{eq_disp2d}
D_{2D} \equiv (\lambdadisp + A) (\lambdadisp + B - iC) - fAB + igAC = 0, \\
A = \chi \kperp^2, \quad B = \frac{\achi\chi \kperp^4}{1+\kperp^2}, \quad C = k_y \frac{1+\vt \kperp^2}{1+\kperp^2}, \quad f = \frac{\vt k_y^2}{\achi\chi^2\kperp^6}, \quad g = \frac{\bchi \vt \kperp^2}{1 + \vt \kperp^2}.
\end{gather}
An example of the solutions of \cref{eq_disp} is given in \cref{fig_linear3d}. It is evident that \cref{eq_disp} is too complicated for a general analytical solution. Thus, we will limit our discussion here to several important limits.

\subsection{Stable waves}
\label{sect_waves}

Setting \(\vt = 0\) and \(\chi = 0\) eliminates the linear instability and damping. The dispersion relation \cref{eq_disp} reduces to
\begin{equation}
\label{eq_disp_waves}
(1+\ksq)\omega_\vk^2 + k_y \omega_\vk - \kpar^2 = 0,
\end{equation}
with solutions
\begin{equation}
\label{eq_disp_waves_solution}
\omega_\vk = \frac{-k_y \pm \sqrt{k_y^2 + 4 \kpar^2(1+\ksq)}}{2(1 + \ksq)}.
\end{equation}
In the limit \(\kpar \ll k_y\) (which, in terms of dimensional wavenumbers, corresponds to \(\kpar L_B \ll k_y \rho_s\)), we find the familiar two-dimensional drift waves \(\omega_\vk = -k_y / (1 + \ksq)\) that result from the magnetic drift. The opposite limit, \(\kpar \gg k_y\), corresponds to equally familiar ion sound waves, modified by ion finite-Larmor-radius (FLR) effects: \({\omega_\vk = \kpar / \sqrt{1 + \ksq}}\). We can undo the normalisations \cref{eq_normalisations} to verify that this is the usual dispersion for the ion sound waves in terms of the dimensional wavenumbers \(\kpar\) and \(\kperp\):
\begin{equation}
\label{eq_sound_dimensional}
\omega_\vk = \pm \frac{c_s \kpar}{\sqrt{1 + \ksq \rho_s^2}}.
\end{equation}
We now briefly recap the 2D cITG instability before turning to the \(\kpar \neq 0\) sITG.

\subsection{Curvature-driven ITG modes}
\label{sect_curvature_modes}

\subsubsection{Instability in 2D}
\label{sect_2dmodes}
The dispersion relation for the unstable 2D (\(\kpar = 0\)) modes is \cref{eq_disp2d}. These modes were studied carefully in \citet{ivanov2020}; let us recap some important points.

The 2D modes exist at large perpendicular scales, viz., \mbox{\(\kperp < \text{min}\lbrace\kperpmaxflr, \kperpmaxcol\rbrace\)}, where the collisionless and collisional cut-offs are given~by
\begin{equation}	
	\label{eq_2d_cutoff}
	\kperpmaxflr^2 = \frac{1+2\sqrt{\vt}}{\vt}, \quad \kperpmaxcol^2 = \sqrt{\frac{\vt}{\achi \chi^2}},
\end{equation}
respectively. As shown in \citet{ivanov2020}, the Dimits threshold in 2D satisfies \mbox{\(\vt \sim \chi\)}. Here, however, we shall be interested in the strongly driven limit of \(\vt \gg \chi\) and \(\vt \gg 1\), for which a saturated state exists only in 3D. In this limit, \cref{eq_2d_cutoff} tells us that the cITG modes exist at (and below) wavenumbers 
\begin{equation}
	\label{eq_2d_cutoff_largekappa}
	\kperp \sim \kperpmaxflr \sim \vt^{-1/4} \ll 1.
\end{equation}
Solving \cref{eq_disp2d} shows that these modes also satisfy
\begin{equation}
	\label{eq_2d_omega_scaling}
	\re{\omk} \sim \im{\omk} \sim \vt^{1/4}.
\end{equation}

\subsubsection{\(\kpar \neq 0\) corrections}
\label{sect_kpar_curvature}

Let us now see how \(\kpar \neq 0\) affects the strongly driven modes at the curvature-driven scales \(\kperp \sim \vt^{-1/4} \ll 1\). At these large perpendicular scales, the effects of collisions are negligible, so we may set \(\chi = 0\). Note that the scaling \(\kperp \sim \vt^{-1/4} \ll 1\) implies \(\vt \kperp^2 \sim \sqrt{\vt} \gg 1\), in which case the dispersion \cref{eq_disp} becomes
\begin{equation}
	\label{eq_disp_cless}
	\omk\left[\omk \left(\omk + \vt\kperp^2k_y\right) + \vt k_y^2\right] = \kpar^2 \left(\omk + \vt k_y\right),
\end{equation}
where the dispersion relation for the curvature-driven 2D modes is the expression in the square brackets on the left-hand side. Using the results in \cref{sect_2dmodes}, we can estimate that for these modes, the left-hand and right-hand sides of \cref{eq_disp_cless} satisfy
\begin{equation}
	\omk\left[\omk \left(\omk + \vt\kperp^2k_y\right) + \vt k_y^2\right] \sim \vt^{3/4}, \quad \kpar^2 \left(\omk + \vt k_y\right) \sim \kpar^2 \vt^{3/4}.
\end{equation}
We can then conclude that for \(\kpar \ll 1\), the solutions are essentially 2D, i.e., the dispersion relation \cref{eq_disp_cless} is well-approximated by \cref{eq_disp2d}, whereas \(\kpar \gg 1\) is expected to introduce qualitative changes to the modes. Let us now investigate the \(\kpar \gg 1\) sITG instability.

\subsection{Collisionless slab-ITG modes}
\label{sect_slabinst}

Let us investigate the linear instability of \modeleqns{} in the absence the magnetic-gradient term \(-\py (\phinorm + \deltaT)\) in \cref{curvy_phi}. We shall see shortly when this is appropriate. For now, we limit ourselves to the collisionless (\(\chi = 0\)) regime (see also \cref{sect_slab_resonance} and \cref{appendix_col_slab}). Then, \cref{eq_disp} becomes
\begin{equation}
    \label{eq_disp_slab}
    \left(\omkh^2 - \frac{\kparh^2}{1 + \kperp^2} \right) \left(\omkh + 1\right) = \frac{2\kperp^2\gammakh^2\omkh^2}{1 + \kperp^2},
\end{equation}
where we have defined \(\omk \equiv \vt k_y \omkh\), \(\kpar \equiv \vt k_y \kparh\), and \(\gammakh^2 \equiv 1/2\kperp^2\).\footnotemark\footnotetext{This maps onto equation (49) of \citet{cowley91} for their \(Q = \Gamma = 0\) under the following change of notation (from ours to theirs): \(\kparh \mapsto k_z / k_y\), \(\omkh \mapsto -\Omega\).} The last of these may seem like an inconvenience now, but will make the following analysis more easily generalisable for our needs in \cref{sect_nl}. Since \cref{eq_disp_slab} is a real cubic in \(\omkh\), it either has three real solutions, so all linear modes are stable waves, or one real and two complex solutions, in which case one of the complex solutions has a positive imaginary part and thus corresponds to a linearly unstable mode. It can be shown (see \cref{appendix_slabdisp}) that \cref{eq_disp_slab} has complex solutions if and only if \(\gammakh^2 > 0 \) and \(\kparh^2 \in (\hat{k}_{\parallel,-}^2, \hat{k}_{\parallel,+}^2)\), where
\begin{equation}
	\label{eq_slab_instab_marginalmodes_with_gamma}	
	\hat{k}_{\parallel,\pm}^2 = \frac{\kperp^4 + 10\kperp^2\gammakh^2(1+\kperp^2)+\kperp^2(4-\kperp^2\gammakh^2) + 2 \pm \kperp\gammakh\left[4(1+\kperp^2) + \kperp^2\gammakh^2\right]^{3/2}}{2(1+\kperp^2)}.
\end{equation}
Substituting \(\gammakh^2 = 1/2\kperp^2\) yields
\begin{equation}
	\label{eq_slab_instab_marginalmodes}
	\hat{k}_{\parallel,\pm}^2 = \frac{9}{8(1+\kperp^2)} \left[ \frac{8\kperp^4}{9} + 4\kperp^2 + 3 \pm 3 \left(1 + \frac{8\kperp^2}{9}\right)^{3/2} \right].
\end{equation}
The marginal modes, i.e., those on the boundary between unstable and oscillatory modes, are given by \(\kparh^2 =\hat{k}_{\parallel,\pm}^2 \); these are shown in \cref{fig_linear_cless_slab}. We now consider two distinct asymptotic limits of \cref{eq_slab_instab_marginalmodes}: \(\kperp \ll 1\) and \(\kperp \gg 1\). 

\subsubsection{Large-scale slab-ITG instability: \(\kperp \ll 1\) modes}
\label{sect_smallkperp}

To lowest order in \(\kperp \ll 1\), \cref{eq_disp_slab} simplifies to
\begin{equation}
	\label{eq_smallkperp_disp_slab}
    \omkh^3 - \kparh^2 \omkh - \kparh^2 = 0.
\end{equation}
This is the well-known sITG dispersion relation without FLR effects and in the absence of a density gradient \citep{cowley91}. In this limit, the instability boundaries \cref{eq_slab_instab_marginalmodes} become
\begin{equation}
	\label{eq_smallkperp_marginalmodes}
	\hat{k}_{\parallel,-} = \frac{2}{3\sqrt{3}}\kperp^3 + \order{\kperp^5}, \quad \hat{k}_{\parallel,+} = \frac{3\sqrt{3}}{2} + \frac{\sqrt{3}}{4}\kperp^2 + \order{\kperp^4}.
\end{equation}
For small \(\kparh\), the linearly unstable solution of \cref{eq_smallkperp_disp_slab} is \mbox{\(\omkh \approx |\kparh^{2/3}| (-1 + i\sqrt{3})/2\)}. Thus, the linear growth rate for small \(\kparh\), or \(\kpar \ll \vt k_y\), is
\begin{equation}
    \label{eq_smallkperpkpar_growthrate}
    \im{\omk} \approx \frac{\sqrt{3}}{2} \left(\vt k_y\kpar^2  \right)^{1/3}.
\end{equation}
This is the most widely recognised expression for the sITG growth rate at long wavelengths, however, it is not the fastest-growing mode at \(k_y \ll 1\). From \cref{eq_smallkperp_marginalmodes}, we know that \cref{eq_smallkperp_disp_slab} has \(\iminline{\omkh} > 0\) solutions up to \(\kparh = \orderinline{1}\), or \(\kpar \sim \vt k_y\). The growth rate of these modes evidently satisfies \(\iminline{\omkh} = \orderinline{1}\), or \(\iminline{\omk} \sim \vt k_y\).

We are now able to confirm that neglecting the magnetic drift in deriving \cref{eq_disp_slab}, and hence \cref{eq_smallkperp_disp_slab}, was appropriate. As we saw in \cref{sect_2dmodes}, the strongly driven \mbox{(\(\vt \gg 1\))} 2D cITG modes satisfy \(k_y \sim \vt^{-1/4}\). At these wavenumbers, the sITG modes exist at scales \(\kpar \sim \vt^{3/4} \gg 1\) (as expected and assumed) and have a growth rate \mbox{\(\iminline{\omk} \sim \vt k_y \sim \vt^{3/4}\)}, which is asymptotically larger than the growth rate \mbox{\(\iminline{\omk}\sim\vt^{1/4}\)} of the cITG modes.

\subsubsection{Small-scale slab-ITG instability: \(\kperp \gg 1\) modes}
\label{sect_largekperp}

\begin{figure}
	\centering
	\includegraphics[scale=0.26]{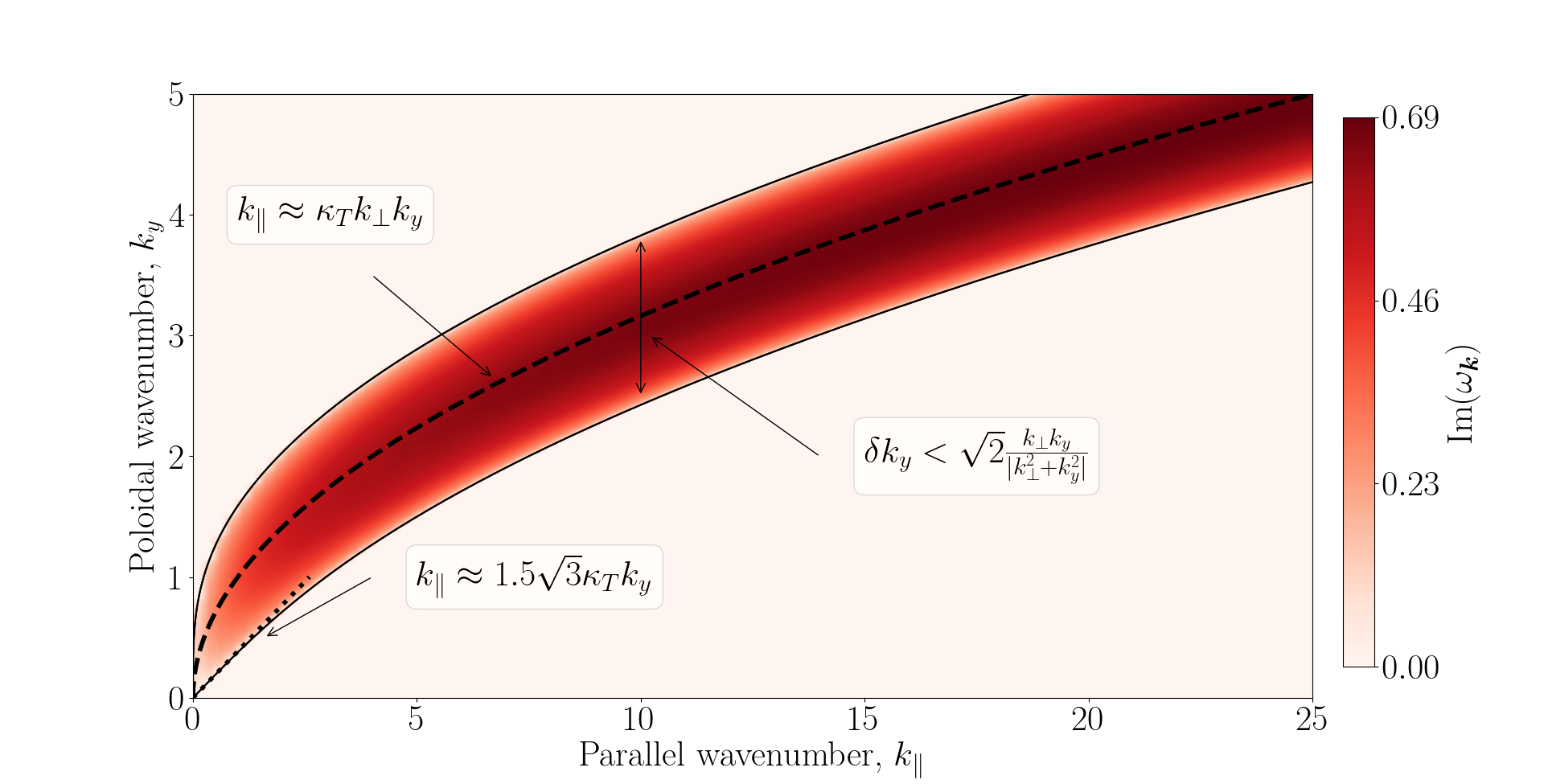}
	\caption{ Linear growth rates for the sITG instability (without the magnetic drift) as a function of parallel (\(\kpar\)) and poloidal (\(k_y\)) wavenumbers for \(k_x = 0\), \(\vt = 1\), \(\chi = 0\). The growth rate along the \(\kpar = \vt \kperp k_y\) line converges to \(\vt / \sqrt{2} \approx 0.7\) for large \(\kpar\). The solid black lines are the instability boundary given by \cref{eq_slab_instab_marginalmodes}. }
	\label{fig_linear_cless_slab}
\end{figure}

Expanding \cref{eq_slab_instab_marginalmodes_with_gamma} for \(\kperp \gg 1\) and using \(\gammakh=\order{\kperp^{-1}}\), we find
\begin{equation}
    \label{eq_largekperp_marginalmodes_general}
    \hat{k}_{\parallel,\pm} = \kperp (1 \pm 2\gammakh) + \order{\kperp^{-1}}.
\end{equation}
Therefore, at small perpendicular scales, the sITG is localised at \mbox{\(\kparh = \pm \kperp\)}, or, equivalently, at
\begin{equation}
	\label{eq_slab_inst_kpar_lowestorder}
	\kpar \approx \pm \vt k_y \kperp.
\end{equation}
In terms of the dimensional wavenumbers, \cref{eq_slab_inst_kpar_lowestorder} tells us that this instability is localised at \(\kpar L_B/2 \approx \pm\vt k_y\kperp \rho_s^2 \), or, equivalently, \(\kpar L_T \approx \pm k_y\kperp\rho_i^2\). For \(\gammakh^2 = 1/2\kperp^2\), \cref{eq_largekperp_marginalmodes_general} is
\begin{equation}
	\label{eq_largekperp_marginalmodes}
	\hat{k}_{\parallel,\pm} = \kperp \pm \sqrt{2} + \order{\kperp^{-1}},
\end{equation}
which implies, for the unstable modes,
\begin{equation}
	\label{eq_largekperp_inst_region}
	\left\lvert \kparh - \kperp \right\rvert = \left\lvert \frac{\kpar}{\vt k_y} - \kperp \right\rvert < \sqrt{2}.
\end{equation}
We can also find the \(k_y\) width of the region of instability at fixed \(\kpar\). Substituting \mbox{\(k_y = \kpar / \vt \kperp + \delta k_y\)} into \cref{eq_largekperp_inst_region} and expanding for \(\delta k_y \ll k_y\), we find
\begin{equation}
	|\delta k_y| < \sqrt{2}\frac{\kperp k_y}{|\kperp^2 + k_y^2|} \leq \frac{\sqrt{2}}{2}.
\end{equation}

To find the growth rate, consider the \(\kperp \gg 1\) limit of \cref{eq_disp_slab} and set \(\kparh = \pm\kperp + \delta\kparh\) and \(\omkh = -1 + \delta \omkh\), where \(\delta \omkh \sim \delta\kparh / \kperp \sim \orderinline{\kperp^{-1}} \ll 1\). Keeping terms of order up to \(\orderinline{\kperp^{-2}}\), \cref{eq_disp_slab} becomes
\begin{equation}
	\label{eq_largekperp_omega}
	\left(\delta \omkh \pm \frac{\delta \kparh}{\kperp} \right) \delta \omkh + \gammakh^2 \approx 0 \implies \delta \omkh \approx -\frac{\delta \kparh}{2\kperp} \pm \sqrt{\frac{\delta \kparh^2}{4\kperp^2} - \gammakh^2}. 
\end{equation}
Thus, in agreement with \cref{eq_largekperp_inst_region}, the instability exists only for \(|\delta\kparh| < 2\kperp\gammakh\), and its growth rate is
\begin{equation}
	\im{\omkh} \approx \sqrt{\gammakh^2 - \frac{\delta \kparh^2}{4\kperp^2}}. 
\end{equation}
The maximum growth rate is then achieved for \(\delta \kparh = 0\), i.e., at \(\kpar=\pm\vt k_y\kperp\), and is given by \(\im{\omkh} \approx \gammakh\). Since \(\gammakh^2 = 1/2\kperp^2\), this is
\begin{equation}
	\label{eq_largekperp_growthrate}
	\im{\omk} \approx \frac{\vt k_y}{\sqrt{2} \kperp} \leq \frac{\vt}{\sqrt{2}}.
\end{equation}
The characteristics of the small-scale sITG instability are summarised in \cref{fig_linear_cless_slab}. 

Note that for \(\vt \gg 1\), the linear growth rate \cref{eq_largekperp_growthrate} of the small-scale sITG modes scales as \(\orderinline{\vt}\) and, therefore, dominates both the curvature-driven modes (\cref{sect_2dmodes}) and the large-scale slab modes (\cref{sect_smallkperp}). This time-scale separation will prove critical for the saturation of strongly driven turbulence (see \cref{sect_scalesep}).

Finally, an important feature of the \(\kperp \gg 1\) sITG modes is the approximate relation \(\deltaT \approx - \phinorm\), or equivalently, \(\pr / \phinorm \ll 1\), where \(\pr = \phinorm + \deltaT\) is the perturbed pressure. Indeed, using \cref{curvy_psi} and \cref{eq_largekperp_omega}, we find
\begin{equation}
	\label{eq_Toverphi_slab_lowestorder}
	\frac{\deltaT_\vk}{\phinorm_\vk} = \frac{\vt k_y}{\omk} = \frac{1}{\omkh} = -1 + \frac{\delta \kparh}{2 \kperp} - i\sqrt{\gammakh^2 - \frac{\delta \kparh^2}{4\kperp^2}} + \order{\kperp^{-2}}
\end{equation}
for the modes with \(\im{\omkh} > 0\). Thus, these modes generally have \(\pr_\vk / \phinorm_\vk \sim \orderinline{\kperp^{-1}} \ll 1\), while the most unstable of them (\(\delta \kparh=0\)) satisfy \(\pr_\vk / \phinorm_\vk \sim \orderinline{\kperp^{-2}}\) and \(\re{\deltaT_\vk / \phinorm_\vk} = -1 + \orderinline{\kperp^{-2}}\). This relationship between \(\deltaT_\vk\) and \(\phinorm_\vk\) will allow us to identify the sITG modes in the saturated state, and will prove useful in understanding their role in maintaining the Dimits state (see \cref{sect_num_evidence} and \cref{sect_turb_stress}).

\subsection{Mechanism of the small-scale slab-ITG instability}
\label{sect_slab_resonance}

The analysis in \cref{sect_largekperp} is somewhat physically opaque. To get a better grasp of the small-scale sITG modes, we can consider the problem from a slightly different angle. Let us subtract the Laplacian \(\delsq\) of \cref{curvy_psi} from \cref{curvy_phi} and rewrite the linear part of the system \modeleqns{} as
\begin{gather}
	\partial_t \left( \dw{\phinorm} - \delsq \phinorm\right) + \pz \upar - \partial_y \pr + \vt \partial_y  \delsq \phinorm + \chi \nabla_\perp^4 \left[(\achi+\bchi)\phinorm - \bchi\pr\right] = 0, \label{eq_curvy_linearphi} \\
	\label{eq_curvy_linearpressure}
	-\pt \delsq \pr + \pz \upar - \py\pr + \chi \del^4(1-\bchi)\pr = -\pt \dw{\phinorm} - \chi \del^4 (\achi + \bchi - 1)\phinorm,\\
	\label{eq_curvy_linearu}
	\pt \upar + \pz \pr - \cchi \chi \delsq \upar = 0,
\end{gather}
where \(\pr = \phinorm + \deltaT\) is the pressure perturbation. Let us first concentrate on the \(\chi = 0\) case, viz., 
\begin{gather}
	\partial_t \left( \dw{\phinorm} - \delsq \phinorm\right) + \pz \upar - \partial_y \pr + \vt \partial_y  \delsq \phinorm = 0, \label{eq_curvy_linearphi_cless} \\
	\label{eq_curvy_linearpressure_cless}
	-\pt \delsq \pr + \pz \upar - \py\pr = -\pt \dw{\phinorm},\\
	\label{eq_curvy_linearu_cless}
	\pt \upar + \pz \pr = 0.
\end{gather}
Observe that the term \(\pt \dw{\phinorm}\) on the right-hand side of \cref{eq_curvy_linearpressure_cless} is asymptotically small in the \(\kperp \gg 1\) limit. Indeed, had we approximated \(\pt (1+\kperp^2)\phinorm \approx \pt \kperp^2\phinorm\) in \cref{curvy_phi}, as we should have done for \(\kperp \gg 1\), the right-hand side of \cref{eq_curvy_linearpressure_cless} would have been zero. In this approximation, \cref{eq_curvy_linearpressure_cless} and \cref{eq_curvy_linearu_cless} decouple from \cref{eq_curvy_linearphi_cless}. Their dispersion relation coincides with the \(\kperp \gg 1\) limit of \cref{eq_disp_waves_solution}, so \cref{eq_curvy_linearpressure_cless} and \cref{eq_curvy_linearu_cless} describe two propagating waves, independent of \(\vt\). Let us call these two modes `pressure waves'.\footnotemark\footnotetext{As discussed in \cref{sect_waves}, such a pressure wave is really a combination of a finite-\(\kpar\) sound wave and a finite-\(k_y\) magnetic-drift wave. The name is chosen because, unlike the diamagnetic wave described by \cref{eq_curvy_linearphi_cless}, a pressure wave carries a finite pressure perturbation.} The third mode is a \(\pr = \upar = 0\) wave, described by \cref{eq_curvy_linearphi_cless}; its frequency in the \(\kperp \gg 1\) limit is \(\omk = -\vt k_y\). We shall call this a `diamagnetic wave' because the restoring force comes from the diamagnetic-drift term \(\vt \py\delsq\phinorm\) in \cref{eq_curvy_linearphi_cless}. 

Since the diamagnetic and pressure waves have, in general, disparate frequencies, the small coupling term \(-\pt \dw{\phinorm}\) in \cref{eq_curvy_linearpressure_cless} can indeed be neglected. However, if the frequencies of these modes happen to coincide, i.e., if they are in resonance, the small coupling term can no longer be neglected. Using \cref{eq_disp_waves_solution} for the frequency of the pressure waves and \(\omk = -\vt k_y\) for the diamagnetic wave, we find that such a resonance occurs when \(\kpar = \vt \kperp k_y\), assuming \(\kperp \gg 1\). Thus, the instability condition \cref{eq_slab_inst_kpar_lowestorder} for collisionless small-scale sITG modes is the resonance condition for the two types of linear modes in the system, viz., pressure waves and diamagnetic waves.

Let us now restore \(\chi \neq 0\). Then, \cref{eq_curvy_linearpressure} shows that for \(\achi + \bchi \neq 1\) (as is generally the case), there is a second coupling mechanism, via the term \(\chi \del^4 (\achi + \bchi - 1)\phinorm\). For \(\omk \sim \vt k_y\), this term is comparable to the collisionless-coupling term \(\pt \dw{\phinorm}\) when \(\omk \sim \vt k_y \sim \chi \kperp^4\), i.e., when 
\begin{equation}
	\label{eq_cless_slab_col_cutoff}
	\kperp \sim \left(\frac{\vt}{\chi}\right)^{1/3} \equiv \kperpcol, 
\end{equation}
assuming \(\kperp \sim k_y\). We find that \(\kperpcol\) is the perpendicular scale at which the collisionless results of \cref{sect_largekperp} are no longer valid as the effects of finite \(\chi\) can no longer be neglected. Na\"ively, one might expect that for \(\kperp > \kperpcol\), collisions will act to damp the sITG instability. However, this turns out not to be the case, and, in fact, the coupling term \(\chi \del^4 (\achi + \bchi - 1)\phinorm\) can mediate a new collisional ITG instability (\(\chi\)ITG) for \(\kperp \gtrsim \kperpcol\) in the absence of the collisionless coupling term \(\pt \dw{\phinorm}\). However, it turns out that in order for \(\chi\)ITG to be non-negligible compared to sITG, very large temperature gradients are required, viz., \(\vt/\chi \gtrsim 830\). Numerically, we shall not investigate such large gradients, so the \(\chi\)ITG instability will not be relevant for us. The detailed treatment of the \(\chi\)ITG instability has been relegated to \cref{appendix_col_slab}.

\FloatBarrier

\section{Nonlinear states of low and high transport}
\label{sect_nl}

We now proceed to study the nonlinear saturated state of \modeleqns{}. We solve these equations using an enhanced version of the code used in \citet{ivanov2020}, whereby \modeleqns{} are solved using a pseudo-spectral algorithm in a triply periodic box of dimensions \(L_x\), \(L_y\), and \(\Lpar\). The linear terms are integrated implicitly in time, while the nonlinear terms are integrated explicitly using the Adams–Bashforth three-step method. This integration scheme is similar to the one implemented in the popular GK code GS2 \citep{kotschenreuther95, dorland2000}. As the 3D model has no dissipation terms that depend on \(\kpar\), we usually include small (compared to the collisional dissipation) parallel hyperviscosity of the form \(\nu \kpar^4\). It is incorporated in the equations by replacing \(\pt \mapsto \pt + \nu \kpar^4\) for all three fields in the model. The value of \(\nu\) is typically chosen to give a maximum parallel hyperviscosity of \(10\%\) of \(\chi k_{\perp, \text{largest}}^2\), where \(k_{\perp, \text{largest}}\) is the largest \(\kperp\) included the simulation. This form of hyperviscosity effectively subtracts \(\nu \kpar^4\) from the growth rate of every mode, but does not alter the linear mode structure, i.e., it does not influence the ratio of Reynolds to diamagnetic stresses given by \(\re{\deltaT_\vk / \phinorm_\vk}\) (see \cref{sect_dimits} and \cref{sect_turb_stress}). Thus, it dissipates energy without perturbing the saturated state either towards or away from the Dimits regime. 

Recall that the 2D model has two distinct nonlinear states --- a Dimits regime, where saturation is achieved with the aid of strong ZFs that quench the cITG instability by shearing the perturbations it produces, and a blow-up regime, where no finite-amplitude saturation is achieved, but amplitudes continue to grow exponentially indefinitely (or at least until numerical efforts become futile). This unphysical blow up is arguably the main limitation of the 2D model, and there are good reasons to believe that it is a consequence of the \(\kpar = 0\) restriction \citep[see \S4.5 of][]{ivanov2020}. This will indeed be corroborated below as we find that the 3D model is able to saturate for all values of \(\vt\) and \(\chi\) that we have investigated numerically.

\begin{figure}
	\centering
	\begin{tabular}{c c}
		\includegraphics[scale=0.27]{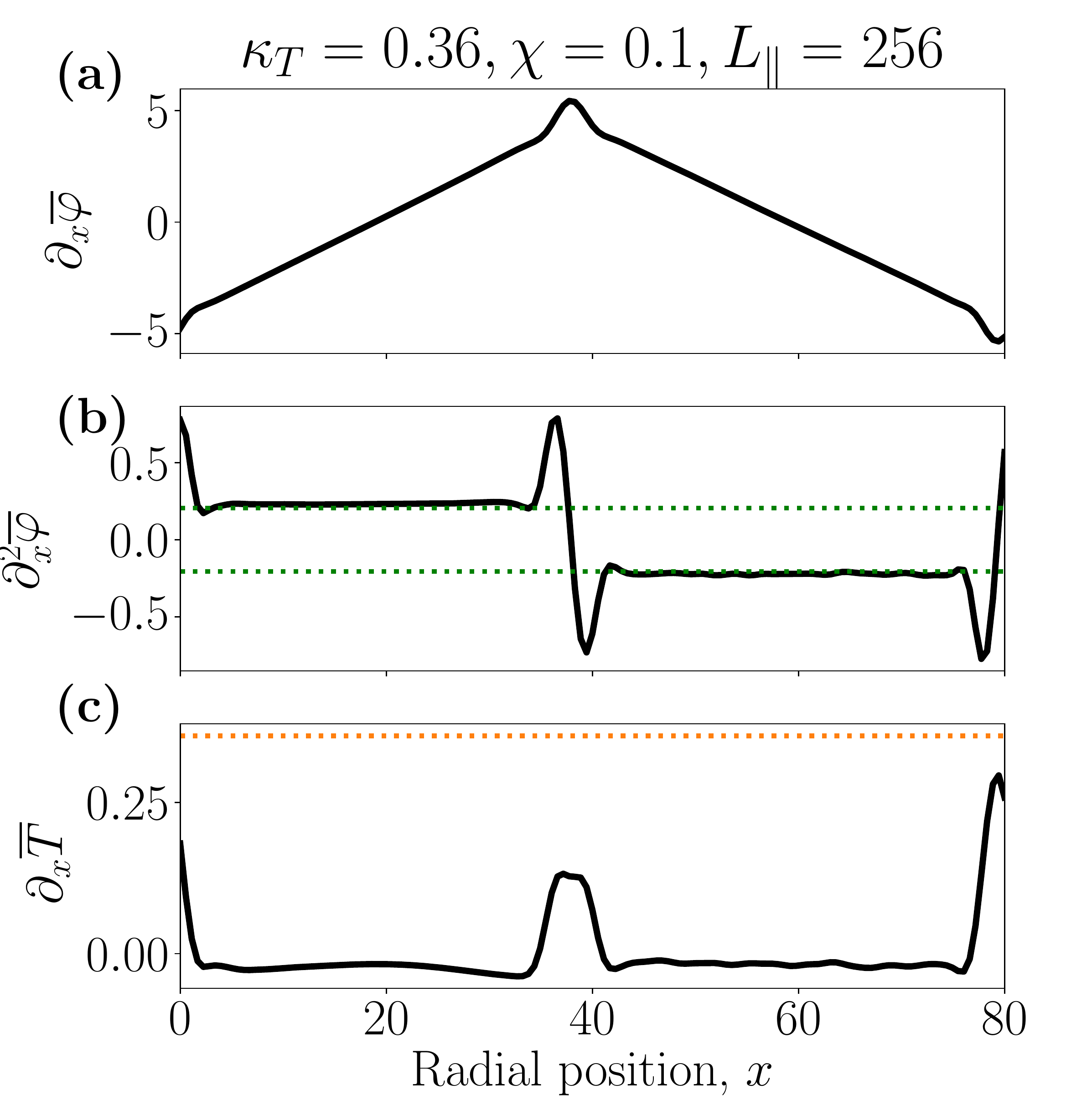} & 	
		\includegraphics[scale=0.27]{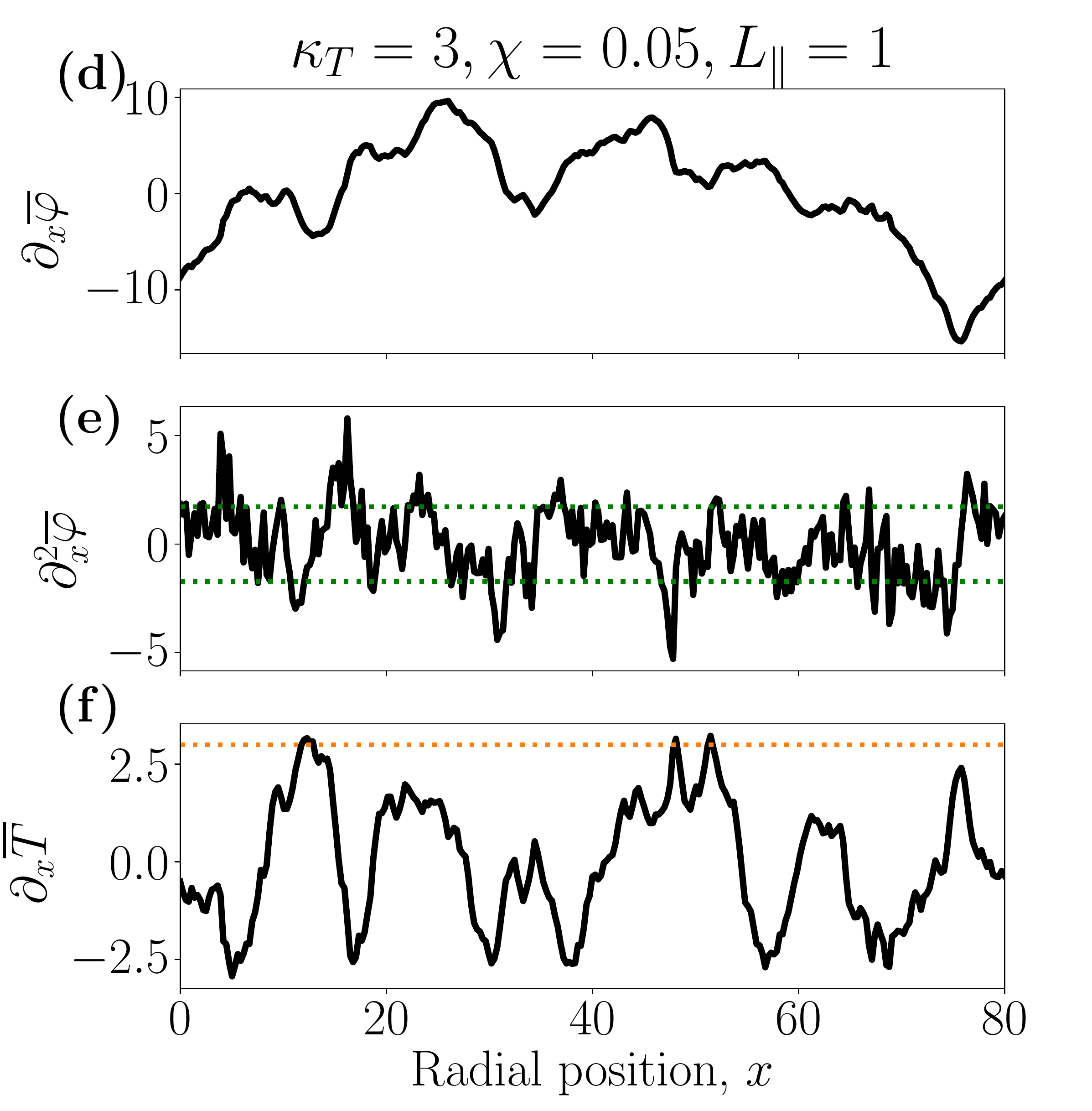}\\
	\end{tabular}
	\caption{ Example of instantaneous radial profiles of perturbations in the 3D Dimits state for \(L_x = L_y = 80\): \textbf{(a)} ZF, \textbf{(b)} zonal shear, \textbf{(c)} zonal temperature gradient. Example of instantaneous radial profiles in strong turbulence:  \textbf{(d)} ZF, \textbf{(e)} zonal shear, \textbf{(f)} zonal temperature gradient. The dotted green lines in (b) and (e) are the largest linear growth rates for the respective simulations. The dotted orange lines in (c) and (f) show the value of \(\vt\), which is equal to minus the normalised equilibrium temperature gradient. Just as in 2D, the zonal shear in the Dimits state is determined by the largest linear growth rate. Strongly turbulent ZFs do not have regions of coherent shear.}
	\label{fig_zf_profiles}
\end{figure}
\begin{figure}
	\centering
	\includegraphics[scale=0.27]{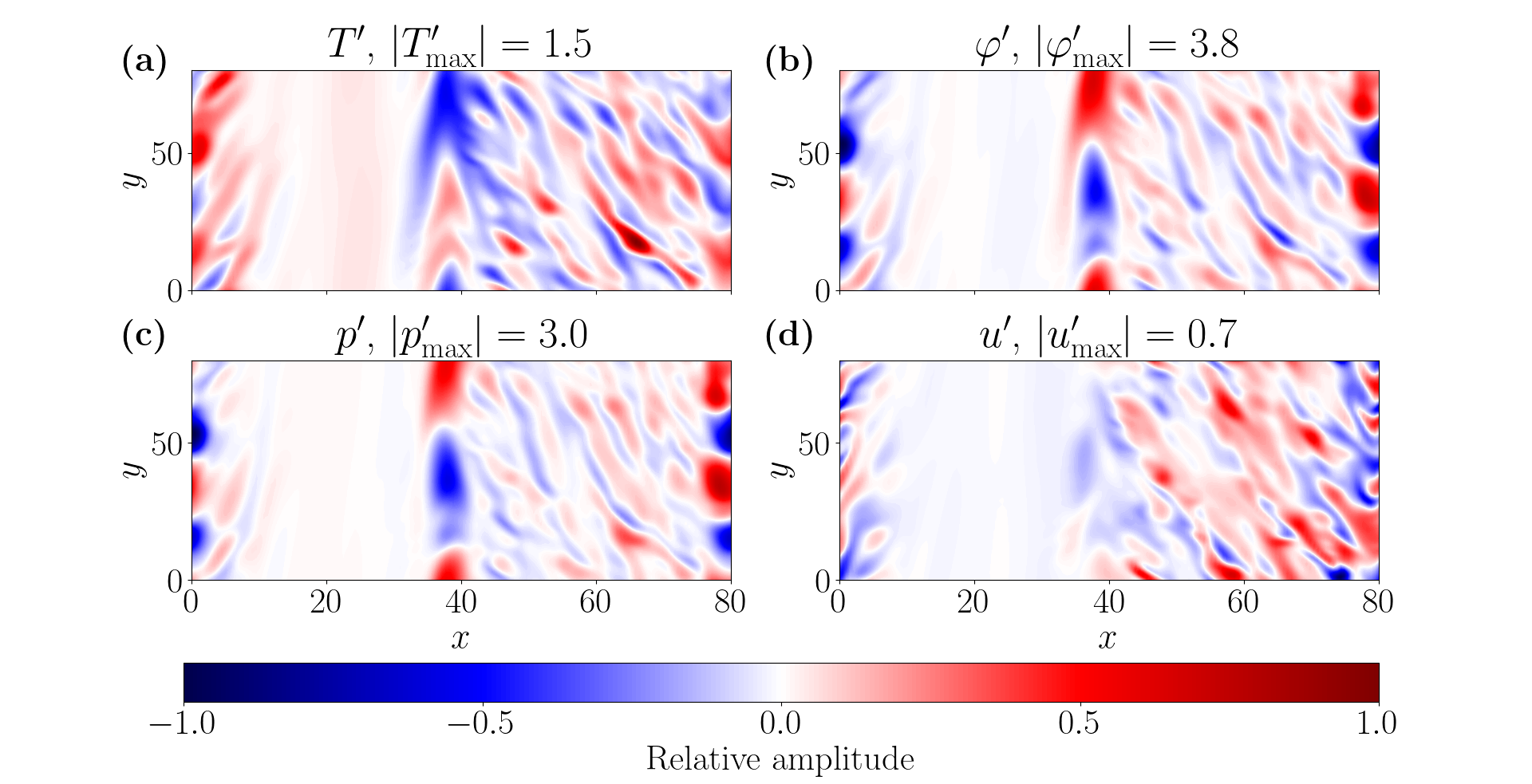}
	\caption{ Snapshots of the perturbed nonzonal \textbf{(a)} temperature \(\dw{\deltaT}\), \textbf{(b)} potential \(\dw{\phinorm}\), \textbf{(c)} pressure \(\dw{p} = \dw{\phinorm} + \dw{\deltaT}\), and \textbf{(d)} parallel velocity \(\dw{\upar}\) in the 3D Dimits state. The colour scale is relative to the maximum absolute amplitude in each panel (given in the panels' titles). We see that ferdinons carry a \(\upar\) perturbation, as well as \(T\) and \(\phinorm\) perturbations. A more detailed view of one of the ferdinons is shown in \cref{fig_ferds_elongation}. These snapshots are from the same simulation as figures \ref{fig_zf_profiles}a--c.}
	\label{fig_snapshot_dimits}
\end{figure}

\FloatBarrier
At low collisionality \citep[which can be argued to be the most relevant case, at least for core turbulence, see][]{ivanov2020}, the Dimits regime of the 3D model is strikingly similar to its 2D counterpart. The saturated state is dominated by quasi-static triangular ZFs that break up the radial domain into regions (shear zones) of constant zonal shear, where turbulence is sheared and thus suppressed (see figures~\ref{fig_zf_profiles}a--c). Localised patches of turbulence remain present at the turning points of the ZFs, where the zonal shear vanishes. 

Periodically, when viscosity has eroded the ZFs and their ability to suppress turbulence has diminished, turbulent bursts are triggered. Just as in 2D, these bursts are foreshadowed by an instability located at the ZF maxima and by the appearance of localised travelling structures produced by this instability (`ferdinons', discovered by \mbox{\citealt{vanwyk2016, vanwyk2017}} in GK simulations with external flow shear). An example of a turbulent burst in the 3D model is shown in \cref{fig_snapshot_dimits}. It is visually indistinguishable from a burst in 2D when viewed as a cross section in the \((x, y)\) plane. We shall discuss the 3D structure of the Dimits regime in detail in \cref{sect_dimits}.

The crucial qualitative change in physics that allowing 3D perturbations brings about is the sITG instability. Recall that  the collisionless small-scale sITG modes live at wavenumbers up to \(\kperpcol \sim (\vt/\chi)^{1/3}\) (see \cref{sect_slab_resonance}). This is in stark contrast with the behaviour of the 2D cITG modes whose cut-off wavenumber \cref{eq_2d_cutoff} scales as \mbox{\(k_{\perp, \text{2D cut-off}} \sim \vt^{-1/4}\)} \citep[see also \S2.6.1. of][]{ivanov2020}. Moreover, the maximal growth rate of the sITG modes \cref{eq_largekperp_growthrate} scales as \mbox{\(\im{\omk} \sim \vt\)}, while that of the cITG modes \cref{eq_2d_omega_scaling} satisfies \(\im{\omega_{\vk, \text{2D}}} \sim \vt^{1/4}\). This implies that there is a natural scale separation between slow, large-scale curvature-driven modes and fast, small-scale sITG modes. Crucially, this scale separation allows small-scale turbulence to be driven both by the equilibrium gradients and by the gradients associated with the large-scale 2D modes (which are themselves generated by the cITG instability). In fact, as we shall see in \cref{sect_slabsec}, the latter type of driving dominates in the saturated state to such an extent that the equilibrium temperature gradient can be turned off for the \(\kpar \neq 0\) modes and the saturated state remains largely unchanged. In other words, the sITG modes are `parasitic' modes, a type of 3D `secondary' instability of the 2D cITG modes. 

Most importantly, in the Dimits state, the small-scale instability can be shown always to favour strong, coherent ZFs. It does so in two ways: by providing an effective positive thermal diffusion for the large-scale modes that would otherwise destabilise the ZFs in 2D (see \cref{sect_largescale_resp}), and by generating momentum transport that is beneficial for the ZFs (i.e., a negative turbulent viscosity for the zonal flow, see \cref{sect_turb_stress}). This makes the 3D Dimits state much more resilient than the 2D one. In fact, we find that the 3D system stays in a Dimits state regardless of the values of the parameters \(\vt\) and \(\chi\), provided the domain is `sufficiently 3D', i.e., provided \(\Lpar\) is large enough and that our numerical simulations have sufficient parallel resolution to resolve the sITG modes (see \cref{sect_breakingdimits}).

We now recap the physical mechanism that gives rise to the Dimits regime and also discuss any qualitative and quantitative changes that the 3D physics brings about. Then, in \cref{sect_slabsec}, we turn to the small-scale sITG instability and its consequences for the saturated state. Finally, in \cref{sect_breakingdimits}, we examine the circumstances that can prevent the system from establishing a Dimits state and force it into the strongly turbulent regime.

\subsection{Dimits regime}
\label{sect_dimits}

\subsubsection{The 2D picture}
\label{sect_dimits2d}

Recall that the 2D Dimits transition is a sharp transition from a finite-amplitude saturated state with strong ZFs to a `blow-up' state dominated by ever-growing streamers \citep{ivanov2020}. The key to understanding this is the equation for the zonal electrostatic potential
\begin{equation}
    \label{eq_zonalphi}
    \pt \zf{\phinorm} + \Pi_\phinorm + \Pi_\deltaT + \Pi_\chi = 0,
\end{equation}
where
\begin{equation}
\label{eq_piphi_and_pit_def}
\Pi_\phinorm \equiv - \zf{(\px \phinorm) (\py \phinorm)}, \qquad \Pi_\deltaT \equiv - \zf{(\px \phinorm) (\py \deltaT)}, \qquad \Pi_\chi \equiv - \chi \px^2 \left(a \zf{\phinorm} - b \zf{\deltaT}\right)
\end{equation}
are the Reynolds, diamagnetic, and diffusive stresses, respectively. Equation~\cref{eq_zonalphi} describes how the ZFs are generated or eroded by turbulence (via the Reynolds and diamagnetic stresses, depending on their sign) and damped by collisional viscosity. We then consider a region of nearly constant zonal shear (a `shear zone') of radial width \(d\) and find that the integral of the total turbulent stress \(\Pi_t = \Pi_\phinorm + \Pi_\deltaT\) over such a region can be written as
\begin{equation}
	\label{eq_pi_t}
    \frac{1}{d}\int dx \Pi_t = -\sum_\vect{k} k_xk_y |\phinorm_\vk|^2 \left[1 + \re{\frac{\deltaT_\vk}{\phinorm_\vk}} \right].
\end{equation}
Thus, the effect of the mode with wavenumber \(\vk\) on the ZFs depends on the ratio \(\re{\deltaT_\vk/\phinorm_\vk}\). Namely, \(\re{\deltaT_\vk/\phinorm_\vk} < -1\) implies that the mode will destabilise the ZFs, while \(\re{\deltaT_\vk/\phinorm_\vk} > -1\) means that the mode will reinforce the ZFs. This observation is based on the fact that sheared (by the ZFs) turbulence is `tilted' and the sign of \(k_x k_y\) is correlated with the sign of the zonal shear in each shear zone. In \mbox{\citet{ivanov2020}}, we derived a simple estimate for the Dimits threshold at large \(\vt\) that was based on applying these ideas to the linear modes of the 2D system. More generally, we argued that the Dimits transition occurred at the threshold of a nonlinear version of the secondary instability --- when sheared by ZFs, turbulence either reinforced these flows and thus a Dimits state was maintained (the Reynolds stress won), or it failed to do so (the diamagnetic stress won) and saturation had to be reached via a different route that did not rely on zonal shear. In the 2D case, no such alternative route for finite-amplitude saturation existed. This description of how a ZF-dominated state was maintained was demonstrated to be accurate by calculating the turbulent viscosity
\begin{equation}
	\label{eq_momcorr_def}
	\nu_t \equiv -\frac{ \avgDeltat{\int_0^{L_x} dx \ \Pi_t S}}{ \avgDeltat{\int_0^{L_x} dx \ S^2}}
\end{equation}
in numerical simulations with an imposed static ZF profile; here \(\avgDeltat{\dots}\) is a saturated-state time average and \(S \equiv \px^2 \zf{\phinorm}\) is the zonal shear. Essentially, \(\nu_t\) is a measure of the correlation between the turbulent stress \(\Pi_t\) and the zonal shear \(S\). We found that \(\nu_t < 0\) on the Dimits side of the threshold, indicating that sheared turbulence was feeding the ZFs, which were shearing it. Accordingly, we also found \(\nu_t > 0\) beyond the threshold, implying that the turbulent stress was actively suppressing the ZFs. 

\subsubsection{The influence of \(\Lpar\) on the Dimits state}
\label{sect_parbox}

\begin{figure}
	\centering
	\includegraphics[scale=0.27]{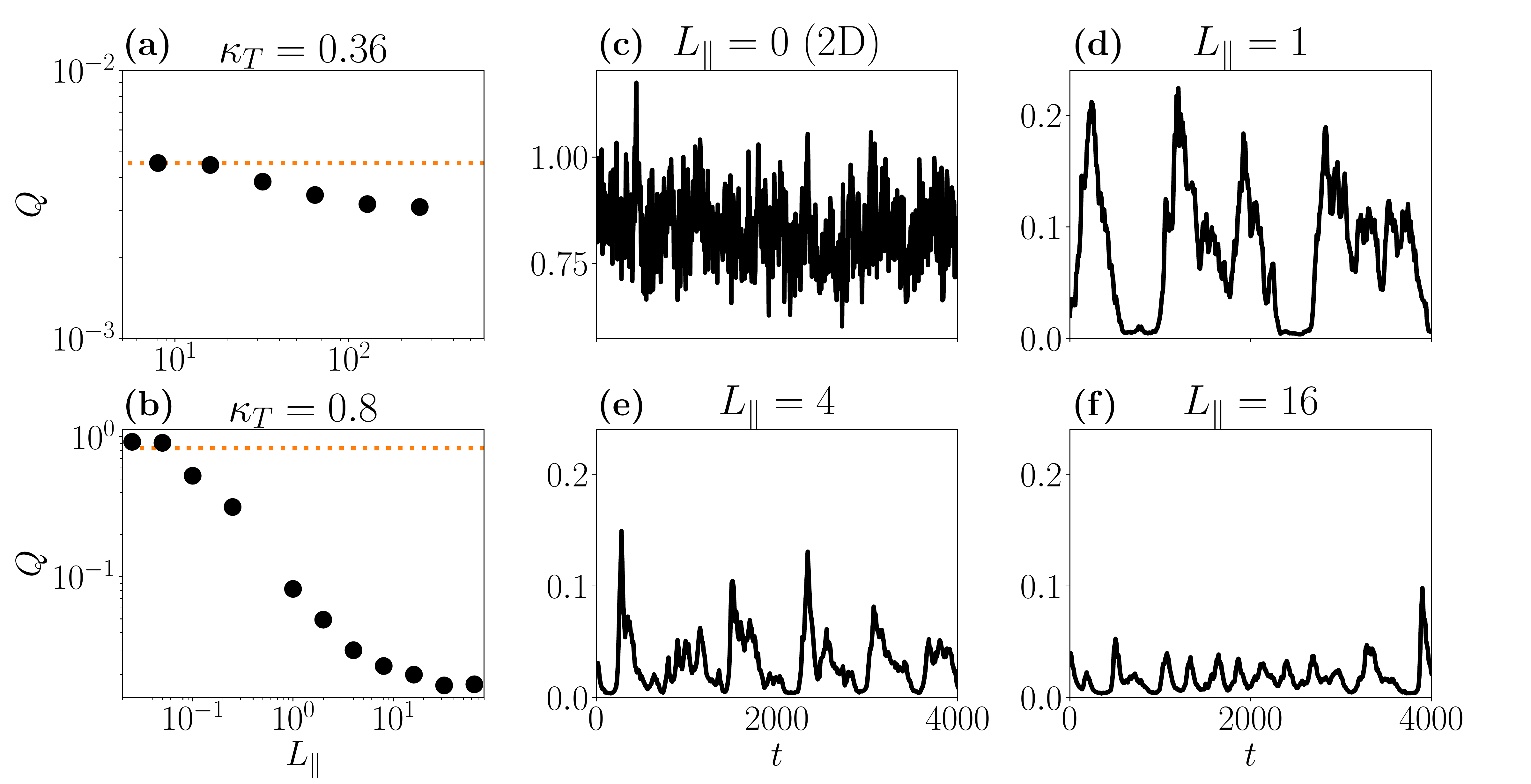}
	\caption{ \textbf{(a)} Dependence of the time-averaged heat flux \(Q\) on the parallel size of the box \(\Lpar\) for \(\chi = 0.1\), \(L_x = L_y = 80\), and \(\vt=0.36\). The orange dotted line shows the time-averaged heat flux for the 2D state (\(\Lpar = 0\)). \textbf{(b)} Same as (a), but with \(\vt = 0.8\). \textbf{(c)}--\textbf{(f)} Time evolution of the heat flux \(Q\) for \(\vt = 0.8\), \(\chi = 0.1\), \(L_x = 80\), \(L_y = 80\) and four different values of \(\Lpar\) (notated on each panel). As \(\Lpar\) increases, the turbulent bursts become more frequent and less violent, and the time-averaged \(Q\) drops. }
	\label{fig_q_vs_lpar}
	\label{fig_q_vs_t_fourplots}
\end{figure}
Taking the limit \(\Lpar \to 0\) effectively restricts our model equations \modeleqns{} to 2D, and thus their saturated Dimits state converges to that of the 2D model. In figures~\ref{fig_q_vs_lpar}a and \ref{fig_q_vs_lpar}b, we show what happens to the turbulent heat flux \(Q\) with increasing \(\Lpar\) for two cases: far below the 2D Dimits threshold (\(\vt = 0.36\), \(\chi = 0.1\)), where turbulent bursts dominate the 2D state, and closer to it (\(\vt = 0.8\), \(\chi = 0.1\)), where the bursts start to overlap in time. As expected, if \(\Lpar\) is small enough, we recover the 2D results. As \(\Lpar\) increases, \(Q\) converges in a monotonic way to a definite 3D value that is smaller than the 2D heat flux. Figures~\ref{fig_q_vs_t_fourplots}c--f show that, for larger values of \(\Lpar\), the turbulent bursts become more frequent, but shorter in duration and lower in amplitude. There are two effects responsible for this --- parallel localisation of turbulence and the development of the `parasitic' small-scale sITG modes. 

\begin{figure}
	\centering
	\includegraphics[scale=0.25]{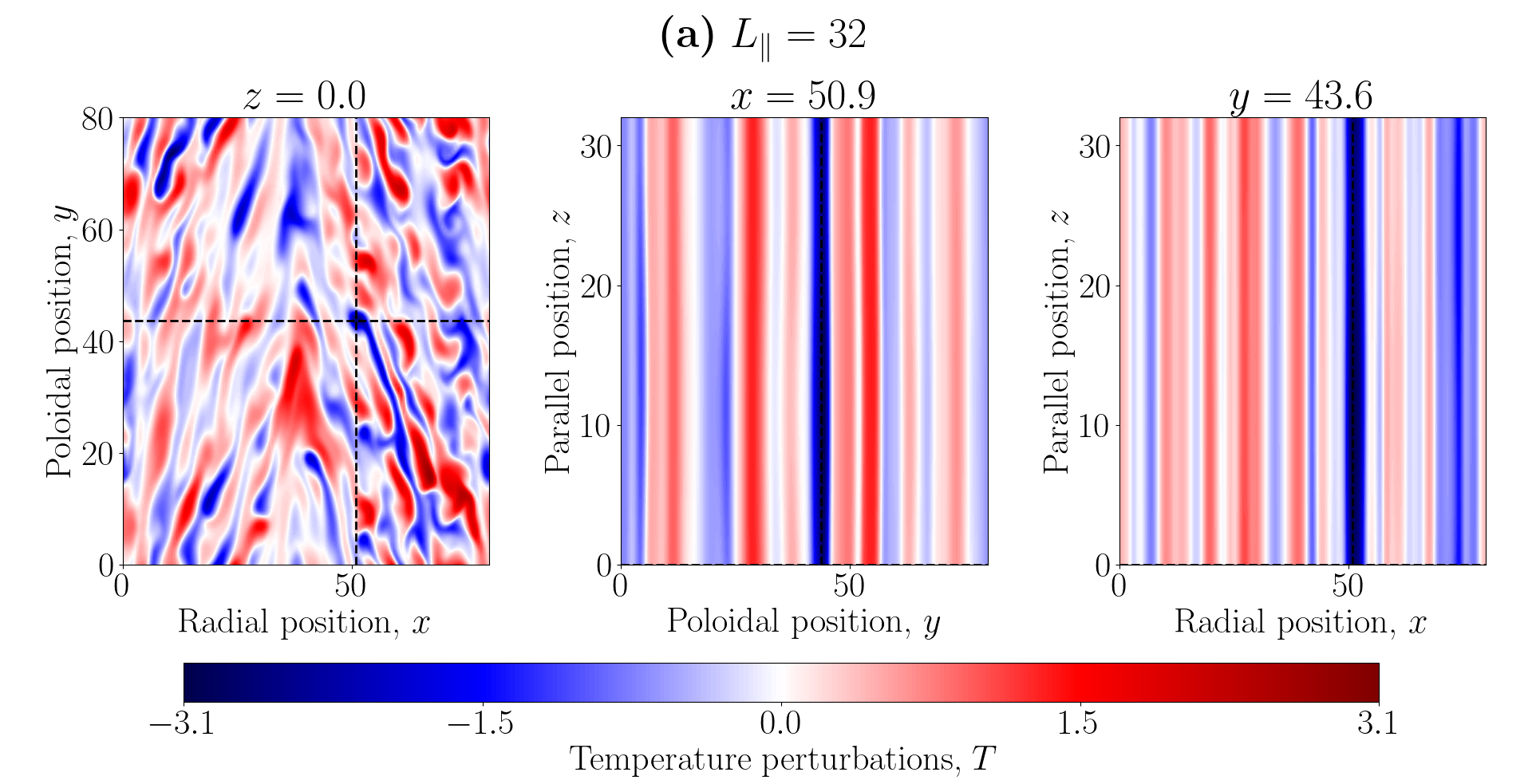}
	\includegraphics[scale=0.25]{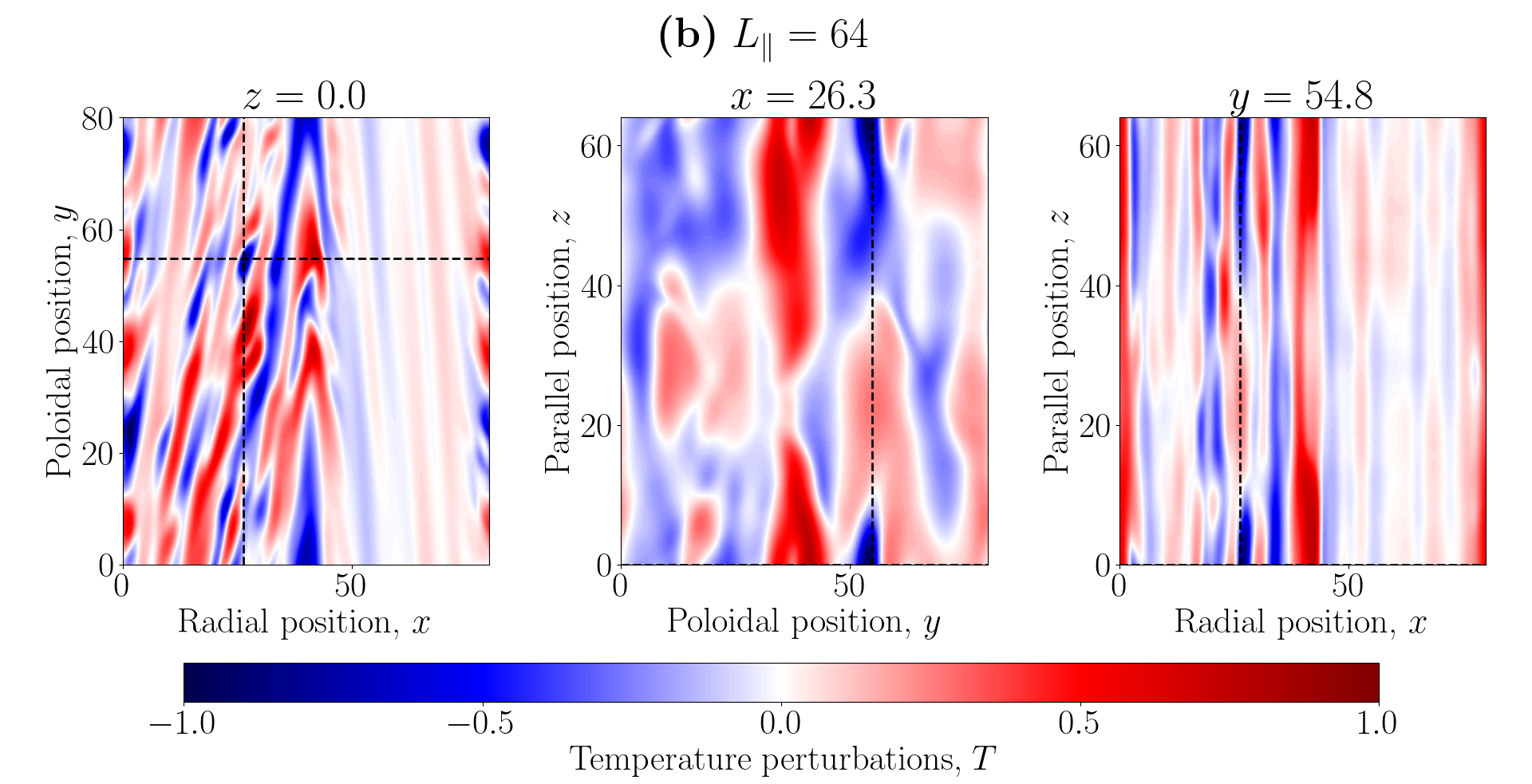}
	\includegraphics[scale=0.25]{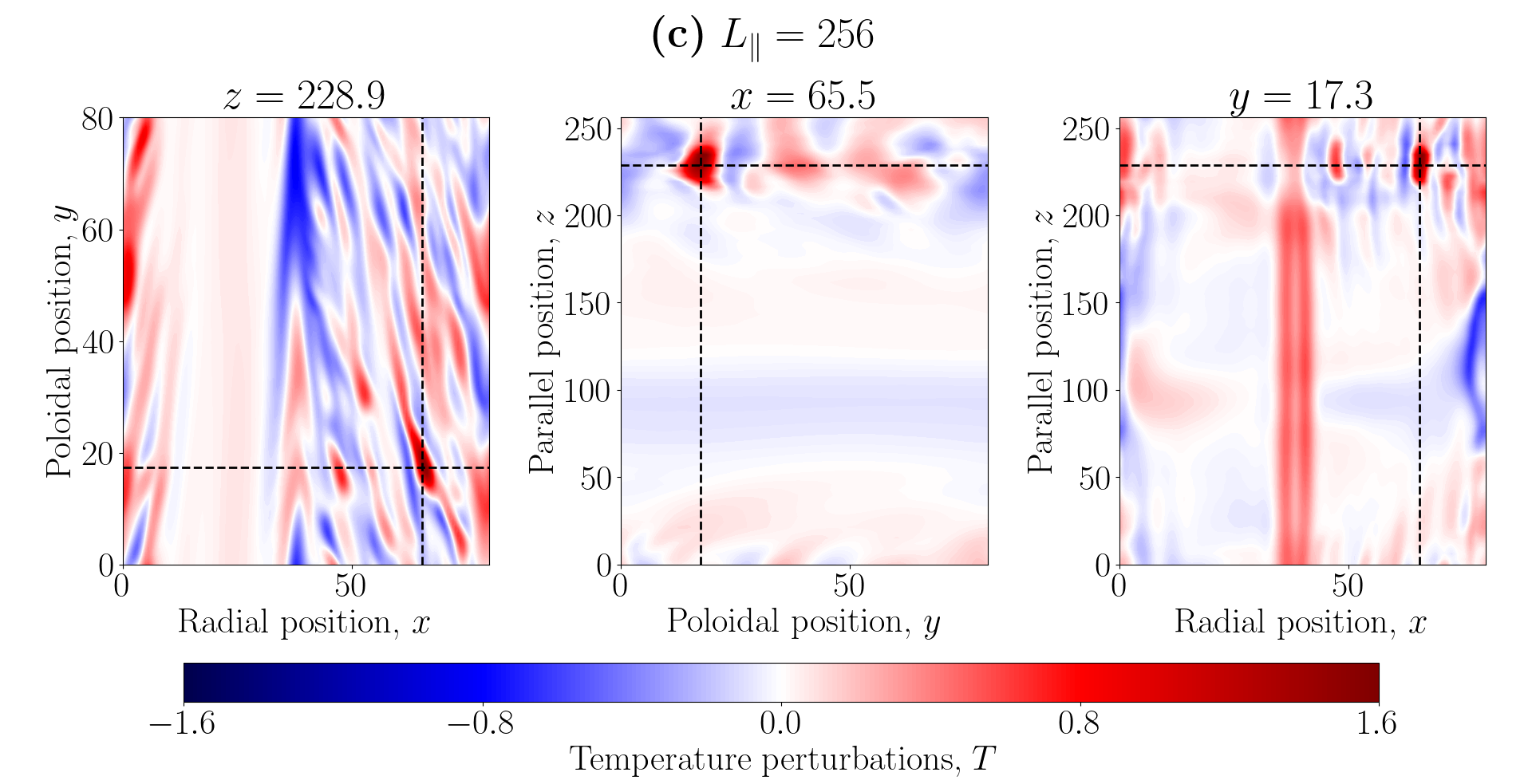}
	\caption{ Snapshots of the 3D temperature perturbations associated with a ferdinon. The plots in each row are cross sections in different planes at the same \(t\) taken from simulations that have the same \(\vt = 0.36\), \(\chi=0.1\), \(L_x = L_y = 80\), but \textbf{(a)} \(\Lpar = 32\), \textbf{(b)} \(\Lpar = 64\), and \textbf{(c)}~\(\Lpar = 256\). The black dashed lines visualise the intersections of the cross-sectional planes. As we increase \(\Lpar\), turbulence loses the ability to stay coherent along the parallel extent of the box and the bursts become localised in \(z\). }
	\label{fig_ferds_elongation}
\end{figure}

Parallel localisation is inevitable because the turbulent nonzonal modes cannot propagate information infinitely quickly along the field lines. As we increase \(\Lpar\) away from~0, we see elongated nonzonal modes that eventually lose the ability to stay coherent along the field lines if \(\Lpar\) is large enough. Figure~\ref{fig_ferds_elongation} shows that the typical Dimits-state ferdinons are not true 2D structures and develop a finite parallel extent if the parallel size of the box allows it. This is in contrast with the ZFs, which do stay perfectly coherent along the entire domain regardless of \(\Lpar\). This puts the ZFs at an advantage because the turbulent stresses in \cref{eq_zonalphi} are parallel averages and so a turbulent burst that is localised to a fraction \(\Delta\Lpar/\Lpar\) of the parallel extent of the box has its turbulent stress diminished by a factor of \(\Delta \Lpar/\Lpar\). As we increase \(\Lpar\), every such localised burst provides a smaller restoring `kick' to the ZFs and so it takes less time for the ZFs to decay to a level that permits the development of a new burst. The turbulent heat flux \(Q\) is also a spatial average of the turbulent fields and it too is diminished for a localised burst. Thus, we expect smaller, more frequent bursts, and this is precisely what is observed. Note that the ability of ZFs to communicate infinitely fast along the field lines is a consequence of the asymptotic limit of small mass ratio and the modified adiabatic electron response \cref{eq_eresponse}, which is itself due to the assumed infinitely fast parallel electron streaming. Therefore, the inclusion of kinetic electron effects in the equations would lead to qualitative changes for a large enough \(\Lpar\). Naturally, this is outside the scope of this work, but is certainly an important consideration for real devices.

\begin{figure}
	\centering
	\includegraphics[scale=0.27]{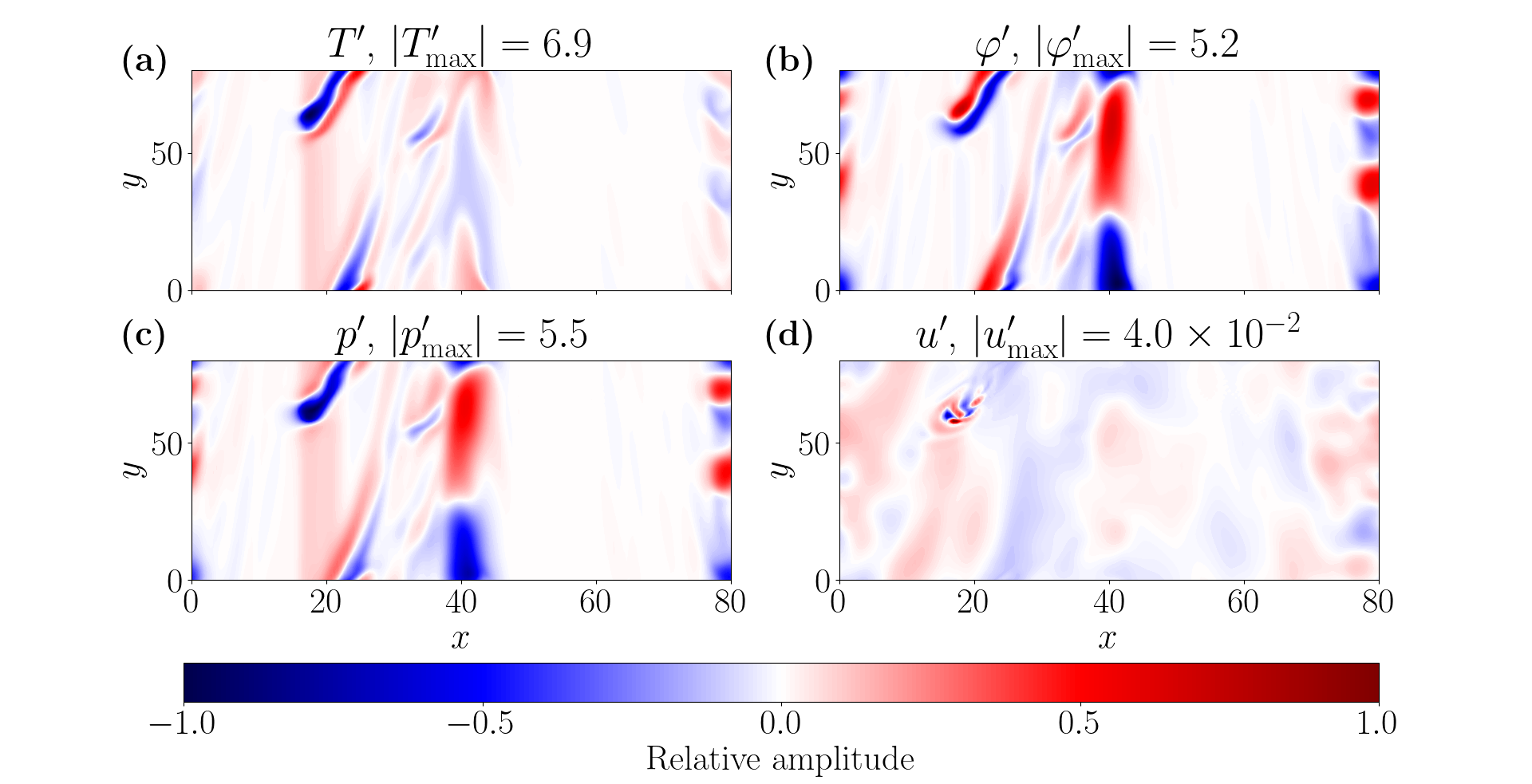}
	\caption{Snapshots of perturbed nonzonal \textbf{(a)} temperature, \textbf{(b)} potential, \textbf{(c)} pressure, and \textbf{(d)} parallel velocity at fixed \(z\) in the Dimits state with parameters \(\vt = 0.8\), \(\chi = 0.1\), \(L_x = L_y = 80\), \(\Lpar = 1\), parallel hyperviscosity \(\nu = 2.4\times10^{-8}\), and Fourier-space resolution \((n_x, n_y, n_z) = (171, 171, 21)\). The colour scale is relative to the maximum absolute amplitude in each panel (given in the panel's title). Small-scale sITG modes driven by the gradients of the ferdinon are evident in panel (d). }
	\label{fig_slabitg_on_ferd}
\end{figure}

Secondly, we find small-scale sITG modes that feed off the perpendicular temperature gradients associated with the ferdinons. The presence of this three-dimensional small-scale `parasitic' instability can be detected via the parallel velocity \(\upar\), because the latter is only involved in the 3D sITG modes and not in the 2D cITG modes. Figure~\ref{fig_slabitg_on_ferd} shows an example of a ferdinon that is `infected' with such small-scale sITG instabilities. As we shall discuss in \cref{sect_slabsec}, the small-scale instability leads to an effective increase in thermal diffusion, and thus an increase in the effective damping at large scales that reduces the large-scale temperature perturbations. This additional damping likely contributes to the reduced \(Q\) of the 3D Dimits state. It also enables saturation at finite amplitude when the Dimits state is broken (\cref{sect_breakingdimits}).

\FloatBarrier

\subsection{The parasitic slab-ITG instability and its role in the saturated state}
\label{sect_slabsec}

\subsubsection{Numerical evidence}
\label{sect_num_evidence}

\begin{figure}
	\centering
	\includegraphics[scale=0.27]{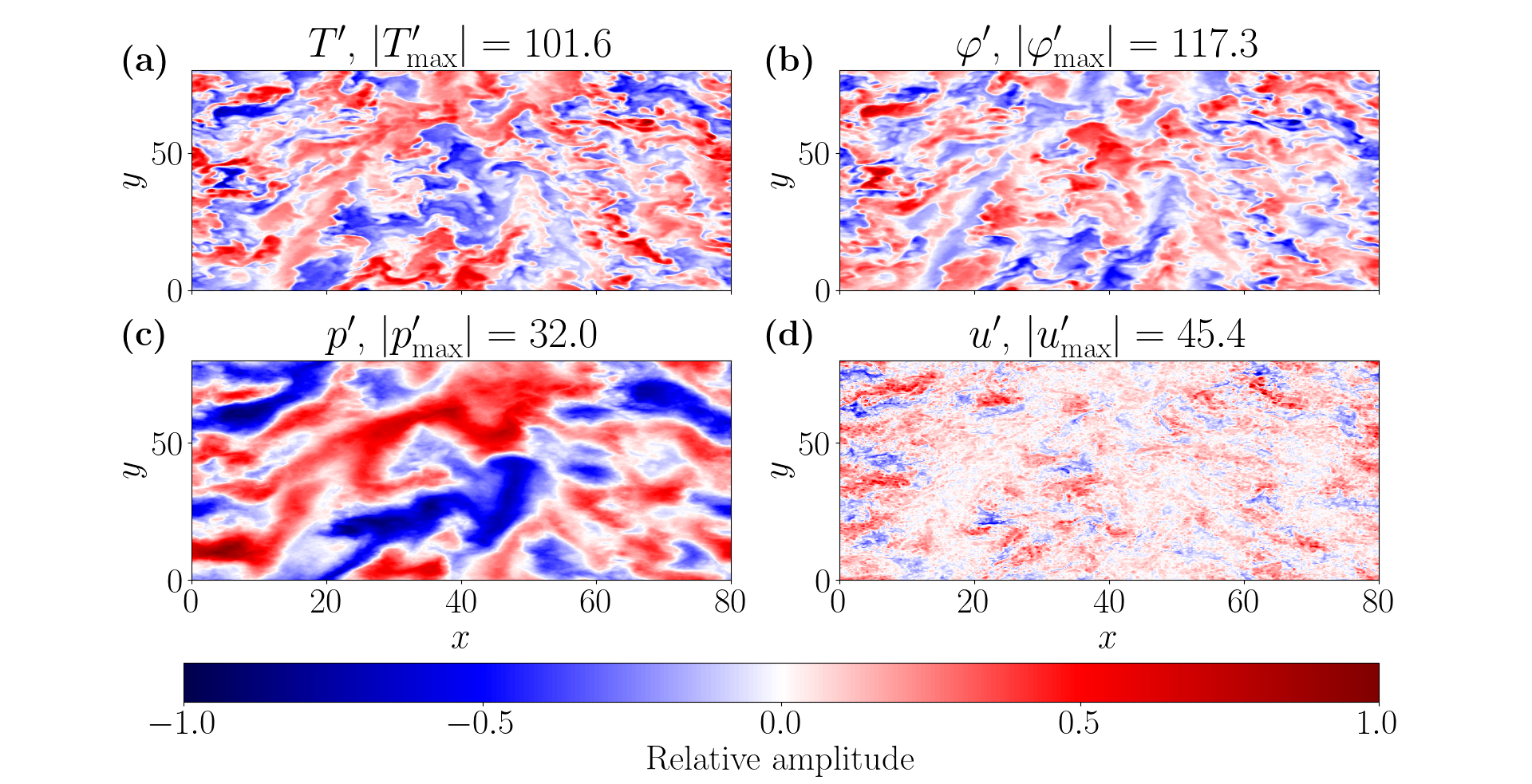}
	\caption{ Snapshots of perturbed nonzonal \textbf{(a)} temperature, \textbf{(b)} potential, \textbf{(c)} pressure, and \textbf{(d)} parallel velocity at fixed \(z\) in the strongly turbulent state with parameters \(\vt = 3\), \(\chi = 0.05\), \(L_x = L_y = 80\), \(\Lpar = 1\), parallel hyperviscosity \(\nu = 1.5\times10^{-10}\), and Fourier-space resolution \((n_x, n_y, n_z) = (285, 285, 83)\).  The colour scale is relative to the maximum absolute amplitude in each panel (given in the panel's title). Time-averaged spectra from the same simulation are shown in \cref{fig_cons_spectra}. }
	\label{fig_strongturb_dw_snapshot}
\end{figure}

\begin{figure}
	\centering
	\includegraphics[scale=0.27]{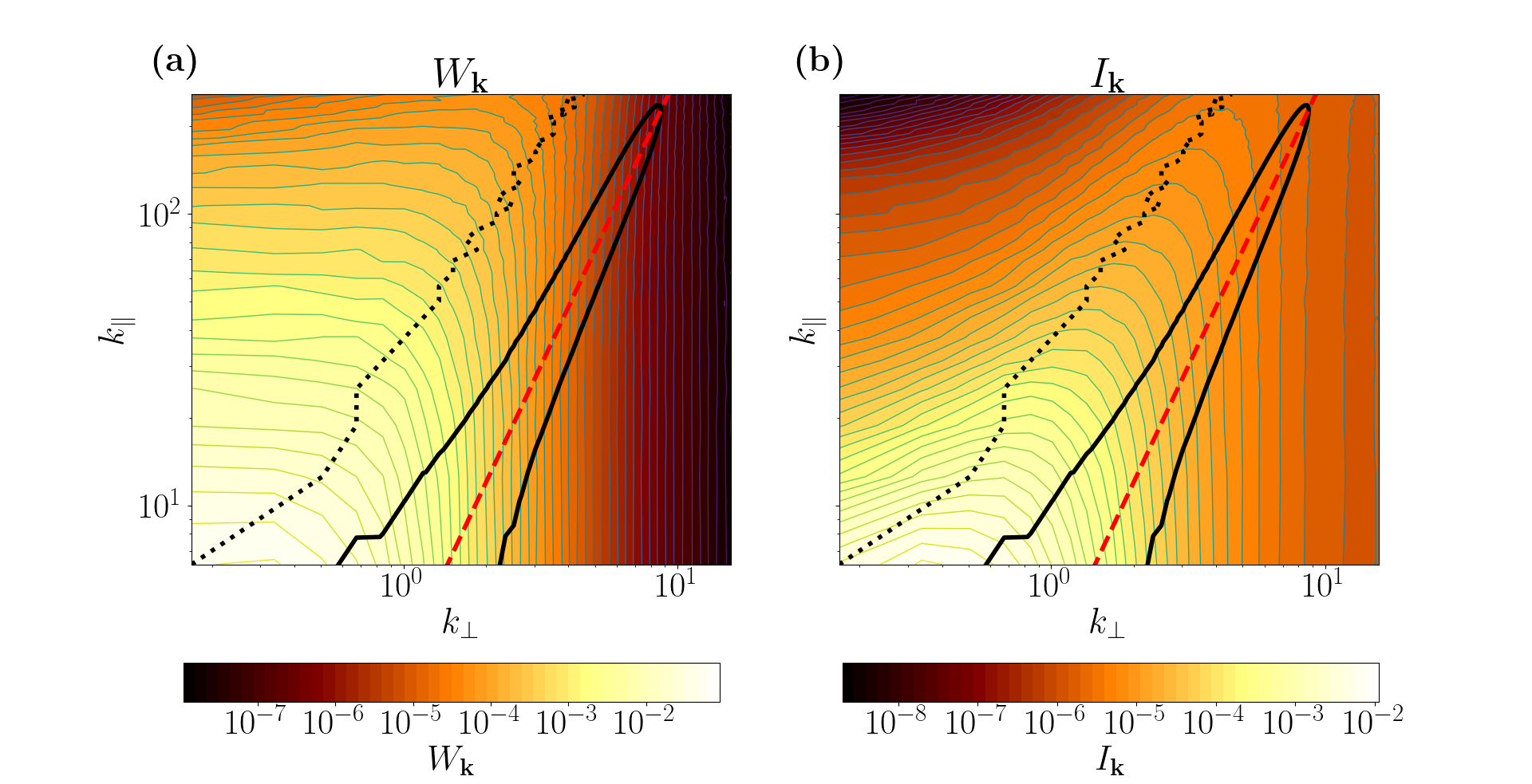}
	\includegraphics[scale=0.27]{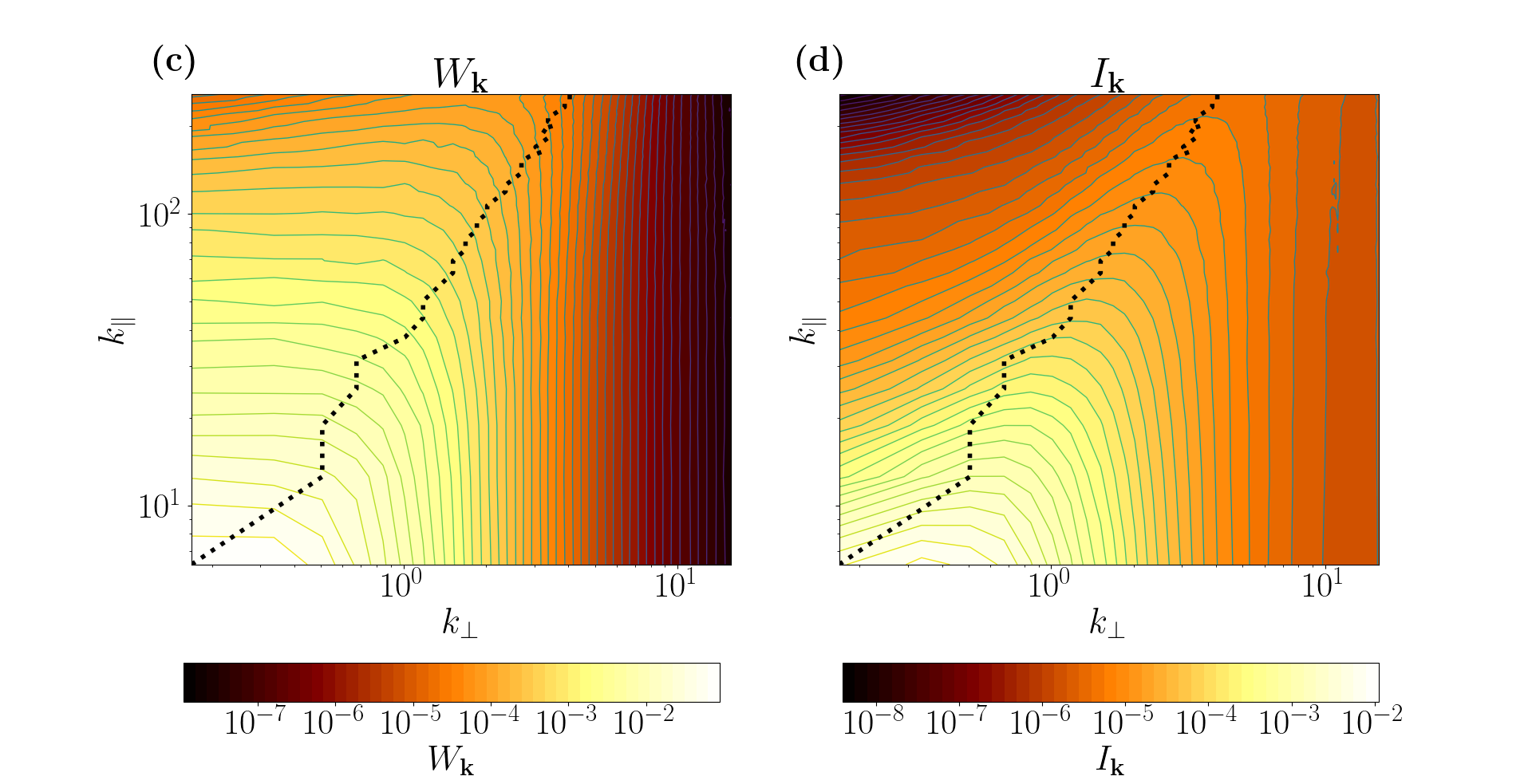}
	\caption{ Time-averaged spectra \textbf{(a)} \(W_\vk\) and \textbf{(b)} \(I_\vk\), defined by \cref{eq_Wk_def} and \cref{eq_Ik_def}, respectively, in the strongly turbulent state with parameters \(\vt = 3\), \(\chi = 0.05\), \(L_x = 80\), \(L_y = 80\), and \(\Lpar = 1\). The solid black lines demarcate the region of linear instability for \(k_x = 0\), and the red dashed line is \(\kpar = \vt \kperp^2\), where the collisionless modes with largest growth rate reside (see \cref{sect_largekperp}). We can see that the largest contributions to the two conserved quantities are offset from the region of linear instability. The dotted black line denotes the peak \(k_{\perp, I}(\kpar)\) of \(I_\vk\) at fixed \(\kpar\). Zonal profiles and cross-sectional snapshots from the same simulation are shown in \cref{fig_zf_profiles} and \cref{fig_strongturb_dw_snapshot}, respectively. The spectra of the saturated state with the same parameters, but with \(\vt\) set to \(0\) for all \(\kpar \neq 0\) modes, are given in \textbf{(c)} and \textbf{(d)}. Turning off the equilibrium gradient for the 3D modes does not alter the spectra noticeably. Snapshots from this modified simulation are shown in \cref{fig_strong_turb_noslab_snapshot}. }
	\label{fig_cons_spectra}
\end{figure}

\begin{figure}
	\centering
	\includegraphics[scale=0.27]{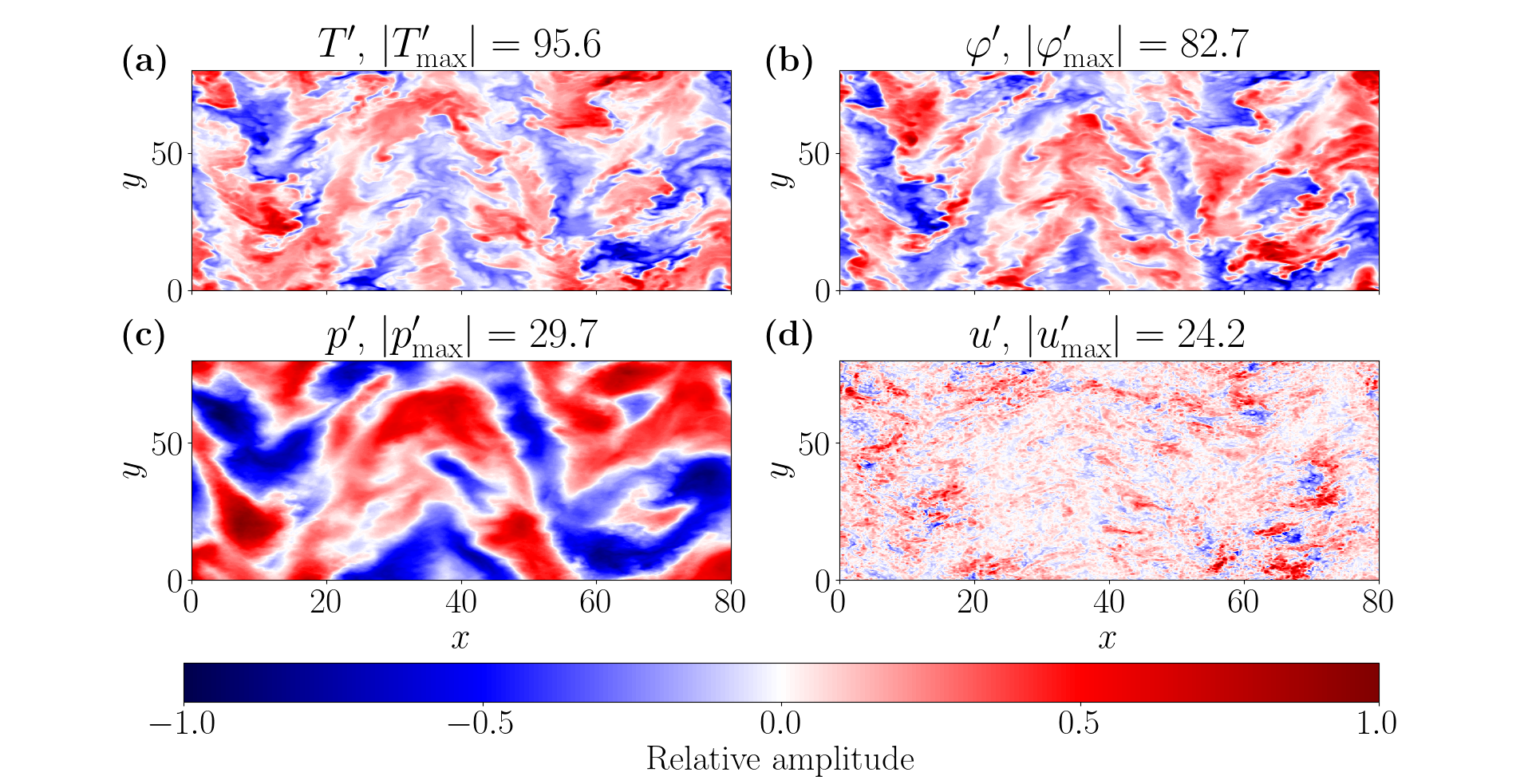}
	\caption{ Same as \cref{fig_strongturb_dw_snapshot}, but for a modified simulation, i.e., with \(\vt\) set to \(0\) for the \(\kpar \neq 0\) modes. Visually, the saturated state is identical to that shown in \cref{fig_strongturb_dw_snapshot}. }
	\label{fig_strong_turb_noslab_snapshot}
\end{figure}

\begin{figure}
	\centering
	\includegraphics[scale=0.27]{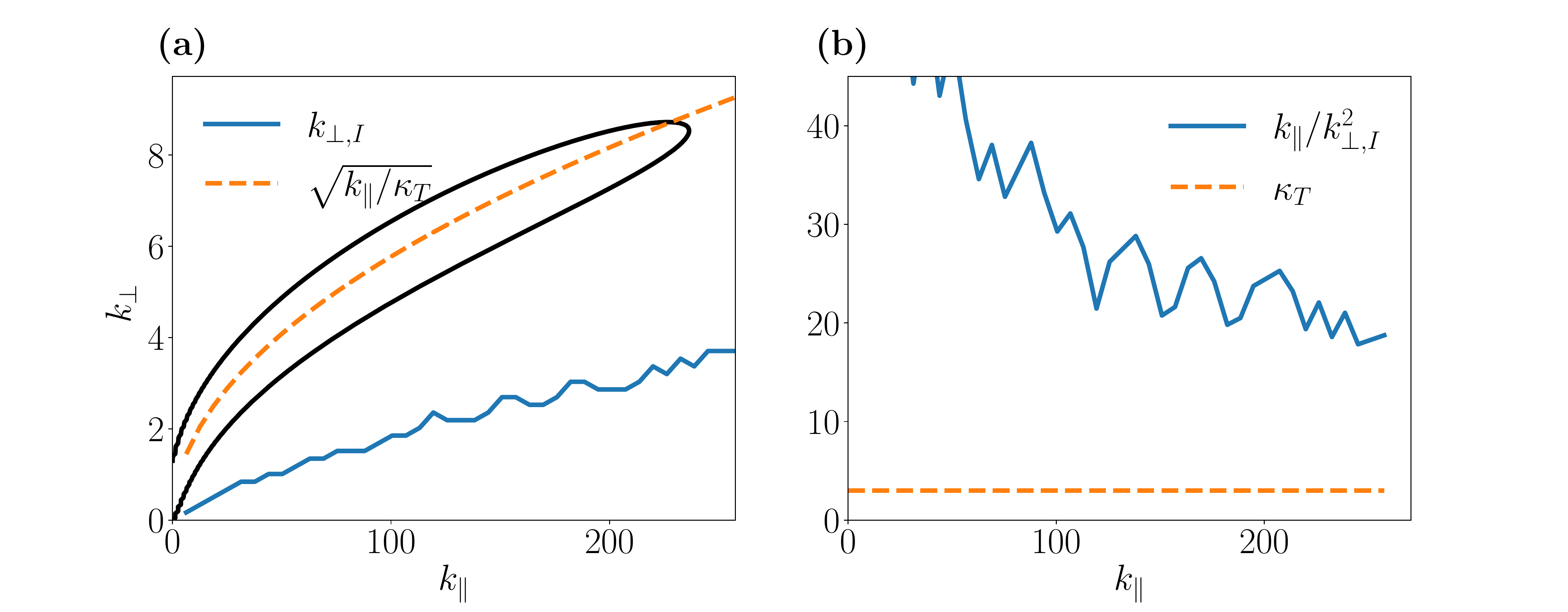}
	\caption{ \textbf{(a)} Comparison of the location of the spectral peak \(k_{\perp, I} (\kpar)\) of \(I_\vk\) (blue line) and the location of the peak of the growth rate of the collisionless linear instability driven by the equilibrium gradient (orange dashed line), given by \(\kpar = \vt \kperp^2\). The black curve circumscribes the region of linear instability, i.e., all \(\im{\omk} \geq 0\) solutions to \cref{eq_disp}  are inside it and outside of it, all solutions satisfy \(\im{\omk} \leq 0\). \textbf{(b)} Comparison of the equilibrium temperature gradient~\(\vt\) and the `effective' temperature gradient \mbox{\(\vteff(\kpar) \equiv \kpar / k_{\perp, I}^2\)}. The data is from the same simulation as shown in \cref{fig_strongturb_dw_snapshot}. The spectra of this simulation are given in \cref{fig_cons_spectra}. This rough estimate of \(\vteff\) being about 5--10 times larger than \(\vt\) is consistent with the calculated growth rate of the parasitic small-scale instability (see \cref{fig_st_smallscalegamma}). }
	\label{fig_kperp_vs_kpar}
\end{figure}

Let us now address the small-scale sITG instability. This instability exists only in the 3D model and is the most important distinction between it and its 2D counterpart. It is the presence of this instability that enables, in 3D and with finite \(\Lpar\), the existence of a strongly turbulent saturated state, i.e., one in which there are no strong, coherent ZFs (the zonal profiles of such a state are shown in figures~\ref{fig_zf_profiles}d--f). We find that the most distinctive feature of this state is the concentration of pressure perturbations at perpendicular scales that are much larger than the typical (small) scales for the perturbations in \(\phinorm\) and \(\deltaT\) (or, to be more precise, the absence of pressure perturbations in the small-scale structure present in \(\phinorm\) and \(\deltaT\)). This is manifest in \cref{fig_strongturb_dw_snapshot}. 

In \cref{sect_largekperp}, we showed that the smallness of the pressure perturbations (compared to the perturbations of the electrostatic potential and temperature) was characteristic of the small-scale (\(\kperp \gg 1\)) sITG instability: see \eqref{eq_Toverphi_slab_lowestorder}. However, the small-scale structure that we see in \cref{fig_strongturb_dw_snapshot} is not produced by the equilibrium-driven instability. In fact, the equilibrium-driven sITG instability is inconsequential in the saturated state. To show this, we ran artificially modified simulations where \(\vt\) was set to \(0\) for all modes with \(\kpar \neq 0\) (this is straightforward to do in our spectral code). This removed the equilibrium-driven linear instability from all 3D (\(\kpar \neq 0\)) modes. Examining the spectra of the two conserved quantities \(W_\vk\) and \(I_\vk\) (see \cref{sect_cons}), we see that turning off the equilibrium temperature gradient for the 3D modes has no noticeable effect on the structure of turbulence (see \cref{fig_cons_spectra}). As \cref{fig_strong_turb_noslab_snapshot} shows, the modified simulations are also visually indistinguishable from the unmodified ones shown in \cref{fig_strongturb_dw_snapshot}. The total heat flux \(Q\) changes by about 20-30\%, likely due to the loss of radial-symmetry breaking for the 3D modes, which are now free to transport heat in either direction equally, so on average, they have zero radial heat flux. The nonlinear interactions between the 2D (\(\kpar = 0\)) modes cannot produce the 3D modes that we see in the modified simulations. Therefore, these 3D modes must be produced by a `parasitic' sITG instability of the 2D fields (into which energy is injected by the equilibrium gradient).

Furthermore, the spectra of \(W_\vk\) and \(I_\vk\) measured in regular simulations are inconsistent with the region of linear instability of the dispersion relation \cref{eq_disp}. Namely, figures~\ref{fig_cons_spectra}a and \ref{fig_cons_spectra}b show that \(W_\vk\) and \(I_\vk\) of the linearly unstable modes of \cref{eq_disp} are orders of magnitude smaller than the corresponding spectral peaks of the two conserved quantities. We can quantify this by using the turbulent spectra to determine the `dominant' perpendicular scale as a function of the parallel scale \(\kpar\). As figures~\ref{fig_cons_spectra}a and \ref{fig_cons_spectra}b show, this is the scale at which \(I_\vk\) peaks and the dependence of \(W_\vk\) on \(\kperp\) changes from flat to steeply declining. To extract this scale, we define \(k_{\perp, I} (\kpar)\) as the \(\kperp\) that maximises \(I_\vk\) at a fixed \(\kpar\). Figure~\ref{fig_kperp_vs_kpar}a shows that \(k_{\perp, I} (\kpar)\) lies outside of the region of linear instability of \cref{eq_disp}. Thus, the 3D structure of the saturated state is not produced by the linear sITG instability driven by the equilibrium gradient.

In \cref{sect_largekperp}, we showed that the equilibrium-driven sITG instability is localised at \mbox{\(\kpar \approx \vt \kperp^2\)}. A similar relationship holds for \(k_{\perp, I} (\kpar)\), viz., \(\kpar \approx \vteff k_{\perp, I}^2\), where \(\vteff\) can be thought of as an effective temperature gradient. Figure~\ref{fig_kperp_vs_kpar}b shows that this \(\vteff\) is several times larger than the equilibrium temperature gradient. As we shall see shortly, \(\vteff\) is actually the gradient of the large-scale 2D temperature perturbations.

\FloatBarrier

\subsubsection{Scale-separated equations for curvature-ITG and slab-ITG modes}
\label{sect_scalesep}

The numerical analysis above leads us to believe that the 3D structure of the saturated state is a consequence of an instability driven not by the equilibrium gradient \(\vt\), but rather by the gradients of the 2D perturbations. Let us attack on the analytical front. As we discussed at the start of \cref{sect_nl}, the 3D sITG modes are naturally scale-separated from the 2D cITG modes both in wavenumber and in frequency. We then introduce the parallel average
\begin{equation}
	\label{eq_slabavg_def}
    \slabavg{f} \equiv \int \frac{dz'}{L_z} f(z').
\end{equation}
This average allows us to split \modeleqns{} into separate equations for the slow 2D modes governed by the cITG instability at large perpendicular scales (\(\kperp \ll 1\)), and for the fast 3D sITG modes, which live at small perpendicular scales (\(\kperp \gg 1\)).\footnotemark\footnotetext{A more accurate analysis should not average over the entire parallel extent of the box, but only over \(l_z\) defined to be larger than the scale of the sITG modes and smaller than the parallel scale of the cITG-like modes. As discussed in \cref{sect_smallkperp}, modes with \(\kpar \lesssim 1\) behave like cITG modes with finite-\(\kpar\) modifications. Here we have taken a cruder approach for the sake of simplifying the analysis. Note, however, that modes with \(\kpar \lesssim 1\) are usually not included in our simulations of strong turbulence for numerical reasons as we need a large maximum \(\kpar\) in order to resolve the sITG instability (see \cref{sect_parres}). Thus, this cruder approach is sufficient for the analysis of the simulations that we report in \cref{sect_largescale_resp}. } We define the small-scale perturbations as \(\slabpert{f} \equiv f - \slabavg{f}\). The large-scale equations are then
\begin{gather}
\partial_t \left( \slabavg{\dw{\phinorm}} - \delsq \slabavg{\phinorm}\right) - \partial_y \left( \slabavg{\phinorm} + \slabavg{\deltaT} \right) + \vt \partial_y  \delsq \slabavg{\phinorm} \nonumber \\ +\pbra{\slabavg{\phinorm}}{\slabavg{\dw{\phinorm}} - \delsq \slabavg{\phinorm}}  
+ \del \bcdot \pbra{\del \slabavg{\phinorm}}{\slabavg{\deltaT}} 
+ \chi \nabla_\perp^4 \left(\achi\slabavg{\phinorm} - \bchi\slabavg{\deltaT}\right) = \nonumber \\
-\slabavg{\pbra{\phinormslabpert}{\dw{\phinormslabpert} - \delsq \phinormslabpert} - \del \bcdot \pbra{\del \phinormslabpert}{\deltaTslabpert}}, \label{curvy_slabavg_phi} \\
\label{curvy_slabavg_psi}
\partial_t\slabavg{\deltaT} +\vt\partial_y \slabavg{\phinorm} +\pbra{\slabavg{\phinorm}}{\slabavg{\deltaT}} - \chi \delsq \slabavg{\deltaT} = -\slabavg{\pbra{\phinormslabpert}{\deltaTslabpert}}, \\
\label{curvy_slabavg_u}
 \pt \slabavg{\upar} + \pbra{\slabavg{\phinorm}}{\slabavg{\upar}} - \cchi \chi \delsq \slabavg{\upar} = -\slabavg{\pbra{\phinormslabpert}{\uparslabpert}}.
\end{gather}
The right-hand sides of \modeleqnsslabavg{} represent the influence of the 3D sITG modes on the large-scale fields. Subtracting \modeleqnsslabavg{} from \modeleqns{}, we find the small-scale equations:
\begin{gather}
\pt \left( \dw{\phinormslabpert} - \delsq \phinormslabpert\right) + \pz \uparslabpert - \py (\phinormslabpert + \deltaTslabpert) + \vt \partial_y  \delsq \phinormslabpert  + \slabpert{\pbra{\phinormslabpert}{\dw{\phinormslabpert} - \delsq \phinormslabpert} }
+ \del \bcdot \slabpert{\pbra{\del \phinormslabpert}{\deltaTslabpert}} 
\nonumber \\ + \chi \nabla_\perp^4 \left(\achi\phinormslabpert - \bchi\deltaTslabpert\right) =
-\Big[\pbra{\slabavg{\phinorm}}{\dw{\phinormslabpert} - \delsq \phinormslabpert}
+ \pbra{\phinormslabpert}{\slabavg{\dw{\phinorm}} - \delsq \slabavg{\phinorm}}
\nonumber \\ + \del \bcdot \pbra{\del \slabavg{\phinorm}}{\deltaTslabpert}
+ \del \bcdot \pbra{\del \phinormslabpert}{\slabavg{\deltaT}}\Big], \label{curvy_slabpert_phi} \\
\label{curvy_slabpert_psi}
\pt\deltaTslabpert +\vt\partial_y \phinormslabpert +\slabpert{\pbra{\phinormslabpert}{\deltaTslabpert}} - \chi \delsq \deltaTslabpert = 
-\left[\pbra{\slabavg{\phinorm}}{\deltaTslabpert} + \pbra{\phinormslabpert}{\slabavg{\deltaT}}\right], \\
\label{curvy_slabpert_u}
\pt \uparslabpert + \pz \left(\phinormslabpert + \deltaTslabpert\right) + \slabpert{\pbra{\phinormslabpert}{\uparslab}} - \cchi \chi \delsq \uparslab = 
-\left[\pbra{\slabavg{\phinorm}}{\uparslabpert} + \pbra{\phinormslabpert}{\slabavg{\upar}}\right].
\end{gather}

In order to simplify the following analysis, we shall assume both temporal and spatial scale separation between \modeleqnsslabavg{} and \modeleqnsslabpert{}, i.e., that the large-scale fields are constant in time in \modeleqnsslabpert{} and that the spatial and temporal scales of \modeleqnsslabpert{} are short compared to the respective scales of \modeleqnsslabavg{}. In particular, we shall assume that the perpendicular scales of the 2D modes are sufficiently large for the derivatives of their gradients to be ignored. This assumption turns out to be equivalent to \(\kperp \qperp \ll 1\), where \(\kperp\) and \(\qperp\) are the typical perpendicular wavenumbers of the 2D and parasitic modes, respectively. According to \cref{eq_2d_cutoff_largekappa} and \cref{eq_cless_slab_col_cutoff}, the linearly unstable modes satisfy \(\kperp \sim \vt^{-1/4}\) and \(\qperp \sim (\vt/\chi)^{1/3}\) in the limit \(\vt \gg \chi\). Therefore, the condition \(\kperp \qperp \ll 1\) is equivalent to \(\chi \gg \vt^{1/4}\). Additionally, recall that the Dimits threshold in 2D is found at \(\vt \sim \chi\) \citep{ivanov2020} and that, as we showed in \cref{sect_dimits}, the 2D Dimits regime is qualitatively unchanged when we include 3D effects. Thus, for the remainder of this section, we shall consider the limit
\begin{equation}
	\label{eq_st_ordering}
	\vt \gg \chi \gg \vt^{1/4} \gg 1,
\end{equation}
which puts us beyond the 2D Dimits transition (i.e., in 2D, such a state blows up). Importantly, we limit our analysis to \(\vt/\chi \ll 830\), in which case the \(\chi\)ITG instability can safely be neglected (see \cref{appendix_col_slab}). The limit \cref{eq_st_ordering} then allows us to simplify \modeleqnsslabavg{} and \modeleqnsslabpert{} significantly and thus to describe the interplay between 2D and parasitic modes analytically. These analytical results agree with our simulations, even though the latter do not strictly conform to \cref{eq_st_ordering}.

\subsubsection{Parasitic slab-ITG instability}
\label{sect_parasitic_inst}

First, we investigate the small-scale sITG instability in the presence of large-scale 2D modes. Linearising \modeleqnsslabpert{} in the limit \cref{eq_st_ordering}, we obtain
\begin{align}
&\left(\pt  + \slabavg{\ve} \bcdot \del \right) \left( \dw{\phinormslabpert} - \delsq \phinormslabpert\right) + \pz \uparslabpert - \py (\phinormslabpert + \deltaTslabpert) \nonumber \\ &\quad  + \vtvect \bcdot \del \delsq \phinormslabpert + \vnvect \bcdot \del \phinormslabpert = - \chi \nabla_\perp^4 (\achi\phinormslabpert - \bchi\deltaTslabpert), \label{curvy_slabpert_phi_lowestorder} \\
\label{curvy_slabpert_psi_lowestorder}
&\left(\pt  + \slabavg{\ve} \bcdot \del \right)\deltaTslabpert +\vtvect \bcdot \del \phinormslabpert = \chi \delsq \deltaTslabpert, \\
\label{curvy_slabpert_u_lowestorder}
&\left(\pt  + \slabavg{\ve} \bcdot \del \right) \uparslabpert + \pz \left(\phinormslabpert + \deltaTslabpert\right) =  \cchi \delsq \uparslabpert,
\end{align}
where the `local-equilibrium' quantities
\begin{align}
	\label{eq_largescale_gradients_def}
    \slabavg{\ve} \equiv \uvect{z} \times \del \slabavg{\phinorm}, \quad \vnvect \equiv - \uvect{z} \times \del \slabavg{\dw{\phinorm}}, \quad \vtvect \equiv \vt \uvect{y} - \uvect{z} \times \del \slabavg{\deltaT}
\end{align}
are the \exb{} advecting flow, the local density gradient, and the total local temperature gradient (large-scale perturbation plus equilibrium), respectively. We assume that \mbox{\(|\vtvect|\sim|\vnvect|\)} (see \cref{sect_largescale_resp}). Note that only the nonzonal electrostatic potential \(\dw{\phinorm}\) gives rise to a density perturbation --- this is a consequence of the modified adiabatic electron response \cref{eq_eresponse}. Note also that we have ignored the large-scale 2D parallel flow \(\slabavg{\upar}\). Since \(\slabavg{\upar}\) is not involved in any linear instability, the only way it could be driven is via the small-scale response, viz., the right-hand side of \cref{curvy_slabavg_u}. In \cref{appendix_upar_largescale}, we show that a small initial \(\slabavg{\upar}\) will decay under the influence of growing small-scale modes. Accordingly, in our numerical simulations, we find that \(\slabavg{u}\) is many orders of magnitude smaller than the other two 2D fields and is irrelevant for the saturated state.

Ignoring collisions (i.e., setting \(\chi = 0\)) and taking the gradients of the large-scale fields to be constant over the small scales at which \modeleqnsslabpertsimple{} hold, we can investigate the small-scale linear instability in a way analogous to what we did in \cref{sect_linear}. In particular, we shall focus on the \mbox{\(\kpar \sim \kperp^2 \gg 1\)} regime analysed in \cref{sect_largekperp}. We look for Doppler-shifted solutions to \modeleqnsslabpertsimple{} of the form \(\phinormslabpert_\vk, \deltaTslabpert_\vk, \uparslabpert_\vk \propto \exp{[-i (\omega_\vk + \slabavg{\ve} \bcdot \vk ) t + i \vk \bcdot \vect{r}]}\). Note that we ignore the shear in the \exb{} flow \(\slabavg{\ve}\). We also ignore the magnetic-drift term \(-\py(\slabpert{\phinorm} + \slabpert{\deltaT})\) in \cref{curvy_slabpert_phi_lowestorder} because it is subdominant for the sITG modes with \(\kperp \gg 1\). The resulting dispersion relation for these modes is
\begin{equation}
    \label{eq_disp_slabpert}
    \left(\omega_\vk^2 - \frac{\kpar^2}{1 + \kperp^2}\right) \left( \omega_\vk + \vtvect \bcdot \vk \right) = \frac{\omega_\vk^2}{1 + \kperp^2} \left[ \left(\vnvect + \vtvect\right) \bcdot \vk \right].
\end{equation}
Since \modeleqnsslabpertsimple{} describe real fields, \cref{eq_disp_slabpert} must be invariant under \(\vk \mapsto -\vk\) and \(\omega_\vk \mapsto -\omega_\vk^*\). We may, therefore, assume that \(\vtvect\bcdot\vk > 0\) without loss of generality. Repeating the arguments of \cref{sect_largekperp}, we define \(\omk \equiv \omkh\vtvect\bcdot\vk\) and \(\kpar \equiv \kparh \vtvect\bcdot\vk\). Then \cref{eq_disp_slabpert} turns out to be formally the same as our old dispersion relation \eqref{eq_disp_slab}, but now with
\begin{equation}
	\label{eq_gammakh_smallscale_slab}
	\gammakh^2 = \frac{(\vnvect+\vtvect)\bcdot\vk}{2\kperp^2\vtvect\bcdot\vk}.
\end{equation}

\begin{figure}
	\centering
	\includegraphics[scale=0.27]{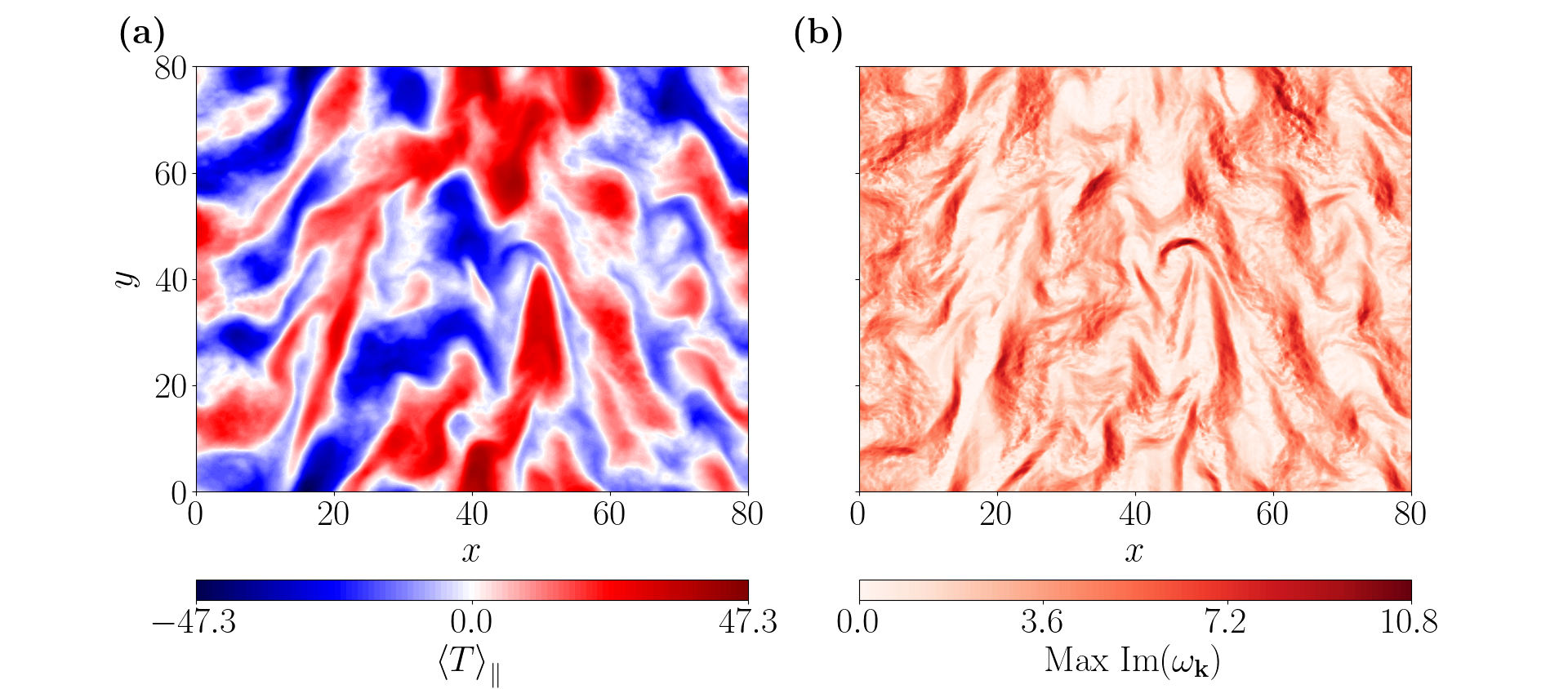} 
	\caption{ \textbf{(a)} Snapshot of the 2D temperature perturbation \(\slabavg{\deltaT}\) in the \((x, y)\) plane. The data is taken from the same \(\vt = 3\), \(\chi = 0.05\) simulation that we showed in \cref{fig_strongturb_dw_snapshot}. The 2D temperature perturbations lack the small-scale structure that was seen in \cref{fig_strongturb_dw_snapshot}a, confirming that the parallel average \cref{eq_slabavg_def} removes small-scale perpendicular structure. \textbf{(b)} Small-scale growth rate in the \((x, y)\) plane. This plot is obtained by finding the maximum growth rate of the full (including collisionality and magnetic curvature) dispersion relation \cref{eq_disp} with the addition of the local temperature and density gradients of the large-scale fields at every point. For this simulation, \(\vt = 3\), and so the largest collisionless growth rate, given by \cref{eq_largekperp_growthrate}, is \(\vt / \sqrt{2} \approx 2.1\). It is thus evident that the influence of the gradients of the large-scale fields dominates over that of the equilibrium gradient \(\vt\) by a factor of \(5\). The `effective' \(\vteff\) that we estimated for the same simulation in \cref{fig_kperp_vs_kpar}b is, indeed, a factor of 5--10 larger that the equilibrium gradient~\(\vt\). }
	\label{fig_st_smallscalegamma}
\end{figure}

Thus, the results of \cref{sect_largekperp} carry over to the parasitic instability described by \cref{eq_disp_slabpert}. In particular, the sITG instability exists if \(\gammakh^2 > 0\), i.e., if \((\vnvect+\vtvect)\bcdot\vk\) and \(\vtvect\bcdot\vk\) have the same sign, and is localised to \(\kpar \approx \pm \vtvect\bcdot\vk \kperp\). Its growth rate is given by
\begin{equation}
	\label{eq_largekpar_secslab_growthrate}
    \im{\omega_\vk} \approx \text{Re} \sqrt{\frac{\vtvect\bcdot \uvect{k} \left(\vnvect + \vtvect\right) \bcdot \uvect{k}}{2}},
\end{equation}
where \(\uvect{k} = \vk / \kperp\). As expected, this is the same as \cref{eq_largekperp_growthrate} if \(\vnvect = 0\) and \(\vtvect = \vt \uvect{y}\). In \cref{fig_st_smallscalegamma}b, we show the maximum growth rate obtained from the numerical solution of the full (with collisionality and magnetic curvature turned back on) dispersion relation~\cref{eq_disp} with the addition of the local temperature and density gradients of the large-scale fields. As expected from the numerical analysis in \cref{sect_num_evidence}, the small-scale instability driven by the large-scale gradients is significantly (\(\approx 5\) times in this case) stronger than the equilibrium-driven instability. This is consistent with the estimate of the effective temperature gradient \(\vteff\) for the sITG instability that we showed in \cref{fig_kperp_vs_kpar}b.

Note that if \(\vnvect \neq 0\), \cref{eq_largekpar_secslab_growthrate} implies that modes with \(\vtvect\bcdot\vk \left(\vnvect + \vtvect\right) \bcdot \vk < 0\) are linearly stable. Is there a \(\vnvect\) that quenches the sITG instability for all \(\vk\)? Suppose \(\vnvect\nparallel\vtvect\). Then we can choose \(\uvect{k}\bcdot\vnvect=0\), but \(\uvect{k}\bcdot\vtvect\neq0\). By \cref{eq_largekpar_secslab_growthrate}, any such \(\uvect{k}\) is an unstable mode. Therefore, to stabilise all modes, we require \(\vnvect\parallel\vtvect\). In this case, it is evident that \(\vnvect\bcdot\uvect{k} = (\vnvect\bcdot\vtvect)(\vtvect\bcdot\uvect{k})/|\vtvect|^2\). Therefore, in order to quench the sITG instability for all \(\vk\), we need  
\begin{equation}
	\label{eq_smallscale_inst_cond}
	\vnvect \parallel \vtvect, \quad \frac{\vnvect\bcdot\vtvect}{|\vtvect|^2} \leq -1.
\end{equation}
We shall now show that the effect of the growing small-scale modes on the large-scale 2D fields can be expressed as an enhanced thermal diffusivity for the latter. 

\subsubsection{Anomalous heat flux due to parasitic slab-ITG modes}
\label{sect_largescale_resp}

We expect that the growth of small-scale sITG modes, which are driven by the gradients associated with the large-scale fluctuations, will check the growth of the amplitudes of the driving large-scale fields. This is an intuitive consequence of the conservation laws \cref{eq_free_e_cons} and \cref{eq_sec_cons}. As the parasitic instability is driven by the nonlinear terms that conserve \(W=\sum_\vk W_\vk\) and \(I =\sum_\vk I_\vk\), an excitation of parasitic small-scale modes should show up as a sink in the large-scale equations. Let us now calculate explicitly the influence of small-scale sITG modes on the large-scale modes and show that this is indeed true. This influence is represented by the terms of the form \(\slabavg{\pbra{.}{.}}\) on the right-hand sides of \modeleqnsslabavg{}.

\begin{figure}
	\centering
	\includegraphics[scale=0.27]{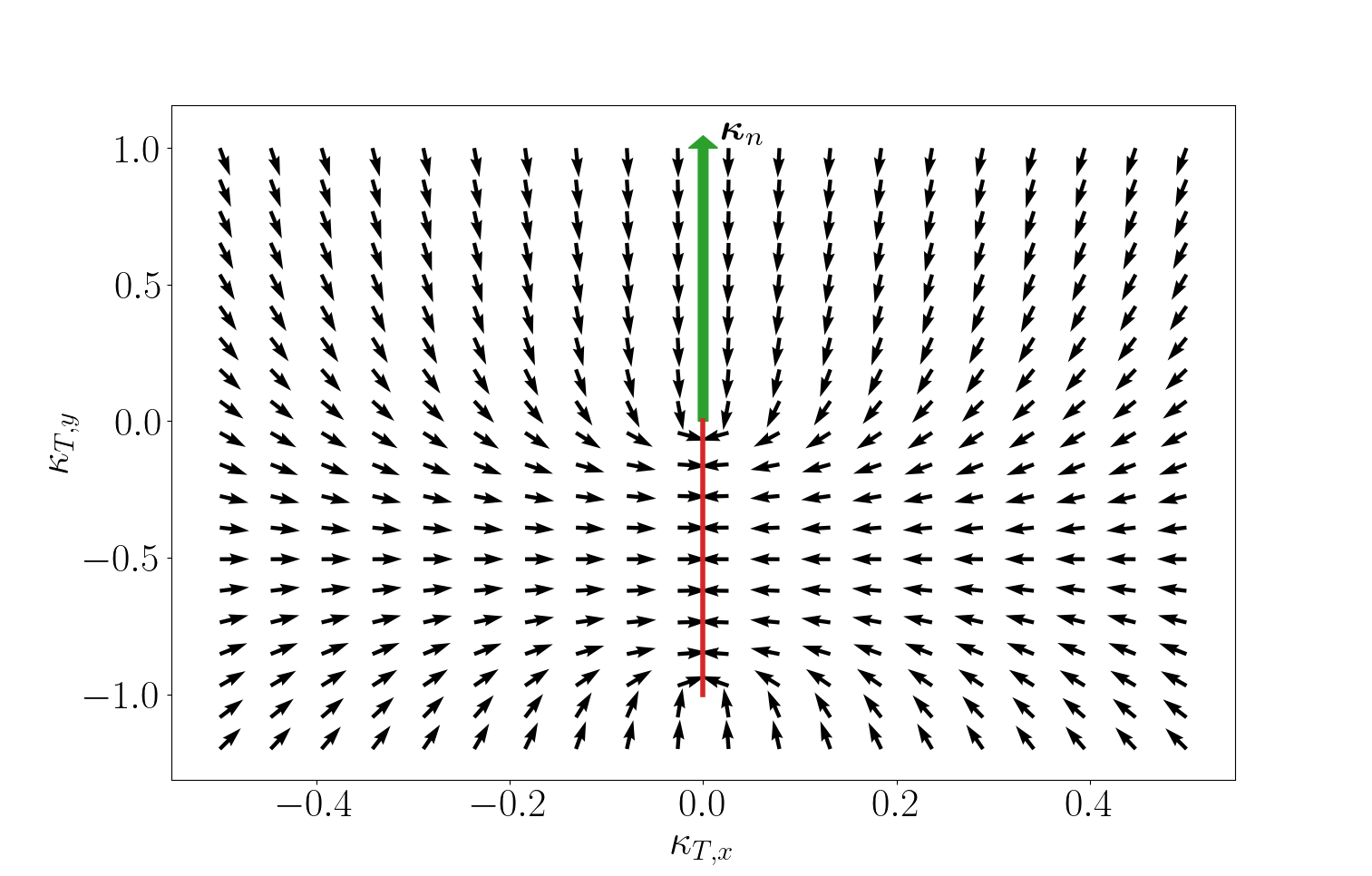}
	\caption{ This plot shows the direction in which the heat flux \cref{eq_smallscaleQ} of the most unstable small-scale mode (\(\vqh = \vqh_\text{max}\)) pushes the temperature gradient \(\vtvect = -\uvect{z}\times\del\slabavg{\deltaT}\). We have chosen a coordinate system in which the large-scale density gradient is \(\vnvect = (0, 1)\), denoted by the green arrow. The red line shows the values of \(\vtvect\) for which the sITG instability has zero growth rate, according to \cref{eq_smallscale_inst_cond}. The black arrows represent the direction of \(-\uvect{z}\times \Qpertslabavg\). We see that \(\Qpertslabavg\) pushes the large-scale temperature gradient \(\vtvect\) towards the linearly stable region. }
	\label{fig_kappaTpush}
\end{figure}

First, consider the temperature equation \cref{curvy_slabavg_psi}. The relevant term is
\begin{equation}
	-\slabavg{\pbra{\phinormslabpert}{\deltaTslabpert}} = -\del \bcdot \slabavg{\left[ (\uvect{z}\times\del\slabpert{\phinorm}) \slabpert{\deltaT} \right]} \equiv -\del \bcdot \Qpertslabavg,
\end{equation}
where \(\Qpertslabavg\) is the turbulent heat flux associated with the small-scale modes. Let us compute it quasilinearly (i.e., assuming that \(\Qpertslabavg\) is determined by the most unstable small-scale modes), assuming scale separation. As stated in \cref{sect_scalesep}, we imagine that the small-scale equations are solved in an infinitesimal (compared to the large scales) box, and thus the parallel average is equivalent to an average over such a small-scale box:
\begin{equation}
	\label{eq_smallscaleQ}
	\Qpertslabavg = -\sum_\vq  i \uvect{z}\times\vq\phinormslabpert_{-\vq} \deltaTslabpert_\vq \approx -\sum_\vq \uvect{z}\times\vqh \sqrt{\frac{(\vtvect + \vnvect)\bcdot\vqh}{2\vtvect\bcdot\vqh}} |\phinormslabpert_\vq|^2,
\end{equation}
where \(\vqh = \vq / q_\perp\) and we have assumed that the sum is dominated by the wavenumbers \(\vq\) corresponding to the largest linear growth rate of the parasitic sITG instability, and so have replaced \(\deltaTslabpert_\vq/\phinormslabpert_\vq\) with the collisionless expression \cref{eq_Toverphi_slab_lowestorder} for the modes with \mbox{\(\kpar = \vtvect\bcdot\vk\kperp\)} that maximise this growth rate. Note that the small-scale fields \(\deltaTslabpert_\vq\) and \(\phinormslabpert_\vq\), and thus \(\Qpertslabavg\) itself, depend implicitly on the position variable of the large-scale equations \modeleqnsslabavg{}.

In order to verify that \(\Qpertslabavg\) does indeed damp the large-scale temperature perturbations \(\slabavg{\deltaT}\), we multiply \cref{curvy_slabavg_psi} by \(\slabavg{\deltaT}\) and integrate over space to find

\begin{align}
	&\pt \int \dr \frac{1}{2} \slabavg{\deltaT}^2 + \text{linear terms} = \int \dr \Qpertslabavg \bcdot \del \slabavg{\deltaT} \nonumber \\
	&\approx - \int \dr \sum_\vq (\uvect{z}\times\vqh)\bcdot \del \slabavg{\deltaT} \sqrt{\frac{(\vtvect + \vnvect)\bcdot\vqh}{2\vtvect\bcdot\vqh}} |\phinormslabpert_\vq|^2 \nonumber \\
	&= - \int \dr \sum_\vq \sqrt{\frac{\vtvect\bcdot\vqh(\vtvect + \vnvect)\bcdot\vqh}{2}} |\phinormslabpert_\vq|^2 = - \int \dr \sum_\vq \im{\omega_\vk} |\phinormslabpert_\vq|^2,
\end{align}
where \(\im{\omega_\vk}\) is the sITG growth rate \cref{eq_largekpar_secslab_growthrate}. Thus, the linearly unstable small-scale modes have a sign-definite effect on \(\slabavg{\deltaT}\): they provide additional dissipation. 

The heat flux \cref{eq_smallscaleQ} depends on \(\slabavg{\deltaT}\) in a nontrivial way. Let us quantify its influence on \(\slabavg{\deltaT}\) by working out its direction as a function of \(\vtvect\). Let us assume that \(\Qpertslabavg\) is dominated by the fastest-growing sITG modes, and let their wavevector direction be~\(\vqh_\text{max}\), so \(\Qpertslabavg\) is parallel to \(\uvect{z}\times\vqh_\text{max}\). In \cref{fig_kappaTpush}, we illustrate the influence on \(\vtvect\) of the contribution to \(\Qpertslabavg\) from the most unstable small-scale modes. As expected, we find that the turbulent heat flux due to the small-scale modes pushes the large-scale gradient~\(\vtvect\) towards the linearly stable configuration \cref{eq_smallscale_inst_cond}. 

Now consider \cref{curvy_slabavg_phi}, the evolution equation for \(\slabavg{\phinorm}\). The relevant nonlinear terms are
\begin{align}
	&\slabavg{\pbra{\phinormslabpert}{\dw{\phinormslabpert} - \delsq \phinormslabpert} + \del \bcdot \pbra{\del \phinormslabpert}{\deltaTslabpert}} = \del \bcdot \slabavg{\pbra{\del \phinormslabpert}{\prslabpert}} \nonumber \\
	&\quad= \del \del \mathbf{:} \sum_\vq (\uvect{z}\times\vq)\vq \left(1 + \renobra{\frac{\deltaTslabpert_\vq}{\phinormslabpert_\vq}}\right) |\phinormslabpert_\vq|^2 \equiv \del \del \mathbf{:} \Pipertslabavg. 
\end{align}
The collisionless calculations of \cref{sect_largekperp} are straightforward to generalise for the collisionless parasitic small-scale instability. They yield the same relation for \(\reinline{\deltaTslabpert_\vq / \phinormslabpert_\vq}\), viz., \mbox{\(1 + \reinline{\deltaTslabpert_\vq / \phinormslabpert_\vq} = \orderinline{\qperp^{-2}}\)}. However, as we shall discuss in \cref{sect_turb_stress}, the presence of nonzero~\(\chi\) alters this to \(1 + \reinline{\deltaTslabpert_\vq / \phinormslabpert_\vq} = \orderinline{\qperp^{-1}}\). Assuming therefore that the dominant parasitic modes satisfy \(1 + \reinline{\deltaTslabpert_\vq/\phinormslabpert_\vq} \lesssim \orderinline{q_\perp^{-1}}\), we find \mbox{\(\del \del \mathbf{:} \Pipertslabavg \lesssim \kperp^2 \qperp |\phinormslab|^2\)}. However, \cref{eq_smallscaleQ} implies \(\del \bcdot \Qpertslabavg \sim \kperp |\phinormslab|^2 \), and so 
\begin{equation}
		\frac{\del \del \mathbf{:} \Pipertslabavg}{\del \bcdot \Qpertslabavg} \lesssim \kperp\qperp \sim \mathcal{O}\left[\left(\frac{\vt^{1/4}}{\chi}\right)^{1/3}\right] \ll 1,
\end{equation}
in line with the assumption on scales formulated at the end of \cref{sect_scalesep}. Therefore, assuming that \(\slabavg{\dw{\phinorm}} \sim \slabavg{\deltaT}\)\footnotemark\footnotetext{While this is in contradiction with the 2D curvature-mode scaling \(\slabavg{\dw{\deltaT}} / \slabavg{\phinorm} \sim \sqrt{\vt} \gg 1\), we do find that \(\slabavg{\dw{\phinorm}} \sim \slabavg{\deltaT}\) in our 3D simulations. This is due to the strong influence of the 3D modes on the dynamical evolution of \(\slabavg{\deltaT}\): see \cref{eq_psi2d_energy_appendix} and the discussion thereafter.} and that they evolve on the same time scale, we conclude that the main effect of the small-scale modes is to provide a feedback to the large-scale temperature in the form of the additional heat flux \(\Qpertslabavg\). 

We can thus summarise the equations that govern the evolution of \(\slabavg{\dw{\phinorm}}\) and \(\slabavg{\deltaT}\) as
\begin{gather}
	\partial_t \slabavg{\dw{\phinorm}} - \partial_y \left( \slabavg{\dw{\phinorm}} + \slabavg{\dw{\deltaT}} \right) + \vt \partial_y  \delsq \slabavg{\dw{\phinorm}} \nonumber \\ +\dw{\pbra{\slabavg{\phinorm}}{\slabavg{\dw{\phinorm}} - \delsq \slabavg{\phinorm}} } 
	+ \dw{\del \bcdot \pbra{\del \slabavg{\phinorm}}{\slabavg{\deltaT}} }
	+ \chi \nabla_\perp^4 \left(\achi\slabavg{\dw{\phinorm}} - \bchi\slabavg{\dw{\deltaT}}\right) = 0 \label{curvy_slabavg_phi_lowestorder}\\
	\label{curvy_slabavg_psi_lowestorder}
	\partial_t\slabavg{\deltaT} +\vt\partial_y \slabavg{\phinorm} +\pbra{\slabavg{\phinorm}}{\slabavg{\deltaT}} - \chi \delsq \slabavg{\deltaT} = -\slabavg{\pbra{\phinormslabpert}{\deltaTslabpert}},
\end{gather}
where the influence of small-scale fields appears only in the temperature equation \cref{curvy_slabavg_psi_lowestorder}. The system of \modeleqnsslabpertsimple{} and \modeleqnsslabavgsimple{} respects the conservation of the two conserved quantities described in \cref{sect_cons} --- this is shown in \cref{appendix_cons}. 

The above reasoning does not apply to the ZFs. Indeed, the equation for \(\zf{\phinorm}\) is
\begin{equation}
	\label{eq_zonalphi_with_parasiticstress}
	\pt \zf{\phinorm} - \zf{\px \slabavg{\phinorm} \py \slabavg{\phinorm + \deltaT}} - \px^2(a\zf{\phinorm} - b\zf{\deltaT}) = \zf{\px \phinormslabpert \py (\phinormslabpert + \deltaTslabpert)} = \zf{\Pipert_{xx}}.
\end{equation}
This shows that the small-scale stress \(\Pipert\) influences the zonal electrostatic potential \(\zf{\phinorm}\) more strongly (by a factor of \(\kperp^{-2}\)) than it does the nonzonal \(\slabavg{\dw{\phinorm}}\). This is a consequence of the electron response \cref{eq_eresponse} and the asymptotically smaller `inertia' (i.e., the factor in front of the time derivative) \(\propto \kperp^2\) of the ZFs compared to the `inertia' \(\propto (1+\kperp^2)\) of the nonzonal \(\dw{\phinorm}\). Thus, the right-hand side of \cref{eq_zonalphi_with_parasiticstress} cannot be ignored. In fact, as we showed in \cref{sect_dimits}, the addition of 3D effects, and hence of parasitic modes, has a profound impact on the stability of the Dimits-state ZFs, viz., the momentum flux \(\Pipert_{xx}\) extends the Dimits state to higher temperature gradients than the 2D system allows. Let us show why this is the case.

\subsubsection{Turbulent stress due to parasitic slab-ITG modes}
\label{sect_turb_stress}

In \citet{ivanov2020}, we obtained a prediction for the critical gradient \(\vtcrittd (\chi)\) above which a Dimits state with strong ZFs could not be sustained. This prediction was based on considerations of the ratio \(\re{\deltaT_\vk/\phinorm_\vk}\) for the linear modes with largest growth rate. As explained in \cref{sect_dimits2d}, this ratio determines the balance of Reynolds and diamagnetic stresses for an individual Fourier mode: if \(\re{\deltaT_\vk/\phinorm_\vk} > -1\), then the Reynolds stress is larger and the mode favours a Dimits state, otherwise its diamagnetic stress is larger and the mode helps suppress the coherent ZFs needed for the Dimits state. In 2D, this ratio is sensitive to both the temperature gradient \(\vt\) and the collisionality \(\chi\), and thus an appropriate balance between these two parameters is required in order to have \(\re{\deltaT_\vk/\phinorm_\vk} > -1\) for the dominant modes and thus to keep the system in the Dimits state. In particular, for \(\vt \gg 1\), the Dimits threshold is given by \(\vt / \chi = \text{const}\).

Let us adopt a similar approach for the fastest-growing small-scale sITG modes located at \(\kpar \approx \vtvect\bcdot\vk \kperp\). Equation~\cref{eq_Toverphi_slab_lowestorder} tells us that these modes satisfy \mbox{\(\re{\deltaT_\vk/\phinorm_\vk} = -1 + \order{\kperp^{-2}}\)} for \(\kperp \gg 1\). Therefore, to lowest order, the sITG modes are Dimits-marginal, i.e., their Reynolds and diamagnetic stresses balance out. This means that the lowest-order collisionless calculations of \cref{sect_largekperp} are insufficient for our needs. While we can extend these calculations to \(\order{\kperp^{-2}}\), ignoring \(\chi\) is, in fact, an unacceptable oversimplification. As we are about to see, nonzero \(\chi\) provides a \(\order{\kperp^{-1}}\) correction to \(\re{\deltaT_\vk/\phinorm_\vk}\) and hence renders any collisionless higher-order corrections irrelevant.

\begin{figure}
	\centering
	\includegraphics[scale=0.26]{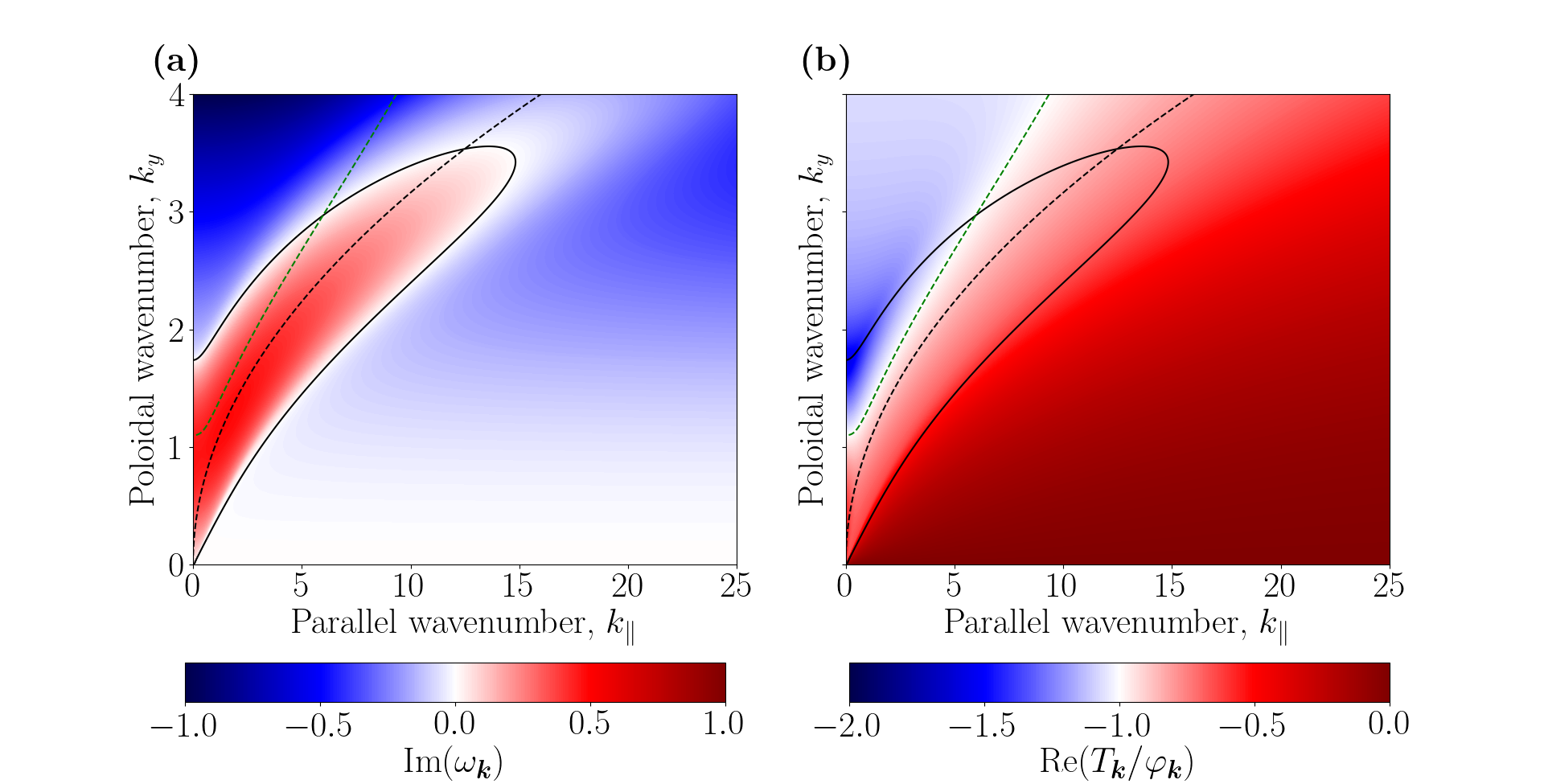}
	\caption{ \textbf{(a)} Linear growth rate and \textbf{(b)} the ratio \(\re{\deltaT_\vk/\phinorm_\vk}\) of the most unstable \mbox{(\(k_x=0\))} modes versus \(\kpar\) and \(k_y\) for \(\vt = 1\) and \(\chi = 0.1\). The green dashed line is \(\re{\deltaT_\vk/\phinorm_\vk} = -1\). The black dashed line is the location of the largest collisionless growth rate \(\kpar = \vt k_y^2\). While the green and black lines would coincide to~\(\orderinline{\kperp^{-2}}\) for the collisionless modes, we see that the addition of collisions shifts the linearly unstable modes towards the Dimits-favourable \(\re{\deltaT_\vk/\phinorm_\vk} > -1\) ratio. }
	\label{fig_linear_col_slab}
\end{figure}

As discussed in the beginning of \cref{sect_nl}, the dominant collisionless sITG modes are found at \(\kperp^3 \sim \vt / \chi\). It turns out that at those scales, it is collisional effects that determine \(\re{\deltaT_\vk/\phinorm_\vk}\). The details of the relevant calculation can be found in \cref{appendix_slabitg_general}. We find that for \(\chi\) ordered as \(\vt \sim \chi \kperp^3\), the most unstable small-scale sITG modes are still located at \(\kpar = \vtvect\bcdot\vk\kperp\) and satisfy 
\begin{equation}
	\label{eq_Toverphi_cols_maintext}
	\frac{\deltaT_\vk}{\phinorm_\vk} = -1 - \sqrt{-\gammakh^2 + \frac{i (\achi + \bchi - 1)\chi \kperp^2}{2\vtvect\bcdot\vk}} + \order{\kperp^{-2}},
\end{equation}
where we take the branch of the square root with a positive imaginary part. Since \(\gammakh^2 > 0\) and \(\vtvect\bcdot\vk > 0\) [as stipulated after \cref{eq_disp_slabpert}], we find that the sign of the real part of the square root in \cref{eq_Toverphi_cols_maintext} is set by the sign of \(\achi + \bchi - 1\). Plugging in the numerical values \(\achi = 9/40\) and \(\bchi = 67/160\), we find \(\achi + \bchi - 1 < 0\), hence the square root has a negative real part and \(\re{\deltaT_\vk/\phinorm_\vk} > -1\). Thus, collisionality always pushes the otherwise Dimits-marginal small-scale sITG modes to side with the Reynolds stress and reinforce the~ZFs.\footnotemark\footnotetext{Note that the same holds for the \(\chi\)ITG modes of \cref{appendix_col_slab}, viz., the sign of \(1 + \re{\deltaT_\vk/\phinorm_\vk}\) is equal to that of \(1 - \achi - \bchi\). And, as \cref{fig_colslab_alphabeta} shows, we always find \(\re{\deltaT_\vk/\phinorm_\vk} > -1\) for our values of \(\achi\) and \(\bchi\).} This is evident in \cref{fig_linear_col_slab}.

The sensitivity of \(\re{\deltaT_\vk/\phinorm_\vk}\) to the numerical factors \(\achi\) and \(\bchi\) allows us to carry out a simple test of the above theory. We pick a simulation that is in the Dimits state in~3D, but above the 2D Dimits threshold, i.e., has \(\vt > \vtcrittd\). We restart this simulation, but set \(\achi = 1\) for all nonzonal modes. Linearly, this increases nonzonal viscosity and reduces growth rates, without affecting zonal physics. Na\"ively, one might expect that with an increased damping of the turbulence, the Dimits state should become `stronger'. However, such reasoning does not take into account the structure of the 3D modes and the change in the balance of Reynolds and diamagnetic stresses stemming from the change of the sign of \(\achi + \bchi - 1\). Indeed, in this numerical experiment, we discover that the Dimits regime is destroyed and strong turbulence sets in, just as the analysis above predicts. This is clear evidence that the most consequential role of collisionality for the Dimits regime of \modeleqns{} is not to dissipate turbulent energy, but rather to regulate the turbulent stress via the ratio \(\re{\deltaT_\vk/\phinorm_\vk}\). This also suggests, for future analysis of the Dimits transition in different models of ITG turbulence, that the Dimits threshold may prove to be sensitive to the details of dissipation effects on the unstable modes, especially if, in the absence of collisions, these modes are Dimits-marginal, i.e., if they satisfy \(\reinline{\deltaT_\vk/\phinorm_\vk} \approx -1\).

Let us also note that in the simple case of sITG modes in slab geometry, a more general calculation that includes kinetic effects is possible. In appendices~\ref{appendix_kinetic_disp}~and~\ref{appendix_stressure}, we derive the kinetic sITG dispersion relation and the kinetic equivalent of \cref{eq_zonalphi}. Then, applying the ideas developed in \citet{ivanov2020} and in this work, we show that ZF-driving small-scale sITG modes are not limited to the cold-ion limit and could play a role outside of the realm of simple fluid approximations. This, of course, can be conclusively confirmed only by appropriate GK simulations.

\subsection{Breaking the Dimits state}
\label{sect_breakingdimits}

Recall that the 2D critical gradient \(\vtcrittd\) was found to be an increasing function of~\(\chi\). Na\"ively, this makes sense on the basis of `more dissipation means less turbulence': one expects that one should be able to compensate for an increase in the drive \(\vt\) by an appropriate increase in \(\chi\) and thus keep the system in the Dimits state. However, this simple picture is false. Collisionality and drive are important for maintaining the Dimits state not because they provide dissipation and injection of energy, but rather because they determine the ratio \(\re{\deltaT_\vk/\phinorm_\vk}\) for the linearly unstable modes. In 2D, this ratio is sensitive to both \(\vt\) and \(\chi\); however, this is not the case in 3D as the small-scale sITG modes always favour the Dimits state. First, their turbulent momentum flux was shown to satisfy \(\re{\deltaT_\vk/\phinorm_\vk} \approx -1\), with collisions pushing this ever so slightly in the Dimits-stable direction of \(\re{\deltaT_\vk/\phinorm_\vk} > -1\) (see \cref{sect_turb_stress}). Secondly, they provide an effective thermal diffusion for the large-scale \(\slabavg{\deltaT}\), which in turn reduces the absolute value of \(\slabavg{\deltaT_\vk} / \slabavg{\phinorm_\vk}\) and partially suppresses the tendency of large-scale modes to destroy the Dimits state (\cref{sect_largescale_resp}). Our numerical simulations show that the combination of the mode structure of the small-scale instability and its influence on large-scale modes proves to be enough to keep the system in the Dimits regime regardless of \(\vt\) and \(\chi\). As \(\vt\) increases beyond the 2D Dimits threshold \(\vtcrittd\), the 2D modes flip the sign of their turbulent momentum flux and start eroding the ZFs, but the small-scale sITG modes are able to provide enough ZF drive to maintain the Dimits state (see \cref{fig_turb_visc_vs_kappaT}). Figure~\ref{fig_flow_diagram}a illustrates the Dimits saturation mechanism.

\begin{figure}
	\centering
	\includegraphics[scale=0.26]{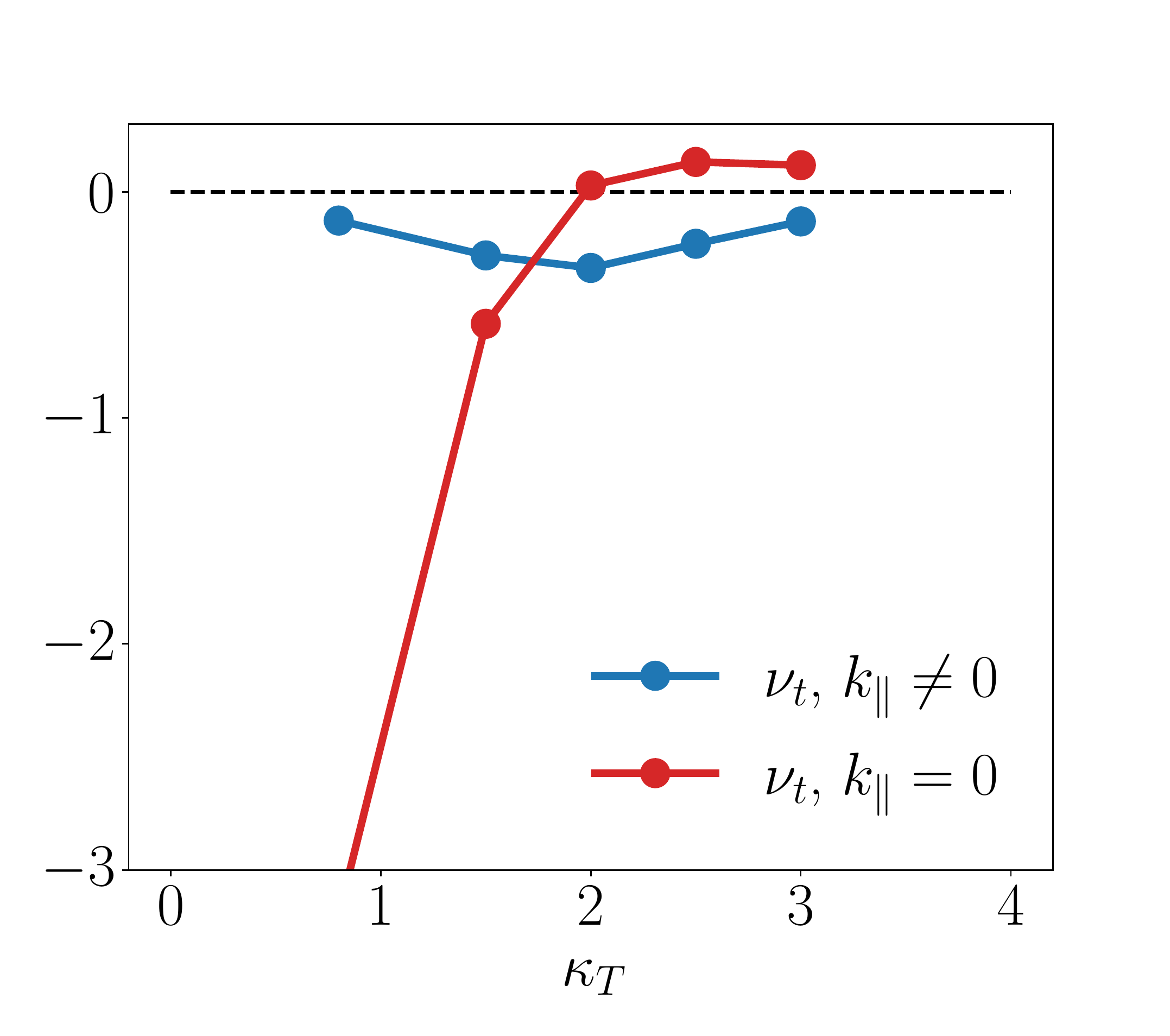}
	\caption{Dependence of the turbulent viscosity \cref{eq_momcorr_def} on the temperature gradient for \(\chi = 0.1\) and \(\Lpar = 1\). The 2D Dimits regime ends at \(\vtcrittd \approx 1\). In 3D simulations, the 2D modes eventually reverse their turbulent viscosity (red), but the 3D sITG modes continue to feed the ZFs through a negative turbulent viscosity (blue). The data is taken from simulations with fixed ZF profiles. }
	\label{fig_turb_visc_vs_kappaT}
\end{figure}
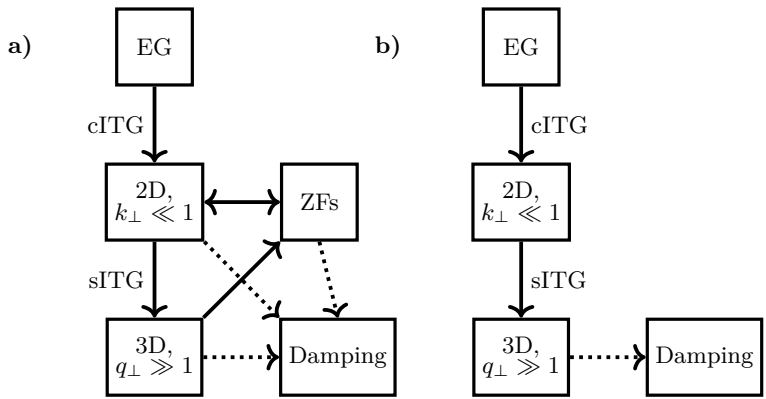
\begin{figure}
	\centering
	\begin{tabular} {c c}
	\begin{tikzpicture}[
	roundnode/.style={rectangle, align=center, draw=black!100, fill=white!100, very thick, minimum size=10mm},
	]
	\node[roundnode]      (EG)                              {EG};
	\node[roundnode]        (C-ITG)       [below=of EG] {2D, \\ $\kperp \ll 1$};
	\node[roundnode]        (ZF)       [right=of C-ITG] {ZFs};
	\node[roundnode]        (S-ITG)       [below=of C-ITG] {3D, \\ $\qperp \gg 1$};
	\node[roundnode]        (VISC)       [right=of S-ITG] {Damping};
	\node[text width=3mm] [left=of EG] {\textbf{a)}};
	
	\draw[->, line width=0.5mm] (EG.south) to node [left] {cITG} (C-ITG.north);
	\draw[<->, line width=0.5mm] (C-ITG.east) -- (ZF.west);
	\draw[->, line width=0.5mm] (C-ITG.south) to node [left] {sITG} (S-ITG.north);
	\draw[->, line width=0.5mm] (S-ITG.north east) -- (ZF.south west);
	\draw[dotted, ->, line width=0.5mm] (C-ITG.south east) -- (VISC.north west);
	\draw[dotted, ->, line width=0.5mm] (S-ITG.east) -- (VISC.west);
	\draw[dotted, ->, line width=0.5mm] (ZF.south) -- (VISC.north);
%

	\node[roundnode]        (S-ITGST)       [right=of VISC] {3D, \\ $\qperp \gg 1$};
	\node[roundnode]        (VISC)       	[right=of S-ITGST] {Damping};
	\node[roundnode]        (C-ITGST)       [above=of S-ITGST] {2D, \\ $\kperp \ll 1$};
	\node[roundnode]      	(EGST)          [above=of C-ITGST] {EG};
	\node[text width=3mm] [left=of EGST] {\textbf{b)}};
	\draw[->, line width=0.5mm] (EGST.south) to node [right] {cITG} (C-ITGST.north);
	\draw[->, line width=0.5mm] (C-ITGST.south) to node [right] {sITG} (S-ITGST.north);
	\draw[dotted, ->, line width=0.5mm] (S-ITGST.east) -- (VISC.west);
\end{tikzpicture}
\end{tabular}
	\caption{Schematic of the flow of energy in \textbf{(a)} the Dimits regime, characterised by strong, turbulence-shearing, staircase ZFs, and \textbf{(b)} strong turbulence, where no such ZFs can be generated or sustained. In the Dimits regime, the equilibrium gradients (EG) inject energy into large-scale modes via the 2D cITG instability. These can then drive ZFs via the secondary instability \citep[see \S2.8 of ][]{ivanov2020} and small-scale perturbations via the parasitic sITG instability (see \cref{sect_slabsec}). In the 2D Dimits regime (\(\vt < \vtcrittd\)), the curvature-driven large-scale modes generate a negative turbulent viscosity on the ZFs and hence reinforce the Dimits state. For \(\vt > \vtcrittd\), the 2D modes erode the ZFs, but the ZF drive of the parasitic modes sustains the ZFs (see \cref{sect_turb_stress}). On the other hand, if a Dimits state cannot be achieved, the energy injected into the large-scale modes is transferred to small scales via the parasitic sITG instability, whence it cascades to even smaller, linearly stable scales where it is taken out of the system.}
	\label{fig_flow_diagram}
\end{figure}

However, the small-scale sITG modes are able to maintain the Dimits state only if the 3D system is `3D enough'. Namely, if we restrict the system in \(z\) by either squeezing it to a small \(\Lpar\) (see \cref{sect_shortlpar}) or by cutting off large-\(\kpar\) modes (see \cref{sect_parres}), we can break the ZF-dominated Dimits regime and push the system into a stongly turbulent state. The former of these methods can be deemed `physical' in the sense that real system can be geometrically limited along the magnetic field, e.g., by magnetic shear. The latter is but a numerical artefact in our cold-ion system; however, parallel transport processes, which were ordered out in \modeleqns{}, do provide a large-\(\kpar\) cut-off for the sITG instability (see \cref{sect_parres}).  

Regardless of how the Dimits state is broken, amplitudes remain finite. The small-scale instability is able to extract energy efficiently from the large-scale (\(\kperp \ll 1\)) fields, into which the cITG instability inputs energy, and to dump it into the small scales \(\kperp \gg 1\) of the sITG instability, whence it cascades to smaller scales, where dissipation can take it out of the system. Figure~\ref{fig_flow_diagram}b shows the flow of energy in the strongly turbulent state.

\FloatBarrier

\subsubsection{Effect of parallel system size on the Dimits state}
\label{sect_shortlpar}

\begin{figure}
	\centering
	\begin{tabular}{c c}
		\includegraphics[scale=0.27]{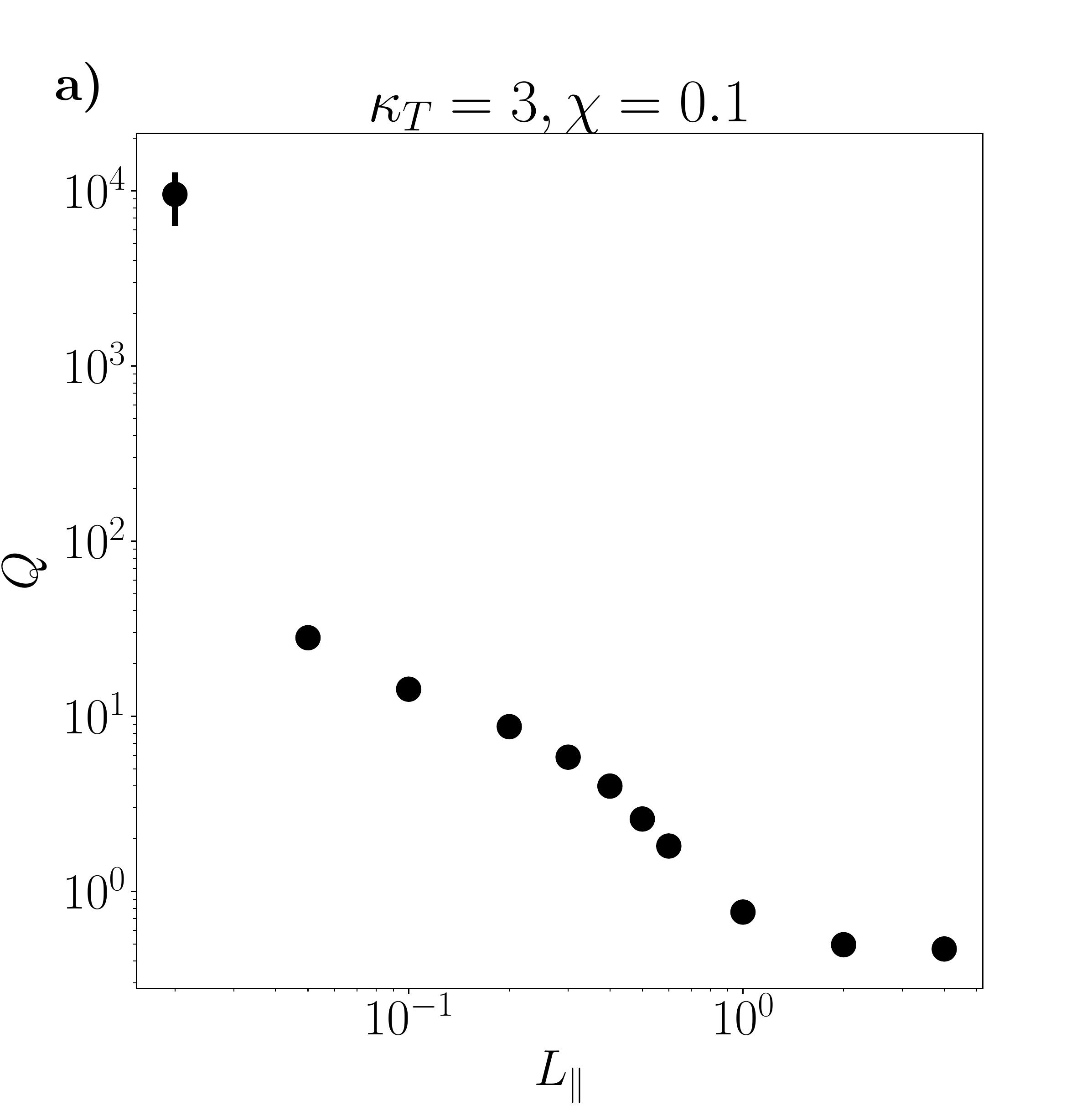} &
		\includegraphics[scale=0.27]{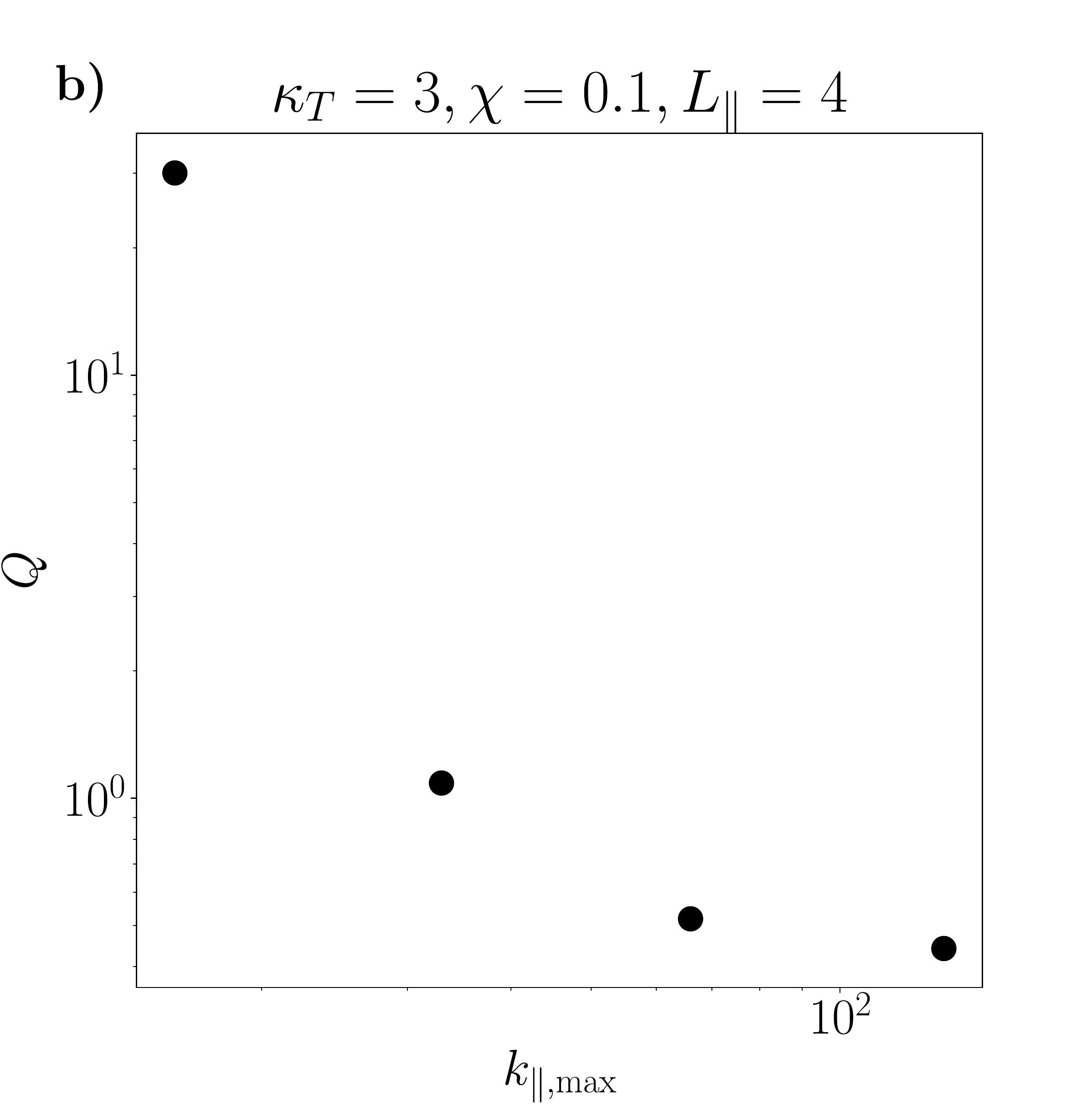}
	\end{tabular}
	\caption{Dependence of the saturated turbulent heat flux \(Q\) on \textbf{(a)} the parallel size of the box \(\Lpar\) and \textbf{(b)} the largest parallel Fourier mode \(k_{\parallel, {\rm max}}\) that is included in the simulation.}
	\label{fig_q_vs_lz}
	\label{fig_q_vs_maxkpar}
\end{figure}

Figure~\ref{fig_q_vs_lz}a shows a typical example of the dependence of the saturated turbulent heat flux \(Q\) on the parallel size of the box \(\Lpar\) for parameters \(\vt\) and \(\chi\) that lie beyond the 2D Dimits regime. For such parameters, the \(\Lpar = 0\) system does not reach finite-amplitude saturation. For \(\Lpar\) large enough, \(Q\) is independent of \(\Lpar\), just as it was for parameters that were within the 2D Dimits threshold (see \cref{sect_dimits}). As \(\Lpar\) is decreased, the ZFs break up and the system enters a strongly turbulent state. In \cref{fig_q_vs_lz}a, this happens for \(\Lpar < 1\). As \(\Lpar\) approaches \(0\), \(Q\) starts to increase rapidly, signifying the approach to the 2D state, where a blow up occurs.

Therefore, for each pair of values of \(\vt\) and \(\chi\), there exists a critical \(\Lparcrit\) such that the system is in the Dimits state for \(\Lpar > \Lparcrit\) and in the strongly turbulent regime for \(\Lpar < \Lparcrit\). It is clear that \(\Lparcrit = 0\) if \(\vt <\vtcrittd\), i.e., if \(\vt\) and \(\chi\) are such that the 2D system is able to reach saturation. The dependence of \(\Lparcrit\) on \(\vt\) and \(\chi\) for \(\vt > \vtcrittd\) is not known at this point, due to the numerical cost of resolving simultaneously both the large \(\kpar\) of the small-scale modes (see \cref{sect_parres}) and the box-sized \(\kpar \sim \Lpar^{-1}\).
 
\subsubsection{Effect of parallel resolution on the Dimits state}
\label{sect_parres}

The scale separation between the large-scale cITG modes and the small-scale sITG modes increases the numerical cost of solving \modeleqns{}. When the parallel resolution, i.e., the largest \(\kpar\) in the simulation, is too small, the Dimits state is destroyed numerically and the system is pushed into a strong-turbulence regime for parameters for which a Dimits state would have existed if given sufficient parallel resolution. This is shown in \cref{fig_q_vs_maxkpar}b. Empirically, we have found that a good rule of thumb is `not to chop the leaves' of the instability, i.e., to make sure that the wavenumbers that lie within the unstable `leaves' at \(\kpar \sim \vt \kperp^2\) (see \cref{fig_linear3d}) are fully included in the simulation.\footnotemark\footnotetext{Of course, this is but a rule of thumb and cannot be entirely accurate because, as discussed in \cref{sect_num_evidence}, the small-scale instability is driven not by \(\vt\), but rather by the gradients of the large-scale fields. In other words, the linear 3D modes shown in \cref{fig_linear3d} are irrelevant for the saturated state. However, we expect that the temperature gradients associated with the saturated large-scale perturbations scale with \(\vt\) and so this rule of thumb is a good heuristic guide for setting up simulations.} This, however, rapidly increases the numerical cost of the simulations. Recall that according to \cref{eq_cless_slab_col_cutoff}, the collisionless sITG instability satisfies \(\kperp \sim (\vt/\chi)^{1/3}\). Therefore, for a fixed~\(\chi\), the dimensional \(\kpar\) of the unstable modes is given by 
\begin{equation}
	\label{eq_cless_kpar_col_cutoff}
	\kpar L_B \sim \vt \kperp^2 \sim \frac{\vt^{5/3}}{\chi^{2/3}}. 
\end{equation}
The number of Fourier modes required to resolve a simulation properly then scales as~\(\vt^{5/3}\), in addition to scaling linearly with \(\Lpar\). This quickly renders numerical efforts futile, even for a fluid code. 

Of course, the infinitely extending `leaves' of the instability in our 3D model will, in reality, be `chopped off' by phenomena that have been ordered out of our equations by \cref{eq_ordering}. For example, \cref{eq_ordering} orders out the parallel thermal diffusion \citep{braginskii65}, but we can nonetheless estimate the dimensional \(\kpar\) at which this effect will become important. This is the parallel scale at which the collisional heat conduction rate \(\vti^2\kpar^2 / \nu_i\) becomes comparable to \(\pt \sim c_s / L_B\) in \modeleqns{}, which happens at 
\begin{equation}
	\label{eq_col_conduction_scale}
	\kpar L_B \sim \sqrt{\frac{L_B}{\sqrt{\tau}\lambda_\text{mfp}}},
\end{equation}
where \(\lambda_\text{mfp} = \vti / \nu_i\) is the mean-free path. In our ordering \cref{eq_ordering}, \(\lambda_\text{mfp} / L_B \sim \tau^{3/2}\), so we find that the collisional heat conduction comes into play at \(\kpar L_B \sim 1/\tau\). Formally, this is outside of the regime \(\kpar L_B \sim 1\) assumed in \modeleqns{}, but physically, we conclude that the Dimits regime could be broken if the collisional cut-off \cref{eq_cless_kpar_col_cutoff} is superseded by the Braginskii scale \cref{eq_col_conduction_scale}, i.e., if 
\begin{equation}
	\frac{L_B}{L_T} \gtrsim \left(\frac{L_B}{\lambda_\text{mfp}}\right)^{7/10}\tau^{-11/20},
\end{equation}
where we used \(\vt \sim \tau L_B/L_T\) and \(\chi \sim L_B\tau^{3/2}/\lambda_\text{mfp}\). In a real fusion device, this condition will not be very difficult to reach, but, in fact, the more relevant mechanism for limiting the parallel wavenumber of the sITG instability is parallel streaming rather than collisional heat conduction. In \cref{appendix_kinetic_disp}, we show that this too imposes a limit on the parallel wavenumber that is \(\orderinline{\tau^{-1}}\) too large to be included in our ordering of~\(\kpar\). Namely, the sITG cut-off is given by 
\begin{equation}
	\kpar^{(c)} L_B = \frac{L_B}{2\sqrt{\pi}(1+\tau)L_T},
\end{equation}
which supersedes the the collisional cut-off \cref{eq_cless_kpar_col_cutoff} if
\begin{equation}
	\frac{L_T}{\lambda_\text{mfp}} \lesssim \tau (1+\tau)^{3/2}.
\end{equation}
Again, such a regime is entirely plausible for a real fusion device.

We conclude that in a more realistic physical regime than the one assumed in the derivation of our model equations \modeleqns{}, the behaviour (or even existence) of parasitic sITG modes may be influenced by parallel thermal diffusion or parallel streaming in a way that breaks the Dimits regime at large enough temperature gradients. 

\section{Discussion}
\label{sect_discussion}

Following our analysis of the Dimits regime and its threshold in the 2D model of \citet{ivanov2020}, we have been able to extend both our model and our understanding of ITG turbulence to 3D. The important qualitative features of the 2D Dimits state, viz., strong coherent ZFs with patch-wise constant shear, turbulent bursts, and localised travelling structures survive the inclusion of 3D physics largely unchanged (see \cref{sect_dimits}). ZFs are generated and destroyed by the Reynolds and diamagnetic stresses of sheared ITG turbulence, respectively. If the Reynolds stress is larger, coherent ZFs can be maintained and the system settles into a low-transport Dimits state. Otherwise, a strongly turbulent, high-transport state arises in which saturation occurs unaided by ZFs. In the 2D model, the ratio of Reynolds to diamagnetic stress is sensitive to the equilibrium parameters --- the temperature gradient \(\vt\) and the ion collisionality \(\chi\) --- and thus an appropriate balance of the two is required in order to keep the system within the Dimits regime. With the inclusion of parallel physics, however, the stresses are modified by the 3D-exclusive sITG instability, which is found always to favour the ZFs (see \cref{sect_turb_stress}). Unless 3D physics is restricted either by a small parallel box size (\cref{sect_shortlpar}) or by insufficient numerical resolution (\cref{sect_parres}), the sITG instability is able to tip the stress balance in the Reynolds direction and a Dimits state is established regardless of the values of the equilibrium parameters.

This 3D sITG instability is found to be scale-separated from the 2D cITG instability (see \cref{sect_scalesep}). In the absence of collisions, the former exists at arbitrarily small perpendicular and parallel scales, while the latter is confined to large scales. This scale separation allows for sITG modes that are predominantly driven not by the equilibrium gradients but rather by the local gradients of large-scale fields, which are themselves driven by the equilibrium gradients (i.e., the sITG instability is parasitic). The nonlinear energy transfer from large-scale to small-scale modes that results from the sITG instability is found to have the form of an effective large-scale thermal diffusion (see \cref{sect_largescale_resp}). The combination of this thermal diffusion and the favourable turbulent stress of the small-scale modes are what makes the 3D Dimits state much more resilient than its 2D counterpart. 

The fact that the Dimits state is governed by essentially the same physical mechanisms in both the 2D and 3D cold-ion \(Z\)-pinch systems gives us not only hope that one day we could understand the Dimits regime of full-blown GK, but also a solid foundation of numerical and analytical work upon which to build such an undertaking. Although there is some numerical evidence of important similarities between these simple systems and GK, e.g., the ferdinon structures seen both by us and by \citet{vanwyk2016, vanwyk2017} in their GK simulations of an experimentally realistic configuration, there is still much unknown. The details of the Dimits state in our 3D model depend on certain peculiar features of cold-ion physics. It is the cold-ion approximation that permits the parasitic small-scale sITG instability that underlies the main differences between the 2D and 3D models. As this is only one asymptotic limit of GK, it is difficult to extrapolate any quantitative predictions. However, it is important to note that the kinetic, \(\tau \sim 1\), dispersion relation also predicts a collisionless sITG instability at arbitrarily large \(\kperp\) (see \cref{appendix_kinetic_disp}), as was already established by \citet{smolyakov2002}. Just as in the cold-ion fluid model, these sITG modes appear to favour a ZF-dominated state (\cref{appendix_stressure}). Thus, the appearance of parasitic modes is not necessarily limited to our cold-ion model and, in certain regimes, could also be a feature of low-collisionality GK. This may also require a careful investigation of GK collisions along the lines of \citet{frei2022}. All of this, combined with the fact that the nature of the Dimits state in the 2D and 3D models is essentially the same, encourages us to carry our ideas over into the vastly more complex world of GK. At this point, it is unknown whether the Reynolds--diamagnetic stress competition is also behind the Dimits transition in GK. One of the prominent alternative ideas is the primary-secondary-tertiary scenario, first proposed by \citet{rogersdorland2000}. Recently, there have been a number of publications discussing the applicability of this paradigm to both fluid and kinetic models \citep{stonge2017, zhu2018_tertiary, zhu2019_dimits, zhu2020jpp, hallenbert2021}. Note that, as we showed in \citet{ivanov2020}, the Dimits transition that we observe cannot be explained by the tertiary instability of ZFs. It is possible that the nature of the transition to high transport in realistic GK simulations is, in fact, not as clear-cut as it is in the simple models, but is rather a combination of both mechanisms, viz., the competition between the stresses and a tertiary instability.


Another important feature that our model lacks is magnetic shear. It is well-known that this can have a significant effect on both the linear instabilities and turbulence levels in realistic-geometry GK simulations \citep{kinsey2006}. Notably, much effort today is being devoted to spherical tokamak designs, which can have large values of field-line-averaged magnetic shear combined with nontrivial variations in the local shear. Therefore, we consider the addition of magnetic shear to our analytical and numerical models to be a key direction for future work.

\section*{Acknowledgements}

The authors would like to thank M. Barnes and S. Tobias for many useful comments. This work has been carried out within the framework of the EUROfusion Consortium and has received funding from the Euratom Research and Training Programme 2014–2018 and 2019–2020 under Grant Agreement No. 633053. The views and opinions expressed herein do not necessarily reflect those of the European Commission. The work of A.A.S. was supported in part by the UK EPSRC Programme Grant EP/R034737/1.

\section*{Declaration of interests}

The authors report no conflict of interest.
\newpage
\appendix
\section{Derivation of the 3D model}
\label{appendix_derivation}

We follow the derivation in Appendix A of \citet{ivanov2020}, but retain the parallel-streaming term in the GK equation. For the sake of brevity, we shall use the notation and definitions of \citet{ivanov2020}. 

The electrostatic ion GK equation is
\begin{align}
\label{eq_curvy_gk}
&\partd{}{t} \left(h - \avgR{\phinorm}F_i\right) + \vpar\pz h  + \frac{\rho_i \vti}{2L_T}\left( \frac{v^2}{v^2_{ti}} - \frac{3}{2} \right)F_i \partd{\avgR{\phinorm}}{Y} - \frac{\rho_i \vti}{L_B} \left( \frac{v^2_\parallel}{\vti^2} + \frac{v^2_\perp}{2 \vti^2} \right)\partd{h}{Y}
\nonumber \\ & \quad + \frac{1}{2} \rho_i \vti \left\{ \avgR{\phinorm}, h \right\}  = \avgR{C_{l}[h]},
\end{align}
closed via the quasineutrality condition and \cref{eq_eresponse}:
\begin{equation}
\label{eq_curvy_qn}
\frac{1}{n_i}\int d^3\vect{v}  \ \avgr{h} = \phinorm + \tau \phinorm'.
\end{equation}
The 2D fluid model was derived in a highly collisional (\(\pt \ll \nu_i\)), cold-ion (\(\tau \ll 1\)), long-wavelength (\(\kperprhoisq \ll 1\)) limit of the ion GK equation that obeys \cref{eq_ordering}. Note that, as discussed in \cref{sect_3dmodel}, in order to retain the sITG instability in the final equations, we need to order \(\pz \sim L_B^{-1}\). Thus, the parallel-streaming term is ordered as
\begin{equation}
	\vpar \pz h \sim \frac{\vti}{L_B} h \ll \pt h \sim  \frac{c_s}{L_B} h \sim \frac{\vti}{L_B \sqrt{\tau}}h,
\end{equation}
i.e., it is one order of \(\sqrt{\tau}\) smaller than the \(\pt h\) term. This means that here we need to expand the distribution function in \(\sqrt{\tau}\), rather than in \(\tau\), as was done in \citet{ivanov2020}. In order to be consistent with the notation of our 2D derivation, we set \mbox{\(h = h^{(0)} + h^{(1/2)} + h^{(1)} + ...\)}, where \(h^{(1/2)} \sim \sqrt{\tau} h^{(0)}\), etc.

\subsection{Lowest-order solution}

To order \(\orderinline{\sqrt{\tau}}\), the ion GK equation \cref{eq_curvy_gk} is dominated by collisions, viz.,
\begin{equation}
	C_l[h^{(0)} + h^{(1/2)}] = 0.
\end{equation}
The solution to this equation is a perturbed Maxwellian distribution \citep{newton2010}:
\begin{equation}
h^{(0)} + h^{(1/2)} = \left[\frac{\delta N}{n_i} + \frac{\delta T}{T_i} \left(\frac{v^2}{\vti ^2} - \frac{3}{2}\right) + \frac{2 \vpar \deltaVpar}{\vti^2}\right]F_i.
\end{equation}
Here \(\delta T / T_i\) will turn out to be just the ion temperature perturbation, while the density-like quantity \(\delta N / n_i\) is
\begin{equation}
	\frac{\delta N}{n_i} = \phinorm + \tau \dw{\phinorm} -\frac{1}{4}\rho_i^2 \delsq \left(\phinorm + \frac{\delta T}{T_i}\right) + \order{\kperprhoisqq\phinorm}.
\end{equation}
For more details, see the derivations in \citet{ivanov2020}. The ordering \(\deltaVpar \sim \tau c_s \phinorm\), which we established using \cref{eq_kpar_ordering_eq}, implies that 
\begin{equation}
	\frac{2 \vpar \deltaVpar}{\vti^2} \sim \frac{\tau c_s \phinorm}{\vti} \sim \sqrt{\tau} \phinorm.
\end{equation}
Therefore, the perturbed parallel flow does not enter into \(h^{(0)}\). We define solution for the distribution function to two lowest orders as
\begin{align}
	\label{eq_h0}
	&h^{(0)} = \left[\frac{\delta N}{n_i} + \frac{\delta T}{T_i} \left(\frac{v^2}{\vti ^2} - \frac{3}{2}\right)\right]F_i, \\
	\label{eq_h12}
	&h^{(1/2)} = \frac{2 \vpar \deltaVpar}{\vti^2}F_i,
\end{align}
and the solubility conditions
\begin{align}
	\intv h^{(n)} = \intv v^2 h^{(n)} = \intv \vpar h^{(n)} = 0
\end{align}
for \(n \geq 1\).

Note that our expansion implies that parallel collisional effects (parallel heat flux and parallel viscosity) enter via \(h^{(3/2)}\) and so are asymptotically too small to appear in any of our fluid equations.

\subsection{Fluid equations}
\label{appendix_fluideq}

We proceed by taking the density, temperature, and parallel-velocity moments of \cref{eq_curvy_gk}. The derivation for the `two-dimensional parts' of the equations for \(\phinorm\) and \(\deltaT\) can be found in \citet{ivanov2020}.

The density moment at fixed particle position, \((1/n_i)\int \dv \ \avgr{.}\), of \cref{eq_curvy_gk} is
\begin{align}
	\label{eq_dens_moment}
	& \frac{\partial}{\partial t} \left( \tau \phinorm' - \frac{1}{2} \rhoidelperpsq \phinorm\right) + \int \dv \ \vpar \pz \avgr{h^{(1/2)}} - \frac{\rhoivti}{L_B} \frac{\partial}{\partial y} \left( \phinorm + \deltaT \right) + \frac{\rhoivti}{2L_T} \frac{\partial}{\partial y} \left( \frac{1}{2} \rhoidelperpsq \phinorm \right)
	\\& \quad + \frac{1}{2} \rhoivti \bigg( \pbra{\phinorm}{\tau\phinorm' - \frac{1}{2} \rhoidelperpsq \phinorm} + \frac{1}{2}\rho_i^2 \boldsymbol{\nabla_\perp} \bcdot \pbra{\boldsymbol{\nabla_\perp}\phinorm}{\deltaT} \bigg) \notag
	\\& \quad = - \frac{1}{2} \chi \rho_i^2 \nabla_\perp^4 (\achi\phinorm - \bchi\deltaT),
\end{align}
where all terms are of order \(\orderinline{\tau h^{(0)}}\). The parallel-velocity moment is, using \cref{eq_h12},
\begin{equation}
	\label{eq_density_par_term}
	\frac{1}{n_i}\int \dv \ \vpar \pz \avgr{h^{(1/2)}} \approx \pz \deltaVpar.
\end{equation}
Combining \cref{eq_density_par_term} with \cref{eq_dens_moment} yields \cref{phiEq}.

Similarly, the temperature moment, \((1/n_i)\int \dv \ v^2/\vti^2 \avgr{.}\), of \cref{eq_curvy_gk} is
\begin{align}
\label{eq_temp_moment}
&\frac{\partial\deltaT}{\partial t} + \frac{1}{n_i}\int \dv \vpar \pz \frac{v^2}{\vti^2} \avgr{h^{(1/2)}}  + \frac{\rhoivti}{2L_T} \frac{\partial\phinorm}{\partial y}  + \frac{1}{2} \rhoivti \pbra{\phinorm}{\deltaT} = \chi \nabla_\perp^2 \deltaT,
\end{align}
where the parallel-streaming term is
\begin{equation}
\label{eq_temp_par_term}
\frac{1}{n_i}\int \dv \vpar \pz \frac{v^2}{\vti^2} \avgr{h^{(1/2)}} = \frac{5}{2} \pz \deltaVpar.
\end{equation}
Hence we obtain \cref{psiEq}.

Finally, we take the parallel-velocity moment, \((1/n_i)\int \dv \ \vpar \avgr{.}\), of \cref{eq_curvy_gk}. The first term is the time derivative of
\begin{equation}
	\label{eq_u_moment_1}
	\frac{1}{n_i}\int \dv \ \vpar \avgr{h - \avgR{\phinorm}F_i} \approx \frac{1}{n_i}\int \dv \ \vpar h^{(1/2)} = \deltaVpar.
\end{equation}
The parallel-streaming term is
\begin{equation}
	\frac{1}{n_i}\int \dv \ v_\parallel^2\pz \avgr{h} \approx  \frac{1}{n_i}\int \dv \ v_\parallel^2\pz h^{(0)} = \frac{1}{2} \vti^2 \pz (\phinorm + \deltaT).
\end{equation}
The temperature-gradient term integrates to 0 because the integrand is odd in \(\vpar\). The magnetic-gradient term is one order of \(L_T / L_B \sim \orderinline{\tau} \ll 1\) smaller than the rest (the magnetic curvature is absent from \cref{eq_temp_moment} for the same reason). The nonlinear term integrates to
\begin{equation}
	\frac{1}{n_i}\int \dv \ \vpar \avgr{\pbra{\avgR{\phinorm}}{h}} \approx \frac{1}{n_i}\int \dv \ \vpar \pbra{\phinorm}{h^{(1/2)}} = \frac{1}{2} \rho_i \vti \pbra{\phinorm}{\deltaVpar}.
\end{equation}
Finally, the parallel-velocity moment of the collisional operator is
\begin{equation}
	\label{eq_u_moment_last}
\frac{1}{n_i}\int \dv \ \vpar \avgr{\avgR{C_{l}[h]}} \approx \frac{1}{n_i}\int \dv \ \vpar C_{l}[h^{(1/2)}] = \cchi \nabla_\perp^2 \deltaVpar,
\end{equation}
where \(\cchi = 9/10\) is a numerical factor \citep[see][]{newton2010}. Putting together \mbox{\cref{eq_u_moment_1}--\cref{eq_u_moment_last}}, we arrive at \cref{uEq}.

\section{Slab-ITG instability condition}
\label{appendix_slabdisp}

\begin{figure}
	\centering
	\includegraphics[scale=0.26]{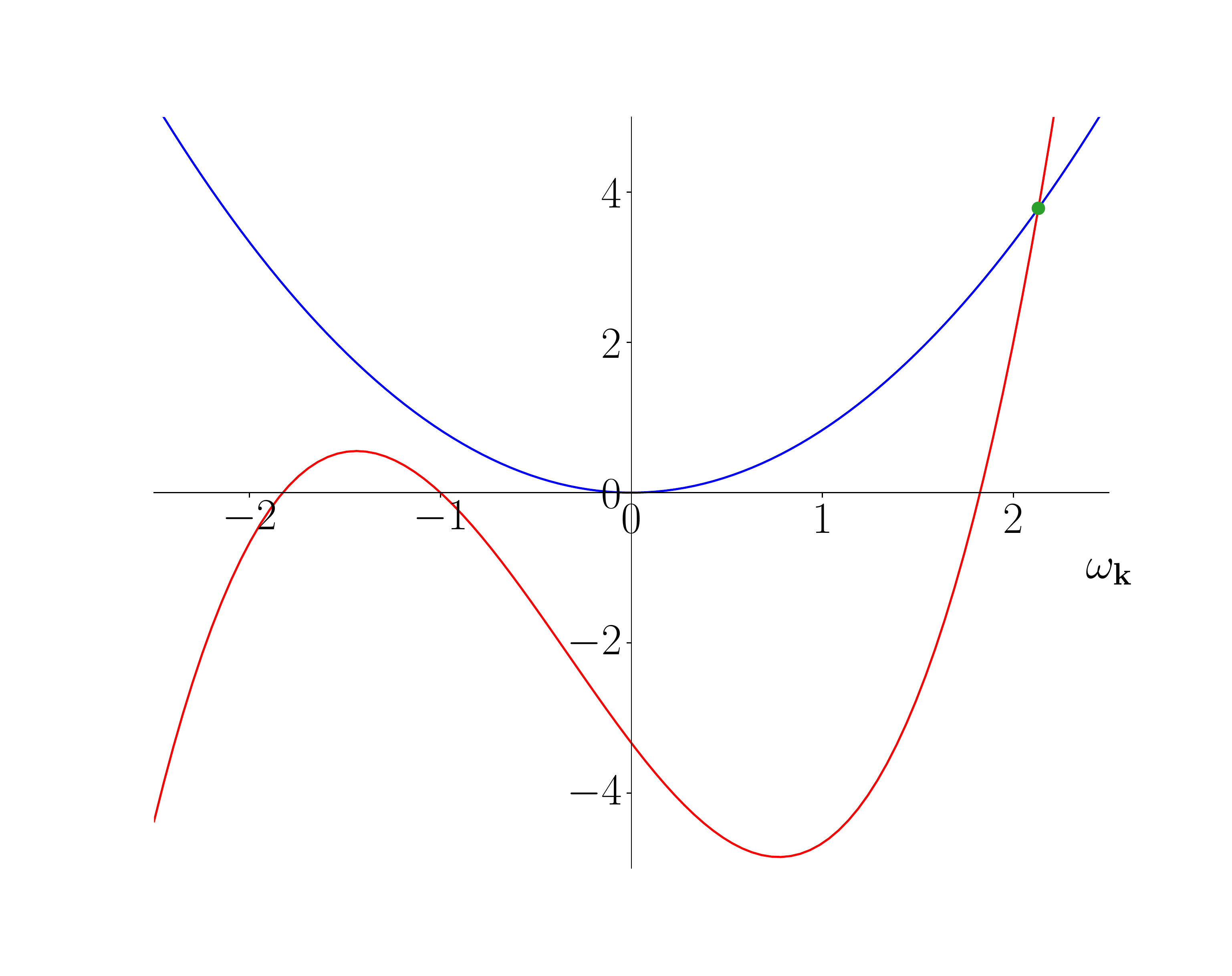}
	\caption{ The left-hand (red) and right-hand (blue) sides of the sITG dispersion relation \cref{eq_disp_slab} for \(\kparh = 2\), \(\kperp^2 = 0.2\). There is only one real solution, so there exists a complex one with positive imaginary part. Thus, there are linearly unstable modes for \(\kparh = 2\), \(\kperp^2 = 0.2\). }
	\label{fig_slab_disp_appendix}
\end{figure}

Here we derive the instability boundaries \cref{eq_slab_instab_marginalmodes_with_gamma} for the dispersion relation \cref{eq_disp_slab}. Note that the left-hand side of \cref{eq_disp_slab} is a cubic polynomial in \(\omkh\) with one positive and two negative roots, while the right-hand side is a simple quadratic propotional to \(\omkh^2\) (see \cref{fig_slab_disp_appendix}). First, if \(\gammakh^2 < 0\), then the right-hand side is a concave parabola and it is geometrically evident that there will always be three intersections of the parabola and the cubic, and so there are no unstable solutions. On the other hand, if \(\gammakh^2 > 0\), then these two curves cross three times if and only if the cubic left-hand side is larger than the quadratic right-hand side at \(\omkh = \omkh^{(0)} < 0\), where the two curves have the same slope. We differentiate \cref{eq_disp_slab} to find that \(\omkh^{(0)}\) is the negative solution to
\begin{equation}
	\label{eq_omkh0_quadratic}
	\left(\omkh^{(0)}\right)^2  +\left[\frac{2}{3} - \frac{4\kperp^2\gammakh^2}{3(1+\kperp^2)}\right] \omkh^{(0)} - \frac{1}{3} \frac{\kpar^2}{1 + \kperp^2} = 0.
\end{equation}
Using \cref{eq_omkh0_quadratic} to substitute for \(\omkh^{(0)}\), the instability condition that the left-hand side of~\cref{eq_disp_slab} be smaller than its right-hand side is then found to be equivalent to
\begin{equation}
	\label{eq_omkh_ineq}
	\omkh^{(0)} >  -\kparh^2 \frac{4(1+\kperp^2)+\kperp^2\gammakh^2}{3\kparh^2(1 + \kperp^2) + \left(1+\kperp^2-2\kperp^2\gammakh^2\right)^2} \equiv \omkh^{\text{min}}.
\end{equation}
Since \(\omkh^{(0)}\) is the negative solution of the quadratic \cref{eq_omkh0_quadratic} and \(\omkh^{\text{min}} < 0\), \cref{eq_omkh_ineq} can be true if and only if the quadratic \cref{eq_omkh0_quadratic} is positive when we substitute \(\omkh^{\text{min}}\) for \(\omkh^{(0)}\). Performing that substitution and simplifying the resulting expression yields a quadratic inequality for \(\kparh^2\):
\begin{align}
	&-(1+\kperp^2)\kparh^4 + \kparh^2 \left[2(1+\kperp^2)^2 + 10\kperp^2\gammakh^2(1+\kperp^2)-\kperp^4\gammakh^4\right] -(1+\kperp^2-2\kperp^2\gammakh^2)^3 > 0.
\end{align}
Its solution is the interval \(\kparh^2 \in (\hat{k}_{\parallel,-}^2, \hat{k}_{\parallel,+}^2)\), where \(\hat{k}_{\parallel,\pm}^2\) are given by~\cref{eq_slab_instab_marginalmodes_with_gamma}.

\section{Collisional slab instability}
\label{appendix_col_slab}

\begin{figure}
	\centering
	\includegraphics[scale=0.26]{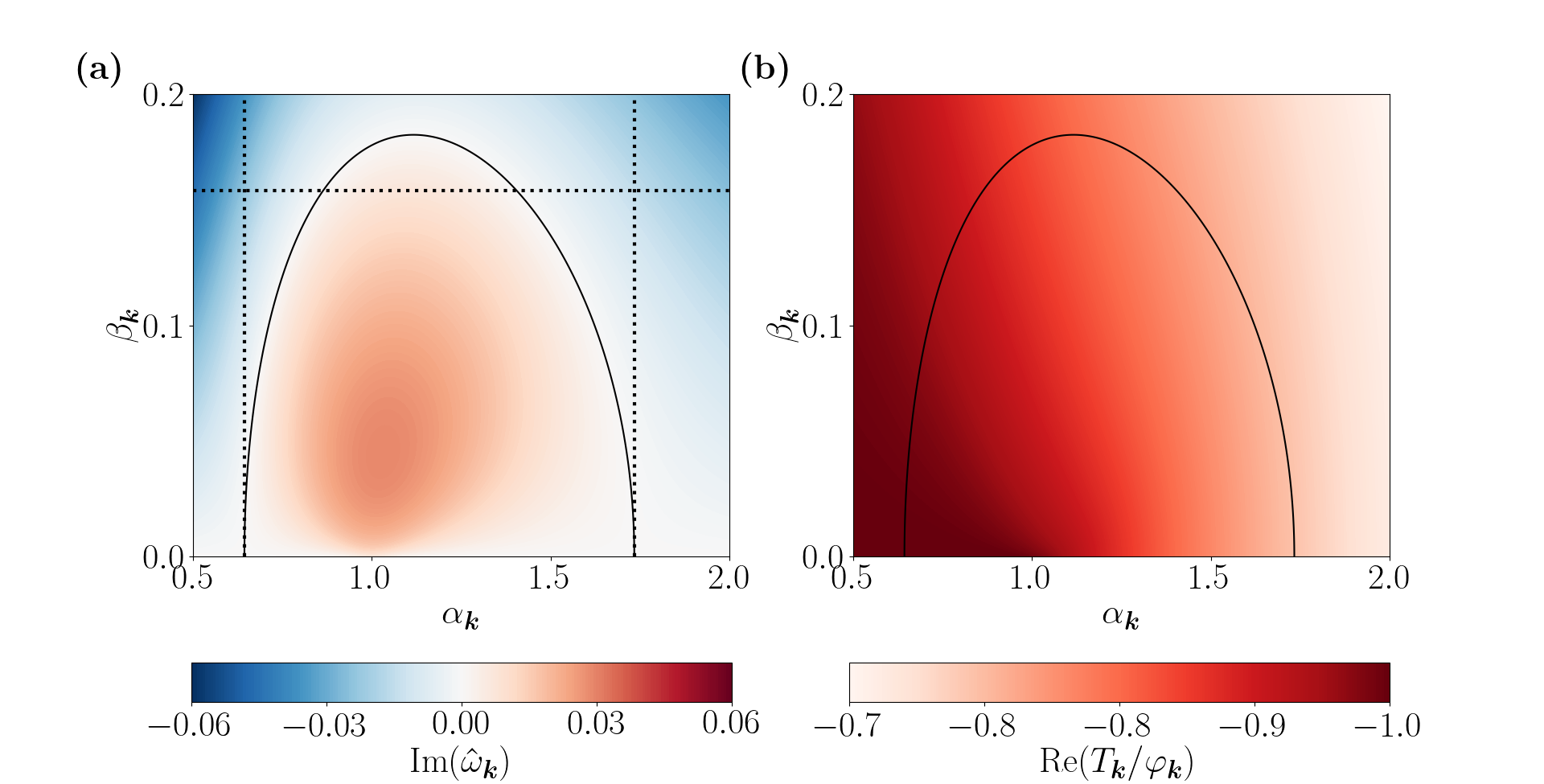}
	\caption{ \textbf{(a)} Largest growth rate \(\iminline{\omkh}\) obtained by solving \cref{eq_disp_colscales}. \textbf{(b)} The ratio \(\reinline{\deltaT_\vk/\phinorm_\vk}\) for the most unstable mode. The solid black line is the stability boundary \(\iminline{\omkh} = 0\). The dotted lines in (a) show the analytic approximations to the \(\ak\) and \(\bk\) instability boundaries, given by \cref{eq_col_alpha_bounds} (for \(\bk \ll 1\)) and \cref{eq_col_beta_bounds} (for \(\bk \sim \lambda \ll 1\)). We see perfect agreement with \cref{eq_col_alpha_bounds}, but a slight discrepancy with \cref{eq_col_beta_bounds}, whose derivation is accurate only under the assumption that \(1 -\achi -\bchi = \lambda \approx 0.36\) is small. All unstable modes lie within \(\reinline{\deltaT_\vk/\phinorm_\vk} > -1\).}
	\label{fig_colslab_alphabeta}
\end{figure}

To simplify the dispersion relation and focus on the \(\chi\)ITG instability promised at the end of \cref{sect_slab_resonance}, let us consider the \(\kperp \gg 1\) limit of \cref{eq_curvy_linearphi}--\cref{eq_curvy_linearu}, i.e., drop the \(\pt\dw{\phinorm}\) term in \cref{eq_curvy_linearphi}, and also drop the collisionless-resonance term \(\pt\dw{\phinorm}\) from the right-hand side of \cref{eq_curvy_linearpressure}. The dispersion relation for the thus simplified equations becomes
\begin{equation}
	(\omkh + 1) (\omkh + i\cchi\bk)(\omkh + i\bk) - \ak (\omkh + 1 + i\bk) + i\bk(\omkh + i\cchi\bk)(\achi \omkh + i\achi\bk - \bchi)=0,
	\label{eq_disp_colscales}
\end{equation}
where we have defined \(\omkh \equiv \omk / \vt k_y\), \(\ak \equiv \kpar^2 / \vt^2k_y^2\kperp^2\), \(\bk \equiv \chi \ksq / \vt k_y\). Note that the five parameters of a Fourier mode, viz., \(\vt\), \(\chi\), and the three components of \(\vk\) have collapsed into only two effective parameters: \(\ak\) and \(\bk\). Thus, we only need to solve~\cref{eq_disp_colscales} in the \((\ak, \bk)\) plane. The solution (in particular, its imaginary part) is shown in \cref{fig_colslab_alphabeta}, alongside the value of \(\reinline{\deltaT_\vk/\phinorm_\vk}\) for the most unstable mode --- a quantity that is crucial for the Dimits regime \citep[see \cref{sect_dimits} and also][]{ivanov2020}. Let us discuss this solution in some easy limits.

First, consider the case of \(\bk \ll 1 \sim \ak\). As \(\bk \sim \kperp\), this limit corresponds to the low-\(\kperp\) end of the wavenumber spectrum of the collisional instability. Note that this is a subsidiary expansion to the \(\kperp \gg 1\) one used to obtain \cref{eq_disp_colscales}. To lowest order in \(\bk\), \cref{eq_disp_colscales} yields
\begin{equation}
	\label{eq_disp_colscales_smallbeta_0}
	(\omkh^2 - \ak) (\omkh + 1) = 0,
\end{equation}
whence \(\omkh \approx \omkh^{(0)} = -1, \pm\sqrt{\ak}\). Letting \(\omkh = \omkh^{(0)} + \delta \omkh\), where \(\delta \omkh / \omkh \sim \bk \ll 1\), we find in the next order
\begin{equation}
	\label{eq_omkh_lowbeta}
	\omkh = -1 + \frac{-i\bk (\achi + \bchi -\ak)}{1-\ak},\ \pm\sqrt{\ak} + \frac{-i\bk\left[\sqrt{\ak} (\cchi + \achi)\pm(\cchi + 1 -\bchi)\right]}{2(\sqrt{\ak}\pm 1)}.
\end{equation}
It is then evident that the \(\omkh^{(0)} = \sqrt{\ak}\) solution is always stable, while the other two give the following condition for instability:
\begin{equation}
	\label{eq_col_alpha_bounds}
	\ak = \frac{\kpar^2}{\vt^2 k_y^2 \kperp^2} \in \left(a+b, \left(\frac{\cchi + 1 - \bchi}{\cchi + \achi}\right)^2\right) \approx (0.64, 1.73),
\end{equation}
the numerical values being valid for \(\achi\), \(\bchi\), and \(\cchi\) as given after \cref{eq_pbra_def}. The instability boundaries in \cref{eq_col_alpha_bounds} agree with \cref{fig_colslab_alphabeta}. Note that for \(\achi + \bchi \to 1\), \cref{eq_col_alpha_bounds} implies that \(\ak \to 1\), i.e., \mbox{\(\kpar \to \vt k_y \kperp\)}, which is precisely the resonance condition we discovered in~\cref{sect_slab_resonance}. In hindsight, this is expected because the collisional coupling term on the right-hand side of~\cref{eq_curvy_linearpressure} goes to zero in the limit \(\achi + \bchi \to 1\). 

Equation \cref{eq_omkh_lowbeta} implies that the growth rate \(\iminline{\omkh}\) of the \(\bk \ll 1\) collisional modes satisfies \(\iminline{\omkh}\sim\bk\). However, these expressions break down when \(\ak = 1 + \orderinline{\bk}\). We now show that this is precisely where the fastest-growing mode resides, similarly to what we found in \cref{sect_largekperp} for the collisionless sITG mode. Setting \(\omkh = -1 + \delta \omkh\), \(\ak = 1 + \delta \ak\), where \(\delta \omkh \sim \delta \ak \sim \sqrt{\bk} \ll 1\), we find from \cref{eq_disp_colscales} that
\begin{equation}
	\label{eq_omkh_lowbeta_fastestmode}
	2\delta\omkh^2 + \delta\ak + i(a+b-1)\bk = 0 \implies \im{\omkh} = \pm \frac{\sqrt{\delta \ak^2 - 8i(\achi+\bchi-1)\bk}}{4},
\end{equation}
which implies that \(\im{\omkh}\) is largest when \(\delta\ak=0\). To see this, note that the imaginary part of the square root of a complex number \(u+iv\) is equal to
\begin{equation}
	\im{\sqrt{u+iv}} = \sqrt{\frac{-u + \sqrt{u^2 + v^2}}{2}},
\end{equation}
which can easily be shown to be a decreasing function of \(u\). Therefore, the growth rate in \cref{eq_omkh_lowbeta_fastestmode} is largest when \(\delta\ak = 0\) and is given by 
\begin{equation}
	\label{eq_growthrate_lowbeta_fastestmode}
	\im{\omkh} = \frac{1}{2}\sqrt{|a+b-1|\bk},
\end{equation}
so it scales as \(\im{\omkh}\sim\sqrt{\bk}\). Note that this growth rate vanishes when \(\achi + \bchi = 1\), i.e., when the instability boundaries \cref{eq_col_alpha_bounds} lie on top of each other. 

The growth rate given by \cref{eq_growthrate_lowbeta_fastestmode} is comparable to the collisionless growth rate \cref{eq_largekperp_growthrate} when \(\vt k_y \sqrt{\bk} \sim \vt\), i.e., when \(\kperp \sim (\vt/\chi)^{1/3}\), where we assumed \(k_y \sim \kperp\). This is precisely the condition \(\kperp \sim \kperpcol\) for the transition from the collisionless to the collisional regime that we found in \cref{sect_slab_resonance}.

In the opposite limit of \(\bk \gg 1\sim\ak\), \cref{eq_disp_colscales} gives
\begin{equation}
	\omkh^3 + i(\cchi + \achi + 1)\bk\omkh^2 - (\cchi+\achi+\achi\cchi)\bk^2\omkh - i\achi\cchi\bk^3 = 0,
\end{equation}
which has three stable solutions: \(\omkh = -i\bk, -i\achi\bk, -i\cchi\bk\). We can therefore conclude that there exists a \(\beta_\text{max} \sim 1\) such that unstable solutions are possible only for \mbox{\(\bk < \beta_\text{max}\)}. A simple analytical estimate for \(\beta_\text{max}\) is obtainable if we make an additional approximation: let \mbox{\(\lambda \equiv 1-\achi-\bchi \ll 1\)} and consider an expansion in small \(\lambda\).\footnotemark\footnotetext{In our case, \(\lambda = 57/160 \approx 0.36\), so the quality of this approximation is marginal.} In this limit, the collisional coupling in \cref{eq_curvy_linearpressure} is small and \cref{eq_col_alpha_bounds} requires \(\ak = 1 + \orderinline{\lambda}\). We let \(\omkh = -1 + \delta\omkh\), \(\ak = 1 + \delta\ak\), where \(\delta\omkh\sim\delta\ak\sim\bk\sim\lambda\ll 1\), and expand \cref{eq_disp_colscales} to \(\orderinline{\lambda}\) to find
\begin{equation}
	\label{eq_col_deltaomega_small_lambda}
	\delta\omkh = \frac{-\delta\ak -i\bk(\achi+\cchi+2) \pm \sqrt{\left[\delta\ak +i\bk(\achi+\cchi+2)\right]^2 -8i\bk(\lambda+\delta\ak) +8(\achi+\cchi)\bk^2}}{4}.
\end{equation}
After some unenlightening algebra, we find that \cref{eq_col_deltaomega_small_lambda} supports unstable solutions for
\begin{equation}
	\label{eq_col_beta_bounds}
	\bk < \beta_\text{max} \approx \frac{\lambda}{2(\achi+\cchi)} = \frac{1-\achi-\bchi}{2(\achi+\cchi)} \approx 0.16,
\end{equation}
which is in reasonable agreement with the numerically determined \(\beta_\text{max} \approx 0.18\). 

Numerically, we find that the fastest-growing mode is located at \(\beta_\text{fastest} \approx 0.04\), \mbox{\(\alpha \approx 1.01\)}, and has a growth rate \(\iminline{\omkh} = \hat{\gamma}_\text{fastest} \approx 0.03\).\footnotemark\footnotetext{The same result can be obtained analytically from \cref{eq_col_deltaomega_small_lambda}.} The dependence of \(\iminline{\omkh}\) on \(\ak\) and \(\bk\) is shown in \cref{fig_colslab_alphabeta}. Undoing the normalisations of \(\ak\), \(\bk\), and \(\omkh\), we find that this collisional instability is localised at \(\kpar \approx \vt k_y \kperp\) (just as the collisionless modes are), is bounded by \(\beta_\text{max} \approx 0.18 > \beta\), and has its largest growth rate
\begin{equation}
	\iminline{\omk} = \vt k_y \gamma_\text{fastest} \approx 0.03 \vt k_y \ \ \text{at} \ \ \frac{\ksq}{k_y} = \frac{\beta_\text{fastest}\vt}{\chi} \approx \frac{0.04\vt}{\chi}.
\end{equation}
As \(\omkh\) depends on \(\vk_\perp\) through \(\bk\), the contours of constant \(\omkh\) in the \((k_x, k_y)\) plane coincide with those of constant \(\bk\). Since \(\bk = \chi \ksq / \vt k_y\), these are circles with radius \(\vt \bk / 2\chi\), centred at \(k_x = 0\) and \( k_y = \vt \bk / 2\chi\). Since \(\omk = \vt k_y \omkh\), the largest \(\iminline{\omk}\) for a given \(\bk\) is found at \(k_x = 0\) and \(k_y = \vt \bk / \chi\). In particular, the most unstable mode has
\begin{equation}
	\label{eq_col_growthrate}
	k_y = \frac{\beta_\text{fastest}\vt}{\chi} \approx \frac{0.04\vt}{\chi}, \quad \im{\omk} \approx 0.0012 \frac{\vt^2}{\chi}.
\end{equation}

The growth rate \cref{eq_col_growthrate} scales quadratically with \(\vt\), unlike the collisionless sITG instabilities considered in \cref{sect_slabinst}, and also diverges as \(\chi \to 0\). Therefore, either for sufficiently large \(\vt\) or sufficiently small \(\chi\), the collisional instability will dominate. However, the small numerical factor in \cref{eq_col_growthrate} means that this collisional mode will be more unstable than the collisionless small-scale sITG mode \cref{eq_largekperp_growthrate} only if
\begin{equation}
	\frac{\vt}{\chi} \gtrsim 830,
\end{equation}
at scales \(k_y \sim 0.04 \vt/\chi \gtrsim 33\). Such a regime is both numerically difficult to access and physically questionable, so for all the rest of the paper, we shall consider only \(\vt/\chi \ll 830\) and ignore the collisional modes. In the absence of collisions, the sITG growth rate asymptotically approaches its maximum value \cref{eq_largekperp_growthrate} as \(\kperp\to\infty\), so we conclude that if \(\vt/\chi \ll 830\), i.e., if the \(\chi\)ITG growth rate is much smaller than the sITG one, then \mbox{\(k_y\sim\kperpcol\sim(\vt/\chi)^{1/3}\)} is also the scale of the fastest-growing sITG mode.

\section{Slab-ITG instability with general gradients and low collisionality}
\label{appendix_slabitg_general}

Here we solve the dispersion relation of \modeleqnsslabpertsimple{} in the \( \kpar \sim \kperp^2 \gg 1\) limit, neglecting the magnetic-drift contributions, and ordering collisionality as \(\chi \kperp^3 \sim \vt \). Ignoring the magnetic-drift term \(-\py \left(\phinorm + \deltaT\right)\) (as it is subdominant for the small-scale sITG modes, see \cref{sect_largekperp}), we find the following dispersion relation
\begin{equation}
	\label{eq_disp_general_appendix_unordered}
	\omega_\vk(1+\kperp^2) - \frac{\kpar^2}{\omega_\vk + i\cchi \chi \kperp^2} \left(1 + \frac{\vtvect\bcdot\vk}{\omega_\vk + i\chi \kperp^2}\right) + \kperp^2 \vtvect\bcdot\vk - \vnvect\bcdot\vk + i\achi\chi\kperp^4 -\frac{i\bchi \kperp^4\vtvect\bcdot\vk}{\omega_\vk + i\chi\kperp^2} = 0.
\end{equation}
As already mentioned in \cref{sect_parasitic_inst}, the dispersion relation must be invariant under \(\vk \mapsto -\vk\) and \(\omega_\vk \mapsto -\omega_\vk^*\), so, without loss of generality, we assume that \(\vtvect\bcdot\vk > 0\). We then write \cref{eq_disp_general_appendix_unordered} as
\begin{equation}
\label{eq_disp_general_appendix}
\left[\omkh^2 (1+\kperp^2) - \kparh^2\right]\left(\omkh + 1\right) = 2 \kperp^2 \gammakh^2 \omkh^2 - i \bk \kperp^2 \left[\frac{\cchi\kparh^2}{\kperp^2}\left(1 + \frac{1}{\omkh}\right) + \frac{\kparh^2}{\kperp^2 \omkh} + \achi \omkh^2 - \bchi\omkh\right],
\end{equation}
where, in addition to the definitions in \cref{sect_slabinst}, we have (re)defined the following quantities:
\begin{equation}
\label{eq_disp_general_appendix_defs}
\gammakh \equiv \sqrt{\frac{\left(\vtvect+\vnvect\right)\bcdot\vk}{2\kperp^2\vtvect\bcdot\vk}}, \quad \bk \equiv \frac{\chi\ksq}{\vtvect\bcdot\vk}.
\end{equation}
Here \(\gammakh\) is the largest collisionless growth rate \cref{eq_largekpar_secslab_growthrate}. 

As we discussed in \cref{sect_largekperp}, the linearly unstable solutions lie close to \(\kparh = \kperp\) and are given by \(\omkh = -1 + \delta \omkh\), where \(\delta \omkh \sim \order{1/\kperp} \ll 1\). Substituting into \cref{eq_disp_general_appendix} \(\omkh = -1 + \delta \omkh\) and \(\kparh = \kperp + \delta \kparh\), where \mbox{\(\delta \omkh / \omkh \sim \delta \kparh / \kparh \sim \order{1/\kperp} \ll 1\)}, we find
\begin{equation}
\label{eq_disp_general_deltaomkh}
\delta \omkh = -\frac{\delta \kparh}{2\kperp} \pm \sqrt{\frac{\delta \kparh^2}{4\kperp^2} - \gammakh^2 + \frac{i (\achi + \bchi - 1)\bk}{2}} + \order{\kperp^{-2}}. 
\end{equation}
As we discussed in \cref{appendix_col_slab}, \(\iminline{\sqrt{u + iv}}\) is a decreasing function of \(u\), where we have taken the square root with a positive imaginary part. Therefore, \cref{eq_disp_general_deltaomkh} attains its largest imaginary part, i.e., the largest growth rate, when \mbox{\(\delta \kparh = 0\)}. Moreover, the sign of~\(\reinline{\sqrt{u + iv}}\) for the branch with \(\iminline{\sqrt{u+iv}}>0\) is determined by the sign of \(v\).

Then, using \cref{curvy_psi}, we obtain
\begin{equation}
	\label{eq_Toverphi_cols_1}
	\frac{\deltaT_\vk}{\phinorm_\vk} = \frac{\vtvect\bcdot\vk}{\omk + i\chi \kperp^2} = \frac{1}{\omkh} + \order{\kperp^{-2}} = -1 -\delta\omkh + \order{\kperp^{-2}},
\end{equation}
where we have dropped the \(i\chi\kperp^2\) term from the denominator because \(\chi\kperp^2 \sim |\vtvect| \kperp^{-1} \sim \omk\kperp^{-2}\) is small. Substituting \cref{eq_disp_general_deltaomkh} into \cref{eq_Toverphi_cols_1} then gives
\begin{equation}
	\label{eq_Toverphi_cols}
	\frac{\deltaT_\vk}{\phinorm_\vk} = -1 - \sqrt{-\gammakh^2 + \frac{i (\achi + \bchi - 1)\bk}{2}} + \order{\kperp^{-2}}
\end{equation}
for the linearly unstable mode with the largest growth rate. This is \cref{eq_Toverphi_cols_maintext}. The sign of the real part of \(\deltaT_\vk/\phinorm_\vk + 1\) for the most unstable mode is, therefore, the same as the sign of~\(\achi + \bchi - 1\).

\section{Quasilinear damping of \(\slabavg{\upar}\)}
\label{appendix_upar_largescale}

Here we show that the parallel velocity of the large-scale 2D perturbations is damped by the parasitic modes excited by them. We shall do so by proving that the norm of \(\slabavg{\upar}\) always decays.

Multiplying \cref{curvy_slabavg_u} by \(\slabavg{\upar}\) and integrating gives
\begin{equation}
	\label{eq_slabavg_u_energy_evo}
	\frac{1}{2} \pt \intr \slabavg{\upar}^2 = -\cchi \chi \intr |\del \slabavg{\upar}|^2 - \intr \slabavg{\upar} \slabavg{\pbra{\phinormslabpert}{\uparslabpert}},
\end{equation} 
where the first term on the right-hand side is negative-definite and corresponds to the collisional damping of the parallel flow, and the second term is the energy transfer from small scales. The latter can be rewritten as 
\begin{equation}
	\label{eq_upar_largescaleresp}
	-\intr \slabavg{\upar} \slabavg{\pbra{\phinormslabpert}{\uparslabpert}} = -\intr \uparslabpert \pbra{\slabavg{\upar}}{\phinormslabpert} = \intr \uparslabpert \vuvect\bcdot\del\phinormslabpert,
\end{equation}
where we have defined the gradient of the large-scale parallel flow as
\begin{equation}
	\vuvect \equiv -\uvect{z}\times\del\slabavg{\upar}. 
\end{equation}
Our objective now is to show that the right-hand side of \cref{eq_upar_largescaleresp} is always negative.

Let us incorporate \(\vuvect\) into \cref{curvy_slabpert_u}. Instead of \cref{curvy_slabpert_u_lowestorder}, we find
\begin{equation}
	\label{curvy_slabpert_u_lowestorder_withlargescaleupar}
	\left(\pt  + \slabavg{\ve} \bcdot \del \right) \uparslabpert + \pz \left(\phinormslabpert + \deltaTslabpert\right) + \vuvect\bcdot\del\uparslabpert =  \cchi \delsq \uparslabpert.
\end{equation}
We now proceed just as we did in \cref{sect_parasitic_inst}, viz., we assume that the large-scale gradients are constant, ignore collisions, and look for Doppler-shifted Fourier modes \(\phinormslabpert_\vk, \deltaTslabpert_\vk, \uparslabpert_\vk \propto \exp{\left[-i \left(\omega_\vk + \slabavg{\ve} \bcdot \vk \right) t + i \vk \bcdot \vect{r}\right]}\).
Combining \cref{curvy_slabpert_phi_lowestorder}, \cref{curvy_slabpert_psi_lowestorder}, and \cref{curvy_slabpert_u_lowestorder_withlargescaleupar}, and going through the algebra yields a dispersion relation that is a modified version of \cref{eq_disp_slabpert}:
\begin{equation}
	\label{eq_disp_slabpert_withlargescaleupar}
	\left(\omkh^2 - \frac{\kparh^2}{1 + \kperp^2} \right) \left(\omkh + 1\right) = \frac{2\kperp^2\gammakh^2\omkh^2}{1 + \kperp^2} + \frac{\kparh\vuvect\bcdot\vk\omkh}{(1+\kperp^2)\vtvect\bcdot\vk},
\end{equation}
where, as before, we assumed \(\vtvect\bcdot\vk > 0\) (see \cref{sect_parasitic_inst}), \mbox{\(\omk = \vtvect\bcdot\vk\omkh\)},\mbox{ \(\kpar = \vtvect\bcdot\vk\kparh\)}, and \(\gammakh\) is given by \cref{eq_gammakh_smallscale_slab}. Again, we shall be concerned with the small-scale limit \mbox{\(\kparh \sim \kperp \gg 1\)}. Motivated by the numerical observation that \(\slabavg{\upar}\) is much smaller than \(\slabavg{\phinorm}\) and \(\slabavg{\deltaT}\) (and hence \(|\vuvect|\) is much smaller than \(|\vnvect|\) and \(|\vtvect|\)), we consider the case when the second term on the right-hand side of \cref{eq_disp_slabpert_withlargescaleupar} is a small correction to the first one, viz., we assume \mbox{\(\vuvect\bcdot\vk/\vtvect\bcdot\vk \ll \kperp^{-1}\)}. In this case, the solution to \cref{eq_disp_slabpert_withlargescaleupar} is given by \(\omkh = -1 + \delta\omkh\), \(\kparh = \kparh^{(0)} + \delta\kparh\), \(\kparh^{(0)} = \pm \kperp\), where \(\delta\omkh\) satisfies
\begin{equation}
	\delta\omkh\left(\delta\omkh + \frac{\delta\kparh}{\kparh^{(0)}}\right) \approx -\left(\gammakh^2 - \frac{\kparh^{(0)} \vuvect\bcdot\vk}{2\kperp^2\vtvect\bcdot\vk}\right).
\end{equation}
The maximum growth rate is attained for \(\delta\kparh = 0\) (see \cref{sect_largekperp}). It is
\begin{equation}
	\label{eq_delta_omkh_withlargescaleupar}
	\delta \omkh \approx i\sqrt{\gammakh^2 - \frac{\kparh^{(0)} \vuvect\bcdot\vk}{2\kperp^2\vtvect\bcdot\vk}}.
\end{equation}
Thus, we see that to lowest order in \(\vuvect\bcdot\vk/\vtvect\bcdot\vk \ll \kperp^{-1}\), the role of \(\vuvect\) is to modify the sITG growth rate in a way that breaks the symmetry between the \(\kparh = \pm \kperp\) branches of the instability (see \cref{fig_linear3d}). Specifically, the \(\kpar\vuvect\bcdot\vk > 0\) branch is stabilised and the \(\kpar\vuvect\bcdot\vk < 0\) one is destabilised.

Now let us return to the equation \cref{eq_upar_largescaleresp} for the large-scale \(\slabavg{\upar}\). Repeating the arguments in \cref{sect_largescale_resp}, we write \cref{eq_upar_largescaleresp} as a sum over small-scale modes:
\begin{equation}
	\label{eq_upar_largescaleresp2}
	\intr \uparslabpert \vuvect\bcdot\del\phinormslabpert \approx -i\sum_\vq\uparslabpert_\vq \vuvect\bcdot\vq \phinormslabpert_\vq^* = \sum_\vq \left[\frac{\qpar \vuvect\bcdot\vq}{i\omega_\vq} \left(1 + \frac{\deltaTslabpert_\vq}{\phinormslabpert_\vq}\right) + \frac{(\vuvect\bcdot\vq)^2}{i\omega_\vq} \right] |\phinormslabpert_\vq|^2,
\end{equation}
where we used the \(\chi = 0\) versions of \cref{curvy_slabpert_psi_lowestorder} and \cref{curvy_slabpert_u_lowestorder_withlargescaleupar} to find the (quasi)linear expression for \(\uparslabpert_\vq/\phinormslabpert_\vq\). Assuming that the small-scale perturbations are dominated by the linearly unstable sITG modes with \(\im{\omega_\vq} > 0\), the second term in the square brackets in \cref{eq_upar_largescaleresp2} is clearly negative-definite. The first term requires some work:
\begin{align}
	\label{eq_upar_largescaleresp2_1}
	&\frac{\qpar \vuvect\bcdot\vq}{i\omega_\vq} \left(1 + \frac{\deltaTslabpert_\vq}{\phinormslabpert_\vq}\right) = \frac{\qpar \vuvect\bcdot\vq}{i\omqh\vtvect\bcdot\vq} \left(1 + \frac{1}{\omqh}\right) \approx \frac{\qpar \vuvect\bcdot\vq}{\vtvect\bcdot\vq}\im{\delta\omqh},
\end{align}
where we expressed \(\deltaTslabpert_\vq / \phinormslabpert_\vq = 1/\omkh\) using \cref{curvy_slabpert_psi_lowestorder}. The linearly unstable modes have \(\im{\delta\omqh} > 0\). Additionally, \cref{eq_delta_omkh_withlargescaleupar} tells us that the modes with largest growth rate have \(\qpar\vuvect\bcdot\vq < 0\), so \cref{eq_upar_largescaleresp2_1} is always negative for these modes. Assuming that \cref{eq_upar_largescaleresp2} is dominated by the most unstable modes, we conclude that it, too, is always negative. So, given a small \(\slabavg{\upar}\), the right-hand side of \cref{eq_slabavg_u_energy_evo} is negative-definite. Therefore, \(\slabavg{\upar} = 0\) is a quasilinearly stable state. 

\section{Scale-separated conservation laws}
\label{appendix_cons}

In deriving the simple model for scale-separated dynamics, which consists of the small-scale system \modeleqnsslabpertsimple{} and the large-scale one \modeleqnsslabavgsimple{}, we made several critical approximations: we ignored all but the lowest-order variation of the large-scale fields in~\modeleqnsslabpertsimple{}, we argued that \(\slabavg{\upar} = 0\), and we showed that the nonlinear terms in~\cref{curvy_slabpert_phi_lowestorder} were subdominant, hence, to lowest order, \cref{curvy_slabavg_phi_lowestorder} did not couple to small scales. Let us show that under these assumptions, the conservation laws of \(W\) and \(I\) still hold in the scale-separated system of \modeleqnsslabpertsimple{} and \modeleqnsslabavgsimple{}. We shall be concerned only with the nonlinear terms in the relevant equations because they are responsible for the interactions of small and large scales.

Let us first check the conservation of the free energy \(W\). Multiplying \cref{curvy_slabpert_psi_lowestorder} by \(\deltaTslabpert\) and integrating gives
\begin{equation}
	\label{eq_w_cons_smallscale}
	\pt \intr \frac{1}{2} \deltaTslabpert^2 + \text{linear terms} = -\intr \deltaTslabpert \vtvect\bcdot\del\phinormslabpert .
\end{equation}
Similarly, multiplying the large-scale temperature equation \cref{curvy_slabavg_psi_lowestorder} by \(\slabavg{\deltaT}\) and integrating gives
\begin{align}
	\label{eq_w_cons_largescale}
	&\pt \intr \frac{1}{2} \slabavg{\deltaT}^2 + \text{linear terms} = -\intr \slabavg{\deltaT} \pbra{\phinormslabpert}{\deltaTslabpert} = -\intr  \deltaTslabpert\pbra{\slabavg{\deltaT}}{\phinormslabpert} \nonumber\\&= \intr \deltaTslabpert \left( -\uvect{z}\times\del\slabavg{\deltaT} \right)\bcdot\del\phinormslabpert = \intr \deltaTslabpert \vtvect\bcdot\del\phinormslabpert.
\end{align}
Adding \cref{eq_w_cons_smallscale} and \cref{eq_w_cons_largescale} then gives
\begin{equation}
	\pt \intr \frac{1}{2} \left(\deltaTslabpert^2 + \slabavg{\deltaT}^2\right) + \text{linear terms} = 0,
\end{equation}
which is precisely the statement of conservation of free energy, see \cref{eq_free_e_cons}.

Let us now check the conservation of the second conserved quantity \(I\). We multiply~\cref{curvy_slabpert_phi_lowestorder} by \(\phinormslabpert+\deltaTslabpert\), \cref{curvy_slabpert_psi_lowestorder} by \(\phinormslabpert + \deltaTslabpert - \delsq\deltaTslabpert\), and \cref{curvy_slabpert_u_lowestorder} by \(\uparslabpert\), sum, and integrate to obtain
\begin{align}
	\label{eq_i_cons_smallscale}
	&\pt \intr\left[ \frac{1}{2} (\phinormslabpert + \deltaTslabpert)^2 + \frac{1}{2} (\del\phinormslabpert + \del\deltaTslabpert)^2 + \frac{1}{2}\uparslabpert^2 \right] + \text{linear terms} \nonumber \\
	&=-\intr\left[ (\phinormslabpert + \deltaTslabpert) \vtvect\bcdot\del\delsq\phinormslabpert + (\phinormslabpert + \deltaTslabpert) \vnvect\bcdot\del\phinormslabpert + (\phinormslabpert + \deltaTslabpert - \delsq\deltaTslabpert)\vtvect\bcdot\del\phinormslabpert \right]  \nonumber \\
	&=-\intr \deltaTslabpert \left(\vtvect+\vnvect\right)\bcdot\del\phinormslabpert,
\end{align}
where we got to the last line from the penultimate one using integration by parts and the periodicity of the spatial domain. We now repeat the same procedure with the large-scale equations: we multiply \cref{curvy_slabavg_phi_lowestorder} and \cref{curvy_slabavg_psi_lowestorder} by \(\slabavg{\phinorm} + \slabavg{\deltaT}\), sum, and integrate to obtain
\begin{align}
	\label{eq_i_cons_largescale}
	&\pt \intr\frac{1}{2} \left(\slabavg{\dw{\phinorm}} + \slabavg{\deltaT}\right)^2  + \text{linear terms} =-\intr\left(\slabavg{\phinorm} + \slabavg{\deltaT}\right) \pbra{\phinormslabpert}{\deltaTslabpert} \nonumber \\
	&= -\intr \deltaTslabpert \pbra{\slabavg{\phinorm} + \slabavg{\deltaT}}{\phinormslabpert}= \intr \deltaTslabpert (\vtvect+\vnvect)\bcdot\del \phinormslabpert.
\end{align}
Therefore, summing \cref{eq_i_cons_smallscale} and \cref{eq_i_cons_largescale}, we obtain the conservation law \cref{eq_sec_cons}, where we have ignored the \(\orderinline{\kperp^2}\) terms in the large-scale contributions to \(I\). Note that due to the large 2D scale \(\kperp \ll 1\), the zonal \(\zf{\phinorm}\) is not included in \cref{eq_i_cons_largescale}.

\begin{figure}
	\includegraphics[scale=0.26]{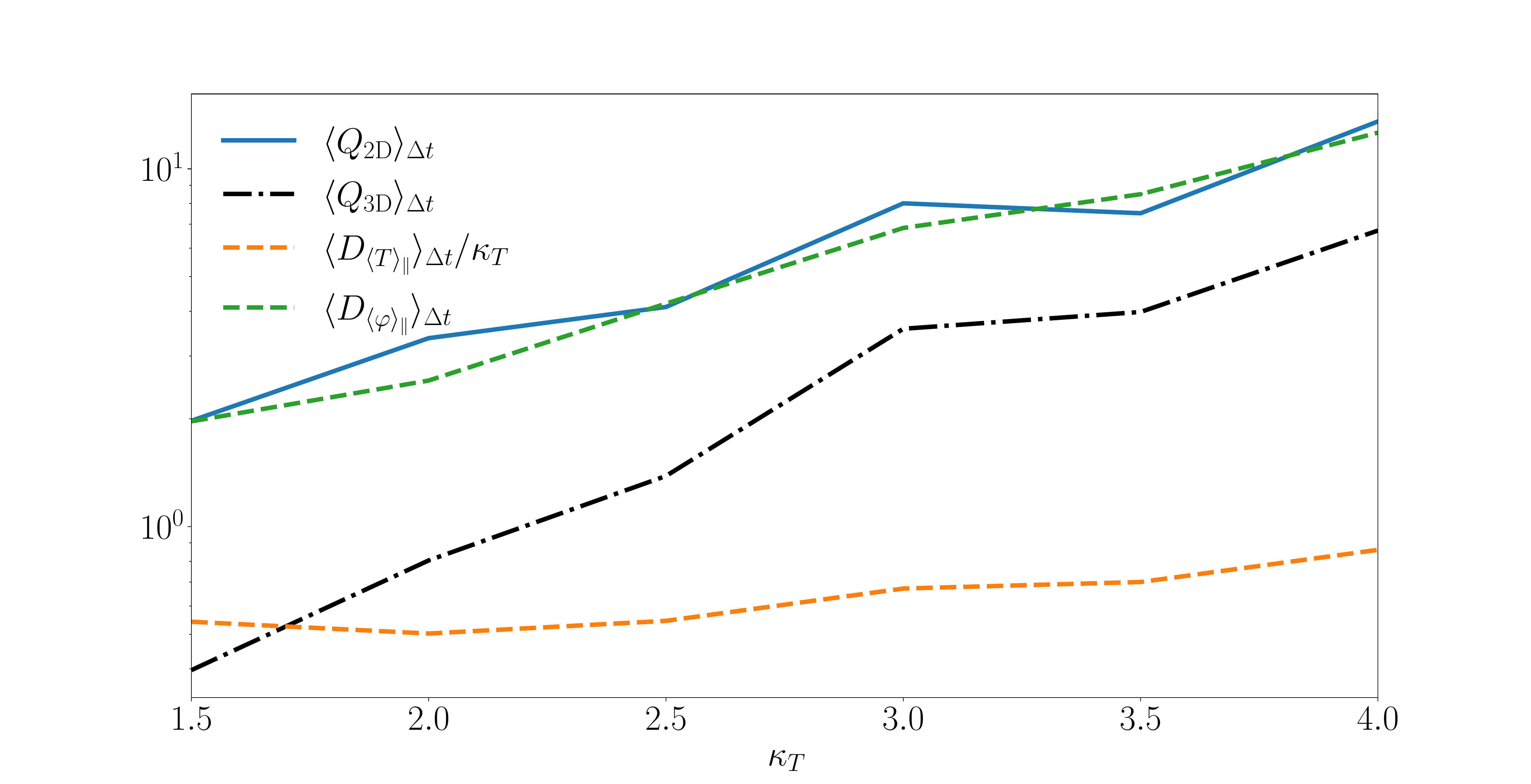}
	\caption{ Plot of the time-averaged 2D heat flux \(Q_\text{2D}\) given by \cref{eq_q2d} (solid blue), the 3D heat flux \(Q_\text{3D} \equiv Q - Q_\text{2D}\) (dash-dotted black), the collisional dissipation in \cref{eq_phi2d_energy_appendix} \(D_{\slabavg{\phinorm}}\) given by~\cref{eq_Dphi} (dashed green), and the collisional dissipation in \cref{eq_psi2d_energy_appendix}, \(D_{\slabavg{\deltaT}}/\vt\) given by \cref{eq_Dpsi} (dashed orange) versus \(\vt\) for \(\chi = 0.05\), \(L_x = L_y = 60\) and \(\Lpar = 0.5\). We see that \(Q_\text{2D}\) is more than twice \(Q_\text{3D}\), and is balanced nearly perfectly by \(D_{\slabavg{\phinorm}}\), i.e., the nonlinear transfer in \cref{curvy_slabavg_phi} is small. On the other hand, \(D_{\slabavg{\deltaT}}/\vt Q_\text{2D} \ll 1\), so the majority of energy injected into \(\slabavg{\deltaT}\) is nonlinearly transferred to 3D modes.}
	\label{fig_eflow_vs_kappaT}
\end{figure}

Since \(\slabavg{\upar} \approx 0\) (see \cref{appendix_upar_largescale}), we can obtain simple conservation laws for the 2D equations directly from \modeleqnsslabavg{} without any additional simplification. Multiplying \cref{curvy_slabavg_phi} by \(\slabavg{\phinorm}\) and \cref{curvy_slabavg_psi} by \(\slabavg{\deltaT}\), and integrating gives
\begin{align}
	\label{eq_phi2d_energy_appendix}
	&\pt \frac{1}{2} \intr \left[\slabavg{\phinorm}^2 + \slabavg{\del \phinorm}^2\right] - Q_\text{2D} + D_{\slabavg{\phinorm}} = -\mathcal{T}_{\phinorm, \text{2D}\rightarrow\text{3D}} \\
	\label{eq_psi2d_energy_appendix}
	&\pt \frac{1}{2} \intr \slabavg{\deltaT}^2 - \vt Q_\text{2D} + D_{\slabavg{\deltaT}} = -\mathcal{T}_{\deltaT, \text{2D}\rightarrow\text{3D}},
\end{align}
where the 2D heat flux \(Q_\text{2D}\), the collisional dissipation terms \(D_{\slabavg{\phinorm}}\) and \(D_{\slabavg{\deltaT}}\), and the energy transfer terms \(\mathcal{T}_{\phinorm, \text{2D}\rightarrow\text{3D}}\) and \(\mathcal{T}_{\deltaT, \text{2D}\rightarrow\text{3D}}\) are given by
\begin{align}
	&Q_\text{2D} = -\intr \slabavg{\phinorm} \py \slabavg{\deltaT} \label{eq_q2d} \\
	&D_{\slabavg{\phinorm}} = \chi \intr \left[\achi \slabavg{\delsq \phinorm}^2 - \bchi \slabavg{\delsq\phinorm}\slabavg{\delsq\deltaT}\right] \label{eq_Dphi} \\
	&D_{\slabavg{\deltaT}} = \chi \intr \slabavg{\del \deltaT}^2 \label{eq_Dpsi} \\
	&\mathcal{T}_{\phinorm, \text{2D}\rightarrow\text{3D}} = \intr \slabavg{\phinorm} \del \bcdot \pbra{\del \phinormslabpert}{\phinormslabpert + \deltaTslabpert} \\
	&\mathcal{T}_{\deltaT, \text{2D}\rightarrow\text{3D}} = \intr \slabavg{\deltaT} \pbra{\phinormslabpert}{\deltaTslabpert}.
\end{align}
In a steady state, \cref{eq_phi2d_energy_appendix} and \cref{eq_psi2d_energy_appendix} imply that \(Q_\text{2D} - D_{\slabavg{\phinorm}} \approx T_{\phinorm, \text{2D}\rightarrow\text{3D}}\) and that \mbox{\(\vt Q_\text{2D} - D_{\slabavg{\deltaT}} \approx T_{\deltaT, \text{2D}\rightarrow\text{3D}}\)}. In other words, the energy injected into the 2D fields (by the \(Q_\text{2D}\) terms) is either dissipated by those fields (through \(D_{\slabavg{\phinorm}}\) and \(D_{\slabavg{\deltaT}}\)) or nonlinearly transferred to the 3D fields. According to our results and analysis in \cref{sect_slabsec}, we expect that the overall energy injection is dominated by the 2D modes, i.e., \(Q \approx Q_\text{2D}\) (within 20-30\%, see \cref{sect_num_evidence}), the nonlinear transfer in \cref{curvy_slabavg_phi} is small, i.e., \(Q_\text{2D} \approx D_{\slabavg{\phinorm}}\), confirming the asymptotic analysis in \cref{sect_largescale_resp}, and the nonlinear transfer in the \(\slabavg{\deltaT}\) equation \cref{curvy_slabavg_psi} is large, i.e., \(\vt Q_\text{2D} \gg D_{\slabavg{\deltaT}}\), which allows \(\slabavg{\deltaT} \sim \slabavg{\phinorm}\) even when the unstable linear 2D modes do not satisfy this. This picture of the saturated state agrees with numerical simulations, as illustrated by \cref{fig_eflow_vs_kappaT}.

\section{Kinetic slab-ITG instability}
\label{appendix_kinetic_disp}

The results in \cref{sect_parres} rely on the existence of the collisionless sITG instability at short parallel and perpendicular wavelengths \(\kpar \sim \vt \kperp^2 \gg 1\). However, the existence of a collisionless sITG instability at infinitely short perpendicular scales is a more general fact that can be established without resorting to a fluid limit. To show this, let us find the sITG dispersion relation directly from the collisionless kinetic equation. 

We begin at \cref{eq_curvy_gk} with zero collisionality (\(\nu_i = 0\)), no magnetic curvature (\(L_B^{-1} = 0\)), and linearised:
\begin{align}
	\label{eq_curvy_gk_slab}
	&\partd{}{t} \left(h - \avgR{\phinorm}F_i\right) + \vpar\pz h  + \frac{\rho_i \vti}{2L_T}\left( \frac{v^2}{v^2_{ti}} - \frac{3}{2} \right)F_i \partd{\avgR{\phinorm}}{Y}  = 0.
\end{align}
We shall consider Fourier modes \(h, \phinorm \propto \exp\left( -i\omega_\vk t + i \vect{k}\bcdot\vect{r} \right)\).
Rearranging \cref{eq_curvy_gk_slab} and using the fact that \(F_i\) is a Maxwellian with density \(n_i\) and thermal speed \(\vti\), we find 
\begin{equation}
	\label{eq_h_linear}
	\frac{h_\vk}{\phinorm_\vk} = \frac{n_i}{\pi^{3/2}\vti^3} \frac{\zeta + \zeta_*\left(\vh^2 - \frac{3}{2}\right)}{\zeta - \sgn{\kpar}\vparh} e^{-\vh^2} J_0\left(\kperp\rho_i\vperph\right),
\end{equation}
where \(\zeta \equiv \omega_\vk/|\kpar|\vti\), \(\zeta_* \equiv \rho_i\vti k_y/2 L_T|\kpar|\vti\), \(\hat{\vect{v}} \equiv \vect{v}/\vti\), \(\sgninline{\kpar}\equiv\kpar/|\kpar|\), and \(J_0\) is the zeroth-order Bessel function of the first kind. Substituting \cref{eq_h_linear} into the quasineutrality condition~\cref{eq_curvy_qn}, we find
\begin{align}
	\label{eq_slab_disp_int}
	1+\tau &= \frac{1}{n_i}\int \dv \ J_0\left(\kperp\rho_i\vperph\right) \frac{h_\vk}{\phinorm_\vk}  \nonumber \\
	&= \frac{1}{\sqrt{\pi}} \intdvparh e^{-\vparh^2} \intdvperph e^{-\vperph^2} \frac{\zeta + \zeta_*\left(\vh^2-\frac{3}{2}\right)}{\zeta - \vparh} J_0^2\left(\kperp\rho_i\vperph\right),
\end{align}
where \(\sgn{\kpar}\) is absorbed into \(\vparh\) and the integral over \(\vparh\) is taken along the Landau contour (i.e., below the pole). After performing the integrals in \cref{eq_slab_disp_int}, we find the sITG dispersion relation:
\begin{equation}
	\label{eq_slab_disp_kinetic}
	\left\lbrace-\left(\zeta -\frac{1}{2}\zeta_*\right)\Gamma_0(\alpha) + \zeta_*\alpha\Gamma_1(\alpha)\right\rbrace Z(\zeta) - \zeta_*\zeta \Gamma_0(\alpha)\left[1 + \zeta Z(\zeta)\right] = 1+\tau,
\end{equation}
where \(\alpha \equiv \kperprhoisq/2\), \(\Gamma_0(\alpha) \equiv I_0(\alpha)e^{-\alpha}\), \(\Gamma_1(\alpha) \equiv \left[I_0(\alpha) - I_1(\alpha)\right]e^{-\alpha}\), \(I_0(\alpha)\) and \(I_1(\alpha)\) are the zeroth- and first-order modified Bessel functions of the first kind, and
\begin{equation}
	Z(\zeta) = \frac{1}{\sqrt{\pi}}\int dz \frac{e^{-z^2}}{z-\zeta}
\end{equation}
is the plasma dispersion function \citep{friedconte61}, with the integral taken along the Landau contour. To express the relevant integral moments of \(J_0^2\), we used the relation
\begin{equation}
	\int_0^{+\infty} d(x^2) \ J_m(bx)J_n(cx) e^{-x^2/a} = aI_0\left(a\frac{b^2+c^2}{4}\right)e^{-abc/2}\delta_{mn},
\end{equation}
for \(a=1\) and \(b=c=\kperp\rho_i\) \citep[][p. 395]{watson66}.

We can verify that the `fluid' dispersion relation \cref{eq_disp_slab} is an asymptotic limit of \cref{eq_slab_disp_kinetic} by expanding the latter for the ordering in \cref{eq_ordering}. Under this ordering, \(\zeta \sim \zeta_* \sim 1/\sqrt{\tau} \sim 1/\sqrt{\alpha} \gg 1\). The large-\(\zeta\) and small-\(\alpha\) expansions of the plasma dispersion function and the Bessel functions are
\begin{equation}
	Z(\zeta) \approx -\frac{1}{\zeta} \left(1 + \frac{1}{2\zeta^2}\right), \quad \Gamma_0(\alpha) = 1 - \alpha + \order{\alpha^2}, \quad \Gamma_1(\alpha) = 1 + \order{\alpha}.
\end{equation}
Using these along with \(\alpha / \tau = \kperp^2\rho_s^2\), and adopting the normalisations \cref{eq_normalisations}, we find that \cref{eq_slab_disp_kinetic} reduces to \cref{eq_disp_slab}.

\begin{figure}
	\includegraphics[scale=0.26]{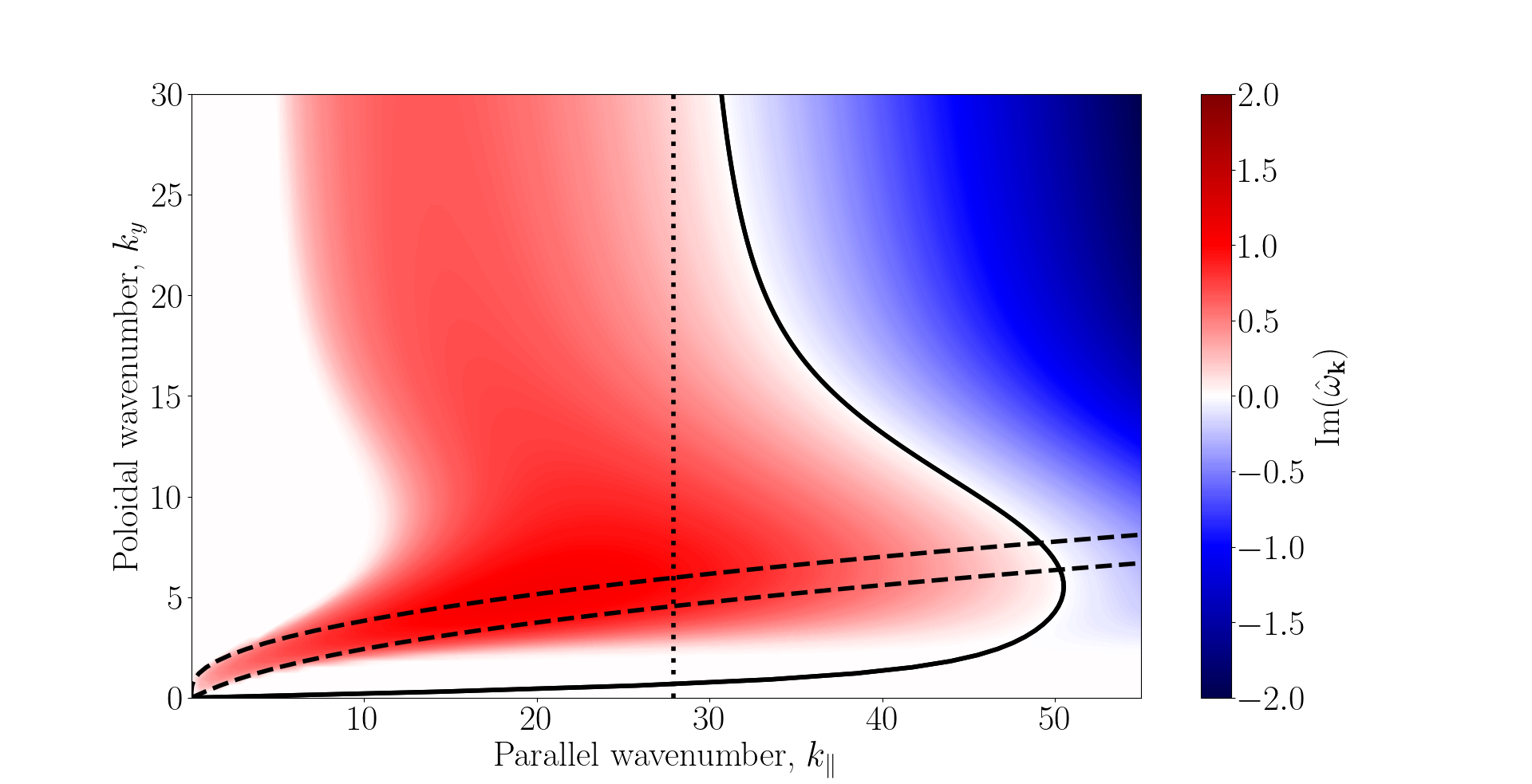}
	\caption{ The linear growth rate \(\iminline{\omk}\) of the kinetic dispersion relation~\cref{eq_slab_disp_kinetic} for \(\tau = 0.01\), normalised as \(\omkh = L_T\omk/c_s\tau\), which is equivalent to normalisation of time in \cref{eq_normalisations} for \(\vt = 1\). The wavenumbers \(k_y\) and \(\kpar\) are also normalised according to \cref{eq_normalisations} with \(\vt=1\). The largest kinetic growth rate is \(\iminline{\omkh^\text{kin}} \approx 1.07\), while the largest cold-ion growth rate, given by \cref{eq_largekperp_growthrate}, is \(\iminline{\omkh^\text{cold}} \approx 0.71\). The vertical dotted line is the critical parallel wavenumber \(\kpar^{(c)}\) for \(\kperp \gg 1\) kinetic modes \cref{eq_kinetic_kparcrit}. The dashed black lines are the cold-ion stability boundary \cref{eq_slab_instab_marginalmodes}. The solid black line is the kinetic stability boundary \cref{eq_kinetic_stab_boundary}. The kinetic sITG instability has a finite growth rate at \(\kperp \to \infty\). }
	\label{fig_kinetic_slab}
	\vspace{1cm}
\end{figure}

We can also find the general stability boundary of \cref{eq_slab_disp_kinetic} in the standard way, by looking for the parameters that allow a \(\iminline{\zeta} = 0\) solution. For such a solution, the only imaginary contributions to \cref{eq_slab_disp_kinetic} come from the terms containing the plasma dispersion function, so the coefficient of \(Z(\zeta)\) must be zero. This gives us a system of two equations:
\begin{align}
	&(1+\tau) + \zeta \zeta_* \Gamma_0(\alpha) = 0, \\
	&\left(\zeta -\frac{1}{2}\zeta_*\right)\Gamma_0(\alpha) - \zeta_*\alpha\Gamma_1(\alpha) +\zeta^2\zeta_*\Gamma_0(\alpha) = 0.
\end{align}
Solving this, we find
\begin{equation}
	\label{eq_kinetic_stab_boundary}
	\zeta_*^2 = \frac{2(1+\tau) \left[1+\tau - \Gamma_0(\alpha)\right]}{\Gamma_0^2(\alpha) + 2\alpha \Gamma_0(\alpha)\Gamma_1(\alpha)}.
\end{equation}

Using \(\omega_* \equiv \rho_i\vti k_y/2 L_T\) and \cref{eq_normalisations}, we can express \(\zeta_*\) and \(\alpha\) in terms of the normalised \(\hat{k}_x\), \(\hat{k}_y\), and \(\kparh\).\footnotemark\footnotetext{Recall that we have been using the normalised \(\hat{k}_x\), \(\hat{k}_y\), and \(\kparh\) throughout this work, but in \cref{sect_3dmodel} dropped the `hats'. These `hats' are not related to the ones in \cref{sect_linear}. } This gives us an analytic expression for the stability boundary. In the limit \(\kperp \gg 1\), i.e., \(\alpha \gg 1\), we can expand \cref{eq_kinetic_stab_boundary} to find that an instability exists for
\begin{equation}
	\label{eq_kinetic_kparcrit}
	\kpar <  \frac{1}{2\sqrt{\pi}(1+\tau)L_T} \equiv \kpar^{(c)}.
\end{equation}
Similarly to the parallel wavenumber at which Braginskii viscosity would kick in (see \cref{sect_parres}), \(\kpar^{(c)}\) exists outside of the cold-ion ordering for our model, as it is asymptotically large in the ordering \cref{eq_ordering}: \(k_\parallel^{(c)} L_B \sim \orderinline{L_B/L_T} \sim \orderinline{\tau^{-1}} \gg 1\). \Cref{fig_kinetic_slab} shows the growth rate obtained from solving \cref{eq_slab_disp_kinetic} numerically, along with the stability boundary of the cold-ion limit \cref{eq_slab_instab_marginalmodes}.

\section{Quasilinear ZF stress of kinetic sITG modes}
\label{appendix_stressure}

The stability of the zonal staircase in the 3D system \modeleqns{} was attributed to the turbulent stress in the presence of strong zonal shear \citep[see \cref{sect_turb_stress}, as well as][]{ivanov2020}. In the fluid limit, this stress was found to be the sum of Reynolds and diamagnetic stresses. Therefore, we were able to conclude that whether a mode with wavenumber \(\vq\) feeds or destroys the ZFs depends on the sign of the quantity \(1 + \reinline{\deltaT_\vq/\phinorm_\vq}\) (see \cref{sect_dimits}). We then found that we could predict the Dimits threshold by investigating the value of this quantity for the fastest-growing linear ITG modes. Here we generalise this approach to kinetic sITG modes. 

We start with the ion GK equation \cref{eq_curvy_gk} and the quasineutrality condition \cref{eq_curvy_qn}. We shall ignore the influence of collisions, so drop the collision operator on the right-hand side of \cref{eq_curvy_gk}. Taking a \((1/n_i)\int \dv\) moment of \cref{eq_curvy_gk} at constant \(\vect{r}\) and integrating over a flux surface, we find
\begin{equation}
	\label{eq_curvy_gk_zonal}
	\pt (\zf{\phinorm - \avgr{\avgR{\phinorm}}}) + \frac{\rho_i \vti}{2n_i} \int \dv \ \zf{\avgr{\pbra{ \avgR{\phinorm}}{h}}} = 0.
\end{equation}
With \(\phinorm\) and \(h\) decomposed into Fourier modes, \(\phinorm_\vk, h_\vk \propto \exp\left(i \vk\bcdot\vect{r} \right)\), \cref{eq_curvy_gk_zonal} gives the following equation for each mode:
\begin{equation}
	\label{eq_curvy_gk_zonal_fourier}
	\pt \left[1-\Gamma_0(\alpha)\right]\phinorm_\vk + \frac{\rho_i \vti}{2n_i} \int \dv \ J_0\left(k\rho_i\vperph\right) \sum_\vq \uvect{z}\bcdot(\vqperp\times\vk)J_0\left(|\vqperp-\vk|\rho_i\vperph\right)\phinorm_{\vk-\vq}h_\vq = 0,
\end{equation}
where \(\vk\) is a zonal wavenumber, i.e., \(\vk = k\uvect{x}\), \(\alpha = k^2\rho_i^2/2\), and the rest of the notation is the same as in \cref{appendix_kinetic_disp}. To make progress, let us assume that the scale of the ZF is much larger than the scale of the modes that contribute to the nonlinear term in \cref{eq_curvy_gk_zonal_fourier}, i.e., that \(k\rho_i \ll \qperp\rho_i\sim 1\). We can then expand
\begin{equation}
	|\vqperp-\vk| = \sqrt{(\vqperp-\vk)^2} = \qperp \left[1 - \frac{\vk\bcdot\vqperp}{\qperp^2} + \order{k^2\rho_i^2}\right],
\end{equation}
whence
\begin{align}
	&J_0\left(|\vqperp-\vk|\rho_i\vperph\right) = J_0\left(\qperp\rho_i\vperph\right) + \qperp\rho_i\vperph\frac{\vk\bcdot\vqperp}{\qperp^2} J_1\left(\qperp\rho_i\vperph\right) + \order{k^2\rho_i^2}, \\
	&J_0\left(\kperp\rho_i\vperph\right) = 1 + \order{k^2\rho_i^2},\\
	&1-\Gamma_0(\alpha) = \frac{1}{2}k^2\rho_i^2 + \order{k^4\rho_i^4} \label{eq_gamma0_lowestorder}.
\end{align}
The integral in \cref{eq_curvy_gk_zonal_fourier} vanishes to the lowest order in \(k\rho_i\), viz., 
\begin{equation}
	\int \dv \ \sum_\vq \uvect{z}\bcdot(\vqperp\times\vk)J_0\left(\qperp\rho_i\vperph\right)\phinorm_{\vk-\vq}h_\vq =\int \dv \ \pbra{\phinorm}{\avgr{h}}_\vk = n_i (1+\tau) \pbra{\phinorm}{\phinorm}_\vk = 0
\end{equation}
by quasineutrality \cref{eq_curvy_qn}. To the next order in \(k\rho_i\), \cref{eq_curvy_gk_zonal_fourier} becomes
\begin{equation}
	\label{eq_curvy_gk_zonal_fourier_lowestorder}
	\frac{1}{2}k^2\rho_i^2 \pt \phinorm_\vk + \frac{1}{4}\rho_i^3 \vti \sum_\vq \uvect{z} \bcdot(\vqperp\times\vk) \vk\bcdot\vqperp \phinorm_{\vk-\vq} \gp_\vq = 0,
\end{equation}
where we have defined the GK perpendicular pressure perturbation
\begin{equation}
	\label{eq_gp_def}
	\gp_\vq \equiv \frac{1}{n_i}\int \dv \ \vperph^2\frac{2J_1\left(\qperp\rho_i\vperph\right)}{\qperp\rho_i\vperph} h_\vq.
\end{equation}
Note that for \(\qperp\rho_i\ll 1\), \(J_1\left(\qperp\rho_i\vperph\right)\approx\qperp\rho_i\vperph/2\), and \cref{eq_gp_def} gives
\begin{equation}
	\gp_\vq \approx \frac{1}{n_i}\int \dv \ \vperph^2 h_\vq = \pr_\vq,
\end{equation}
where \(\pr_\vq\) is the pressure perturbation in the cold-ion model.

Fourier transforming \cref{eq_curvy_gk_zonal_fourier_lowestorder} back to real space, we find
\begin{equation}
	\label{eq_curvy_gk_zonal_lowestorder}
	\pt\zf{\phinorm} + \frac{1}{2}\rho_i \vti \zf{\px\gp\py\phinorm} = 0.
\end{equation}
Therefore, \(\zf{\px\gp\py\phinorm}\) is the GK version of the turbulent stress \(\Pi_t\) (see \cref{sect_dimits2d}) for large-scale~ZFs, but \(\rho_i\)-scale turbulence. In the limit of large-scale turbulence (\(\qperp \rho_i \ll 1\)), \cref{eq_curvy_gk_zonal_lowestorder} reduces to
\begin{equation}
	\label{eq_curvy_gk_zonal_lowestorder_largescaleturb}
	\pt\zf{\phinorm} + \frac{1}{2}\rho_i \vti \zf{\px\pr\py\phinorm} = 0.
\end{equation}
Note that \cref{eq_curvy_gk_zonal_lowestorder_largescaleturb} is not the same as \cref{eq_zonalphi} with \(\Pi_\chi = 0\). This is expected because \cref{eq_zonalphi} was obtained in the limit \(\kperp \rho_i \sim \qperp \rho_i \ll 1\), where \(\vk\) and \(\vq\) were the typical wavenumbers of zonal and nonzonal modes, respectively. In contrast, taking \(\qperp \rho_i \ll 1\) in \cref{eq_curvy_gk_zonal_lowestorder} is actually the limit \(\kperp \rho_i \ll \qperp \rho_i \ll 1\). The difference between \cref{eq_zonalphi} and \cref{eq_curvy_gk_zonal_lowestorder_largescaleturb} is an exact derivative, viz., \(\px (\zf{\phinorm\py\pr})\), and so does not influence the integrated momentum transport in a shear zone of radial width \(d\):
\begin{equation}
	\frac{1}{d}\int dx \zf{\px\gp\py\phinorm} = -\sum_\vq q_xq_y |\phinorm_\vq|^2 \re{\frac{\gp_\vq}{\phinorm_\vq}},
\end{equation}
where the correspondence with \cref{eq_pi_t} is evident. Therefore, following the arguments in \cref{sect_dimits2d} and \citep{ivanov2020}, we can conclude that the effect on the ZFs of a mode with wavenumber \(\vq\) depends on the sign of \(\reinline{\gp_\vq/\phinorm_\vq}\), viz., modes with \(\reinline{\gp_\vq/\phinorm_\vq > 0}\) feed momentum into the ZFs, while those with \(\reinline{\gp_\vq/\phinorm_\vq} < 0\) take momentum away from the ZFs.

Let us calculate \(\reinline{\gp_\vq/\phinorm_\vq}\) for the kinetic sITG modes that we found in \cref{appendix_kinetic_disp}. From~\cref{eq_gp_def}, we find
\begin{align}
	\label{eq_gp_1}
	\frac{\gp_\vq}{\phinorm_\vq} &= \frac{2}{n_i}  \int \dv \frac{\vperph}{\qperp\rho_i} J_1\left(\qperp\rho_i\vperph\right) \nonumber \\
	&= \frac{1}{\sqrt{\pi}} \intdvparh e^{-\vparh^2} \intdvperph e^{-\vperph^2} \frac{\zeta + \zeta_*\left(\vh^2-\frac{3}{2}\right)}{\zeta - \vparh} \frac{2\vperph}{\qperp\rho_i} J_1\left(\qperp\rho_i\vperph\right)J_0\left(\qperp\rho_i\vperph\right),
\end{align}
where we have substituted the linear GK expression \cref{eq_h_linear} for \(h\), \(\hat{\vect{v}} = \vect{v} / \vti\), and \(\zeta\) and \(\zeta_*\) are defined equivalently to those after \cref{eq_h_linear}, but with \(\vk\) replaced by \(\vq\). Notice that
\begin{equation}
	\frac{2\vperph}{\qperp\rho_i} J_1\left(\qperp\rho_i\vperph\right)J_0\left(\qperp\rho_i\vperph\right) = -\frac{1}{\qperp\rho_i} \partd{}{(\qperp\rho_i)} J_0^2\left(\qperp\rho_i\vperph\right) = -\partd{}{\alpha}J_0^2\left(\qperp\rho_i\vperph\right),
\end{equation}
where \(\alpha = \qperp^2\rho_i^2/2\). Therefore, \cref{eq_gp_1} can be written as
\begin{equation}
	\label{eq_gp_2}
	\frac{\gp_\vq}{\phinorm_\vq} = -\partd{}{\alpha} \frac{1}{\sqrt{\pi}} \intdvparh e^{-\vparh^2} \intdvperph e^{-\vperph^2} \frac{\zeta + \zeta_*\left(\vh^2-\frac{3}{2}\right)}{\zeta - \vparh} J_0^2\left(\qperp\rho_i\vperph\right),
\end{equation}
where the partial derivative with respect to \(\alpha\) is taken at constant \(\zeta_*\). We have already calculated the expression in \cref{eq_gp_2} that needs to be differentiated --- this is precisely the left-hand side of \cref{eq_slab_disp_kinetic}. Taking the derivative, we obtain
\begin{equation}
	\label{eq_gp}
	\frac{\gp_\vq}{\phinorm_\vq} = \left[ -\left(\zeta - \frac{1}{2}\zeta_*\right)\Gamma_1(\alpha) - \zeta_* \Gamma_0(\alpha) + 2\zeta_*\alpha\Gamma_1(\alpha) \right] Z(\zeta) - \zeta\zeta_*\Gamma_1(\alpha)\left[1+\zeta Z(\zeta)\right].
\end{equation}

\begin{figure}
	\centering
	\includegraphics[scale=0.26]{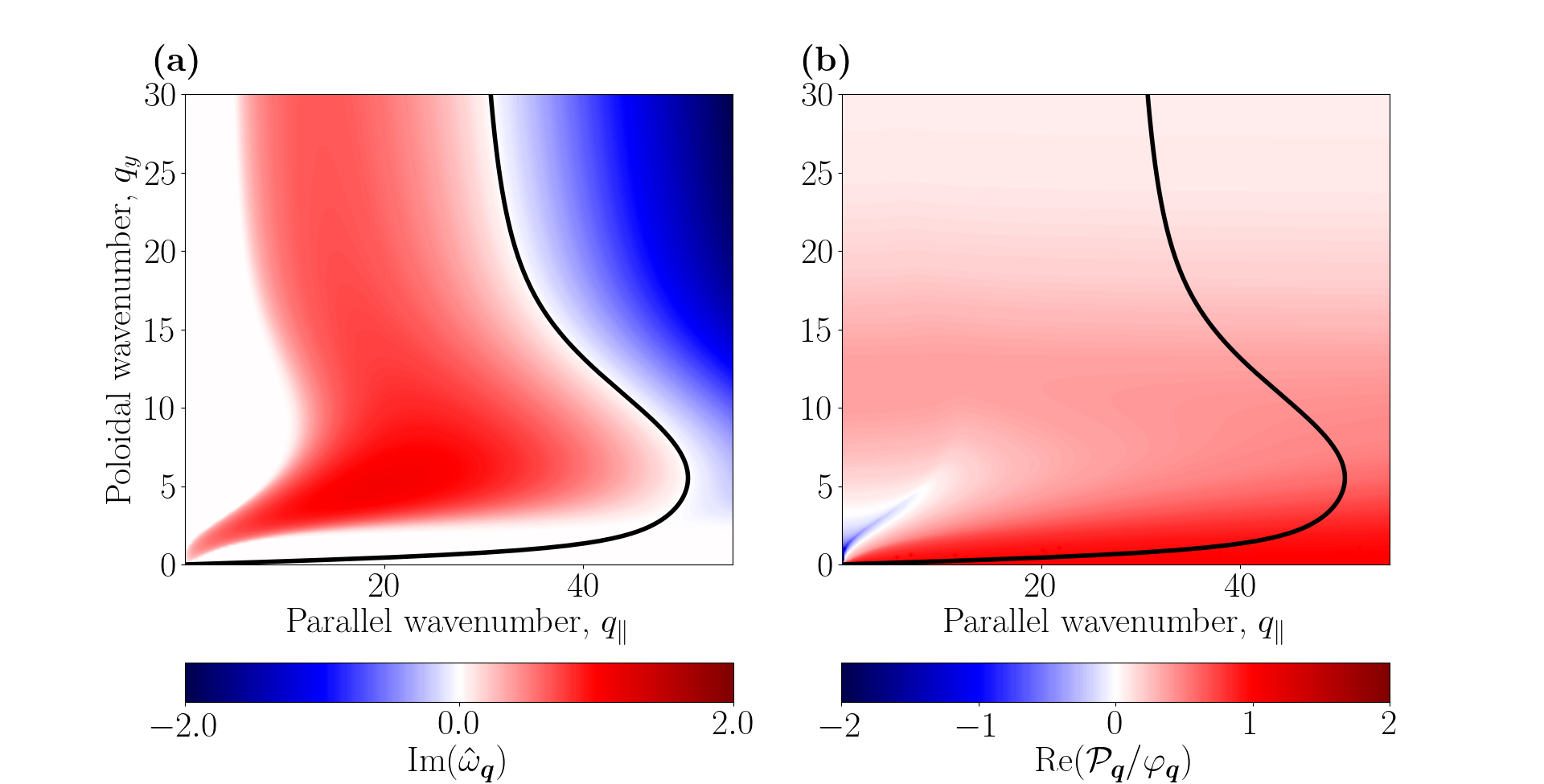}
	\caption{ \textbf{(a)} Growth rate \(\iminline{\omqh}\), normalised as \(\omqh = L_T\omega_\vq/c_s\tau\), which is equivalent to normalisation of time in \cref{eq_normalisations} for \(\vt = 1\). \textbf{(b)} The ratio \(\reinline{\gp_\vq/\phinorm_\vq}\), given by \cref{eq_gp}. For both panels, \(\tau = 0.01\) and \(q_x = 0\). The wavenumbers \(q_y\) and \(\qpar\) are also normalised according to \cref{eq_normalisations} with \(\vt=1\). The solid black lines show the kinetic stability boundary \cref{eq_kinetic_stab_boundary}. It is evident that the vast majority of sITG modes, including the dominant ones, support the ZFs, i.e., satisfy \(\reinline{\gp_\vq/\phinorm_\vq} > 0\).}
	\label{fig_stressure_tau001_kx0}
\end{figure}

\begin{figure}
	\centering
	\includegraphics[scale=0.26]{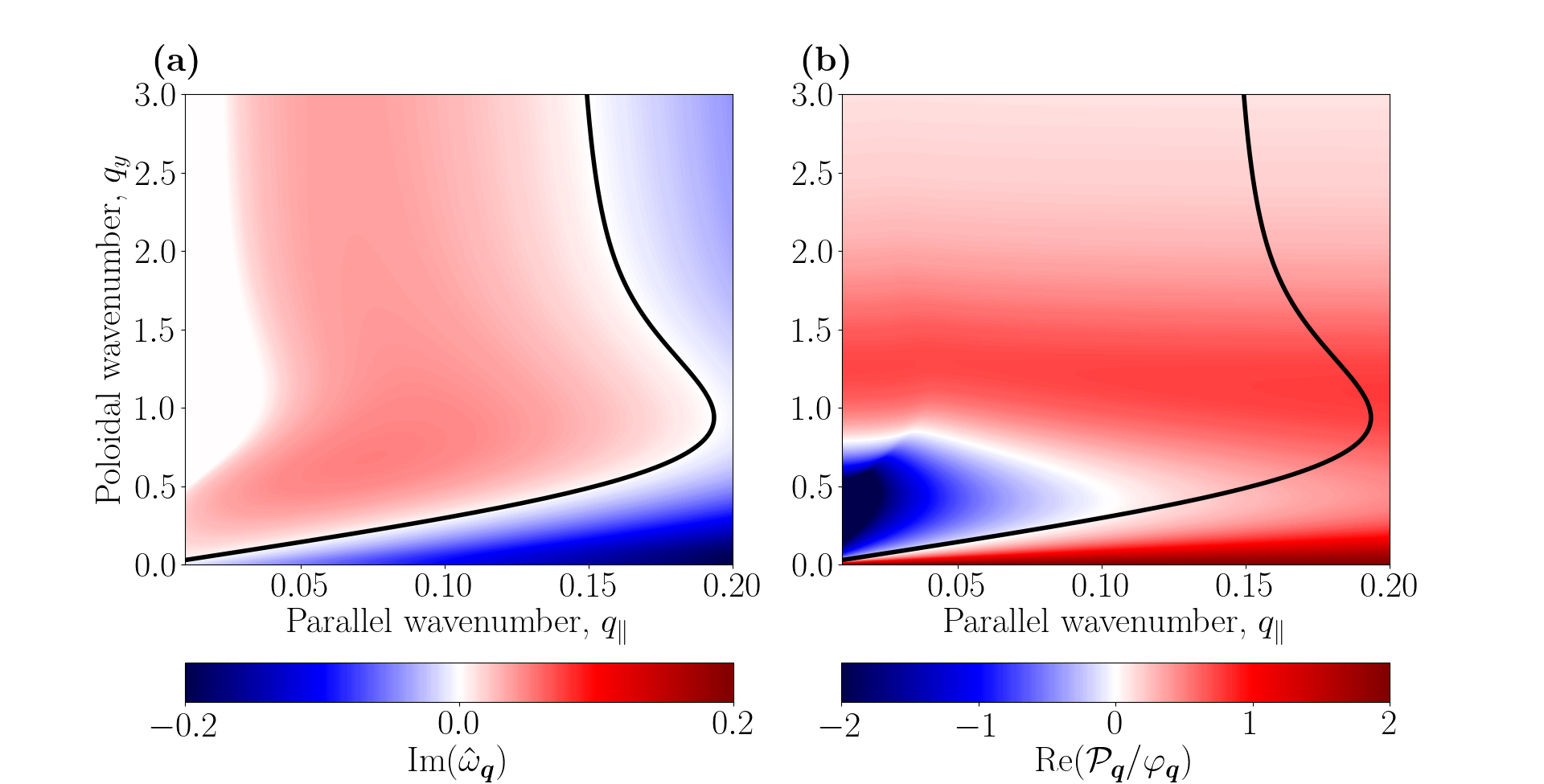}
	\caption{ Same as \cref{fig_stressure_tau001_kx0}, but for \(\tau = 1\). Evidently, the small-scale sITG modes at \(q_y \gg 1\) have \(\reinline{\gp_\vq/\phinorm_\vq} > 0\), i.e., they support the ZFs. However, it is no longer obvious whether the dominant sITG modes support or destroy the ZFs. }
	\label{fig_stressure_tau1_kx0}
\end{figure}

We determine \(\reinline{\gp_\vq/\phinorm_\vq}\) for the sITG modes by solving  for \(\zeta\) using the dispersion relation \cref{eq_slab_disp_kinetic} and then substituting for \(\zeta\) into \cref{eq_gp}. As we can see in figures~\ref{fig_stressure_tau001_kx0}~and~\ref{fig_stressure_tau1_kx0}, both for \(\tau \ll 1\) and \(\tau \sim 1\), the small-scale sITG instability drives the ZFs. The role of the dominant sITG modes (i.e., those with the largest growth rate) in the cold-ion limit is clear --- they support the ZFs, just as they do in the fluid model. However, as \(\tau\) approaches \(1\), it is difficult to discern their effect on the ZFs without the knowledge of the spectrum of the fluctuation amplitudes at the relevant wavenumbers. However, it appears that the number of ZF-destabilising modes increases with increasing \(\tau\). This suggests that the Dimits threshold might be sensitive to the value of \(\tau\). 
\FloatBarrier

\clearpage

\bibliographystyle{jpp}

\bibliography{bib}

\begin{thebibliography}{38}
\expandafter\ifx\csname natexlab\endcsname\relax\def\natexlab#1{#1}\fi
\def\au#1{#1} \def\ed#1{#1} \def\yr#1{#1}\def\at#1{#1}\def\jt#1{\textit{#1}}
  \def\bt#1{#1}\def\bvol#1{\textbf{#1}} \def\vol#1{#1} \def\pg#1{#1}
  \def\publ#1{#1}\def\arxiv#1{#1}\def\org#1{#1}\def\st#1{\textit{#1}}

\bibitem[Barnes {\em et~al.\/}(2011)Barnes, Parra \& Schekochihin]{barnes2011}
{\sc \au{Barnes, M}, \au{Parra, FI} \& \au{Schekochihin, AA}} \yr{2011}
  \at{Critically balanced ion temperature gradient turbulence in fusion
  plasmas}.  \jt{{Phys. Rev. Lett.}}  \bvol{107},  \pg{115003}.

\bibitem[Braginskii(1965)]{braginskii65}
{\sc \au{Braginskii, SI}} \yr{1965}  \at{Transport processes in a plasma}.
  \jt{Reviews of Plasma Physics}  \bvol{1},  \pg{205}.

\bibitem[Coppi {\em et~al.\/}(1967)Coppi, Rosenbluth \& Sagdeev]{coppi67}
{\sc \au{Coppi, B}, \au{Rosenbluth, MN} \& \au{Sagdeev, RZ}} \yr{1967}
  \at{Instabilities due to temperature gradients in complex magnetic field
  configurations}.  \jt{Phys. Fluids}  \bvol{10},  \pg{582}.

\bibitem[Cowley {\em et~al.\/}(1991)Cowley, Kulsrud \& Sudan]{cowley91}
{\sc \au{Cowley, SC}, \au{Kulsrud, RM} \& \au{Sudan, R}} \yr{1991}
  \at{Considerations of ion‐temperature‐gradient‐driven turbulence}.
  \jt{Phys. Fluids B}  \bvol{3},  \pg{2767}.

\bibitem[Diamond {\em et~al.\/}(2005)Diamond, Itoh, Itoh \& Hahm]{diamond2005}
{\sc \au{Diamond, PH}, \au{Itoh, S-I}, \au{Itoh, K} \& \au{Hahm, TS}} \yr{2005}
   \at{Zonal flows in plasma{\textemdash}a review}.  \jt{{Plasma Phys. Control.
  Fusion}}  \bvol{47},  \pg{R35}.

\bibitem[Dif-Pradalier {\em et~al.\/}(2010)Dif-Pradalier, Diamond, Grandgirard,
  Sarazin, Abiteboul, Garbet, Ghendrih, Strugarek, Ku \& Chang]{pradalier2010}
{\sc \au{Dif-Pradalier, G}, \au{Diamond, PH}, \au{Grandgirard, V}, \au{Sarazin,
  Y}, \au{Abiteboul, J}, \au{Garbet, X}, \au{Ghendrih, Ph}, \au{Strugarek, A},
  \au{Ku, S} \& \au{Chang, CS}} \yr{2010}  \at{On the validity of the local
  diffusive paradigm in turbulent plasma transport}.  \jt{Phys. Rev. E}
  \bvol{82},  \pg{025401}.

\bibitem[Dif-Pradalier {\em et~al.\/}(2017)Dif-Pradalier, Hornung, Garbet,
  Ghendrih, Grandgirard, Latu \& Sarazin]{difpradalier2017}
{\sc \au{Dif-Pradalier, G}, \au{Hornung, G}, \au{Garbet, X}, \au{Ghendrih, Ph},
  \au{Grandgirard, V}, \au{Latu, G} \& \au{Sarazin, Y}} \yr{2017}  \at{The
  {ExB} staircase of magnetised plasmas}.  \jt{Nuclear Fusion}  \bvol{57},
  \pg{066026}.

\bibitem[Dorland \& Hammett(1993)]{dorland93}
{\sc \au{Dorland, W} \& \au{Hammett, GW}} \yr{1993}  \at{Gyrofluid turbulence
  models with kinetic effects}.  \jt{Phys. Fluids B}  \bvol{5},  \pg{812}.

\bibitem[Dorland {\em et~al.\/}(2000)Dorland, Jenko, Kotschenreuther \&
  Rogers]{dorland2000}
{\sc \au{Dorland, W}, \au{Jenko, F}, \au{Kotschenreuther, M} \& \au{Rogers,
  BN}} \yr{2000}  \at{Electron temperature gradient turbulence}.  \jt{Phys.
  Rev. Lett.}  \bvol{85},  \pg{5579}.

\bibitem[Drake {\em et~al.\/}(1988)Drake, Guzdar \& Hassam]{drake88}
{\sc \au{Drake, JF}, \au{Guzdar, PN} \& \au{Hassam, AB}} \yr{1988}
  \at{Streamer formation in plasma with a temperature gradient}.  \jt{Phys.
  Rev. Lett.}  \bvol{61},  \pg{2205}.

\bibitem[Fox {\em et~al.\/}(2017)Fox, van Wyk, Field, Ghim, Parra, Schekochihin
  \& Team]{mfox_sym}
{\sc \au{Fox, MFJ}, \au{van Wyk, F}, \au{Field, AR}, \au{Ghim, Y-c}, \au{Parra,
  FI}, \au{Schekochihin, AA} \& \au{Team, MAST}} \yr{2017}  \at{Symmetry
  breaking in {MAST} plasma turbulence due to toroidal flow shear}.  \jt{Plasma
  Phys. Control. Fusion}  \bvol{59},  \pg{034002}.

\bibitem[Frei {\em et~al.\/}(2022)Frei, Hoffmann \& Ricci]{frei2022}
{\sc \au{Frei, BJ}, \au{Hoffmann, ACD} \& \au{Ricci, P}} \yr{2022}  \at{Local
  gyrokinetic collisional theory of the ion-temperature gradient mode}.
  \jt{{J.~Plasma Phys.}}  \bvol{88},  \pg{905880304}.

\bibitem[Fried \& Conte(1961)]{friedconte61}
{\sc \au{Fried, BD} \& \au{Conte, SD}} \yr{1961} {\em The Plasma Dispersion
  Function\/}.  \publ{Academic Press}.

\bibitem[Guzdar {\em et~al.\/}(1983)Guzdar, Chen, Tang \& Rutherford]{guzdar83}
{\sc \au{Guzdar, PN}, \au{Chen, L}, \au{Tang, WM} \& \au{Rutherford, PH}}
  \yr{1983}  \at{Ion‐temperature‐gradient instability in toroidal plasmas}.
   \jt{Phys. Fluids}  \bvol{26},  \pg{673}.

\bibitem[Hallenbert \& Plunk(2021)]{hallenbert2021}
{\sc \au{Hallenbert, A} \& \au{Plunk, GG}} \yr{2021}  \at{Predicting the dimits
  shift through reduced mode tertiary instability analysis in a strongly driven
  gyrokinetic fluid limit}.  \jt{{J.~Plasma Phys.}}  \bvol{87},
  \pg{905870508}.

\bibitem[Hammett {\em et~al.\/}(1993)Hammett, Beer, Dorland, Cowley \&
  Smith]{hammett93}
{\sc \au{Hammett, GW}, \au{Beer, MA}, \au{Dorland, W}, \au{Cowley, SC} \&
  \au{Smith, SA}} \yr{1993}  \at{Developments in the gyrofluid approach to
  tokamak turbulence simulations}.  \jt{{Plasma Phys. Control. Fusion}}
  \bvol{35},  \pg{973}.

\bibitem[Ivanov {\em et~al.\/}(2020)Ivanov, Schekochihin, Dorland, Field \&
  Parra]{ivanov2020}
{\sc \au{Ivanov, PG}, \au{Schekochihin, AA}, \au{Dorland, W}, \au{Field, AR} \&
  \au{Parra, FI}} \yr{2020}  \at{Zonally dominated dynamics and {Dimits}
  threshold in curvature-driven {ITG} turbulence}.  \jt{{J.~Plasma Phys.}}
  \bvol{86},  \pg{855860502}.

\bibitem[Kinsey {\em et~al.\/}(2006)Kinsey, Waltz \& Candy]{kinsey2006}
{\sc \au{Kinsey, JE}, \au{Waltz, RE} \& \au{Candy, J}} \yr{2006}  \at{The
  effect of safety factor and magnetic shear on turbulent transport in
  nonlinear gyrokinetic simulations}.  \jt{Physics of Plasmas}  \bvol{13},
  \pg{022305}.

\bibitem[Kobayashi \& Rogers(2012)]{kobayashi2012}
{\sc \au{Kobayashi, S} \& \au{Rogers, BN}} \yr{2012}  \at{The quench rule,
  {Dimits} shift, and eigenmode localization by small-scale zonal flows}.
  \jt{Phys. Plasmas}  \bvol{19},  \pg{012315}.

\bibitem[Kotschenreuther {\em et~al.\/}(1995)Kotschenreuther, Rewoldt \&
  Tang]{kotschenreuther95}
{\sc \au{Kotschenreuther, M}, \au{Rewoldt, G} \& \au{Tang, WM}} \yr{1995}
  \at{Comparison of initial value and eigenvalue codes for kinetic toroidal
  plasma instabilities}.  \jt{Comput. Phys. Commun.}  \bvol{88},  \pg{128}.

\bibitem[Newton {\em et~al.\/}(2010)Newton, Cowley \& Loureiro]{newton2010}
{\sc \au{Newton, SL}, \au{Cowley, SC} \& \au{Loureiro, NF}} \yr{2010}
  \at{Understanding the effect of sheared flow on microinstabilities}.
  \jt{{Plasma Phys. Control. Fusion}}  \bvol{52},  \pg{125001}.

\bibitem[Parra {\em et~al.\/}(2011)Parra, Barnes \& Peeters]{parra2011}
{\sc \au{Parra, FI}, \au{Barnes, M} \& \au{Peeters, AG}} \yr{2011}  \at{Up-down
  symmetry of the turbulent transport of toroidal angular momentum in
  tokamaks}.  \jt{Phys. Plasmas}  \bvol{18},  \pg{062501}.

\bibitem[Pogutse(1968)]{pogutse68}
{\sc \au{Pogutse, OP}} \yr{1968}  \at{Magnetic drift instability in a
  collisionless plasma}.  \jt{Plasma Physics}  \bvol{10},  \pg{649}.

\bibitem[Rath {\em et~al.\/}(2016)Rath, Peeters, Buchholz, Grosshauser,
  Migliano, Weikl \& Strintzi]{rath2016}
{\sc \au{Rath, F}, \au{Peeters, AG}, \au{Buchholz, R}, \au{Grosshauser, SR},
  \au{Migliano, P}, \au{Weikl, A} \& \au{Strintzi, D}} \yr{2016}
  \at{Comparison of gradient and flux driven gyro-kinetic turbulent transport}.
   \jt{Phys. Plasmas}  \bvol{23},  \pg{052309}.

\bibitem[Rath \& Sridhar(1992)]{rath92}
{\sc \au{Rath, S} \& \au{Sridhar, S}} \yr{1992}  \at{Core instability of
  elliptic drift vortices}.  \jt{Phys. Fluids B}  \bvol{4},  \pg{1367}.

\bibitem[Ricci {\em et~al.\/}(2006)Ricci, Rogers \& Dorland]{ricci2006}
{\sc \au{Ricci, P}, \au{Rogers, BN} \& \au{Dorland, W}} \yr{2006}
  \at{Small-scale turbulence in a closed-field-line geometry}.  \jt{Phys. Rev.
  Lett.}  \bvol{97},  \pg{245001}.

\bibitem[Rogers {\em et~al.\/}(2000)Rogers, Dorland \&
  Kotschenreuther]{rogersdorland2000}
{\sc \au{Rogers, BN}, \au{Dorland, W} \& \au{Kotschenreuther, M}} \yr{2000}
  \at{Generation and stability of zonal flows in ion-temperature-gradient mode
  turbulence}.  \jt{Phys. Rev. Lett.}  \bvol{85},  \pg{5336}.

\bibitem[Rudakov \& Sagdeev(1961)]{rudakov61}
{\sc \au{Rudakov, LI} \& \au{Sagdeev, RZ}} \yr{1961}  \at{On the instability of
  a nonuniform rarefied plasma in a strong magnetic field}.  \jt{Dokl. Akad.
  Nauk SSSR}  \bvol{138},  \pg{581}.

\bibitem[Schekochihin {\em et~al.\/}(2009)Schekochihin, Cowley, Dorland,
  Hammett, Howes, Quataert \& Tatsuno]{schekochihingk2009}
{\sc \au{Schekochihin, AA}, \au{Cowley, SC}, \au{Dorland, W}, \au{Hammett, GW},
  \au{Howes, GG}, \au{Quataert, E} \& \au{Tatsuno, T}} \yr{2009}
  \at{Astrophysical gyrokinetics: kinetic and fluid turbulent cascades in
  magnetized weakly collisional plasmas}.  \jt{Astrophys. J. Suppl. S.}
  \bvol{182},  \pg{310}.

\bibitem[Smolyakov {\em et~al.\/}(2002)Smolyakov, Yagi \&
  Kishimoto]{smolyakov2002}
{\sc \au{Smolyakov, AI}, \au{Yagi, M} \& \au{Kishimoto, Y}} \yr{2002}
  \at{Short wavelength temperature gradient driven modes in tokamak plasmas}.
  \jt{Phys. Rev. Lett.}  \bvol{89},  \pg{125005}.

\bibitem[St-Onge(2017)]{stonge2017}
{\sc \au{St-Onge, DA}} \yr{2017}  \at{{On non-local energy transfer via zonal
  flow in the Dimits shift}}.  \jt{{J.~Plasma Phys.}}  \bvol{83},
  \pg{905830504}.

\bibitem[Villard {\em et~al.\/}(2013)Villard, Angelino, Bottino, Brunner,
  Jolliet, McMillan, Tran \& Vernay]{villard2013}
{\sc \au{Villard, L}, \au{Angelino, P}, \au{Bottino, A}, \au{Brunner, S},
  \au{Jolliet, S}, \au{McMillan, BF}, \au{Tran, TM} \& \au{Vernay, T}}
  \yr{2013}  \at{Global gyrokinetic ion temperature gradient turbulence
  simulations of {ITER}}.  \jt{Plasma Phys. Control. Fusion}  \bvol{55},
  \pg{074017}.

\bibitem[Watson(1966)]{watson66}
{\sc \au{Watson, GN}} \yr{1966} {\em A Treatise on the theory of Bessel
  functions\/}, 2nd edn.  \publ{CUP}.

\bibitem[van Wyk {\em et~al.\/}(2017)van Wyk, Highcock, Field, Roach,
  Schekochihin, Parra \& Dorland]{vanwyk2017}
{\sc \au{van Wyk, F}, \au{Highcock, EG}, \au{Field, AR}, \au{Roach, CM},
  \au{Schekochihin, AA}, \au{Parra, FI} \& \au{Dorland, W}} \yr{2017}
  \at{{Ion-scale turbulence in MAST: anomalous transport, subcritical
  transitions, and comparison to BES measurements}}.  \jt{{Plasma Phys.
  Control. Fusion}}  \bvol{59},  \pg{114003}.

\bibitem[van Wyk {\em et~al.\/}(2016)van Wyk, Highcock, Schekochihin, Roach,
  Field \& Dorland]{vanwyk2016}
{\sc \au{van Wyk, F}, \au{Highcock, EG}, \au{Schekochihin, AA}, \au{Roach, CM},
  \au{Field, AR} \& \au{Dorland, W}} \yr{2016}  \at{Transition to subcritical
  turbulence in a tokamak plasma}.  \jt{{J.~Plasma Phys.}}  \bvol{82},
  \pg{905820609}.

\bibitem[Zhu {\em et~al.\/}(2018)Zhu, Zhou \& Dodin]{zhu2018_tertiary}
{\sc \au{Zhu, H}, \au{Zhou, Y} \& \au{Dodin, IY}} \yr{2018}  \at{On the
  {Rayleigh–Kuo} criterion for the tertiary instability of zonal flows}.
  \jt{Phys. Plasmas}  \bvol{25},  \pg{082121}.

\bibitem[Zhu {\em et~al.\/}(2020{\natexlab{{\em a\/}}})Zhu, Zhou \&
  Dodin]{zhu2019_dimits}
{\sc \au{Zhu, H}, \au{Zhou, Y} \& \au{Dodin, IY}} \yr{2020{\natexlab{{\em
  a\/}}}}  \at{Theory of the tertiary instability and the {Dimits} shift from
  reduced drift-wave models}.  \jt{Phys. Rev. Lett.}  \bvol{124},  \pg{055002}.

\bibitem[Zhu {\em et~al.\/}(2020{\natexlab{{\em b\/}}})Zhu, Zhou \&
  Dodin]{zhu2020jpp}
{\sc \au{Zhu, H}, \au{Zhou, Y} \& \au{Dodin, IY}} \yr{2020{\natexlab{{\em
  b\/}}}}  \at{Theory of the tertiary instability and the dimits shift within a
  scalar model}.  \jt{{J.~Plasma Phys.}}  \bvol{86},  \pg{905860405}.

\end{thebibliography}

\end{document}